\documentclass[journal,twocolumn]{IEEEtran} 
\pdfoutput=1
\usepackage{graphicx} \graphicspath{{./figures/}} \DeclareGraphicsExtensions{.pdf}
\usepackage[cmex10]{amsmath} 
\usepackage{amssymb} 
\usepackage{amsthm} 
\usepackage{flushend} 
\usepackage{mathtools} 
\usepackage{mdwmath} 
\usepackage{mdwtab} 
\usepackage{eqparbox}  \allowdisplaybreaks
\usepackage{algorithm,algorithmic} 
\usepackage[tight,footnotesize]{subfigure} \DeclarePairedDelimiter{\ceil}{\lceil}{\rceil} 
\newcommand{\argmin}{\operatornamewithlimits{arg\,min}} 
\newcommand{\argmax}{\operatornamewithlimits{arg\,max}} 
\newcommand{\abs}[1]{\left\lvert{#1}\right\rvert}

\newcommand{\norm}[1]{\left\lVert{#1}\right\rVert}

\newcommand{\inProd}[2]{\langle{#1},{#2}\rangle} 
\newcommand{\g}{\mid}

\newcommand{\mc}[1]{\mathcal{#1}} 
\newcommand{\mbb}[1]{\mathbb{#1}} 
\newcommand{\mbf}[1]{\mathbf{#1}} 
\newcommand{\st}[2]{\stackrel{#1}{#2}}

\newcommand{\pb}[1]{\left({#1}\right)}

\newcommand{\sqb}[1]{\left[{#1}\right]}

\newcommand{\cb}[1]{\left\{{#1}\right\}}

\newcommand{\nn}{\nonumber}

\newtheorem{theorem}{Theorem} 
\newtheorem{lemma}{Lemma} 
\newcommand{\Expec}[1]{\textbf{E}\left({#1}\right)}

\newcounter{algocount}
\def\mycaption{\relax}

\newenvironment{algo}[1]{\refstepcounter{algocount}
\renewcommand\thetable{\arabic{algocount}}
\renewcommand{\tablename}{Algorithm}
    \if\relax\detokenize{#1}\relax
    \gdef\mycaption{\relax}
    \else
    \gdef\mycaption{#1}
    \fi
    \addtocounter{table}{-1}
\begin{table}[ht]}{ 
 \caption{\mycaption}
\end{table}}

\begin{document} 
\title{High Resolution Radar Sensing with Compressive Illumination} 
\author{\IEEEauthorblockN{Nithin Sugavanam, Siddharth Baskar and Emre Ertin}\\
\IEEEauthorblockA{Electrical and Computer Engineering Department, \\The Ohio State University}} 
\maketitle 
\begin{abstract}
We present a compressive radar design that combines multitone linear frequency modulated (LFM) waveforms in the transmitter with a classical stretch processor and sub-Nyquist sampling in the receiver. The proposed compressive illumination scheme has fewer random elements resulting in reduced storage and complexity for implementation than previously proposed compressive radar designs based on stochastic waveforms. We analyze this illumination scheme for the task of a joint range-angle of arrival estimation in the
multi-input and multi-output (MIMO) radar system. We present recovery guarantees for the proposed illumination technique. We show that for a sufficiently large number of modulating tones, the system achieves high-resolution in range and successfully recovers the range and angle-of-arrival of targets in a sparse scene. Furthermore,
we present an algorithm that estimates the target range, angle of arrival, and scattering coefficient in the continuum. Finally, we present simulation results to illustrate the recovery performance as a function of system parameters. 
\end{abstract}
\IEEEpeerreviewmaketitle

\section{Introduction} Radar imaging systems acquire information about the scene of interest by transmitting pulsed waveforms and analyzing the received backscatter energy to form an estimate of the range, angle of arrival, Doppler velocity, and amplitude of the reflectors in the scene. These range profiles from multiple pulses and multiple antenna elements can be processed jointly to solve many inference tasks, including detection, tracking, and classification~\cite{RadarSigProc}. We analyze a coherent MIMO radar system with closely separated antennas, such that the angle of arrival of each scattering center in the scene is approximately the same for all phase-centers. The main advantage of coherent MIMO radar is its ability to synthesize a sizeable virtual array with fewer antenna elements for improved spatial processing. Additionally, MIMO radar systems with multiple transmit and receive elements employing independent waveforms on transmitter provide spatial processing gains by exploiting the diversity of channels between the target and radar~\cite{MIMORadarSigProc,Ertin_SDR_frontEnd_2015}. 

\begin{figure}
	\centering 
	\includegraphics[width=\linewidth]{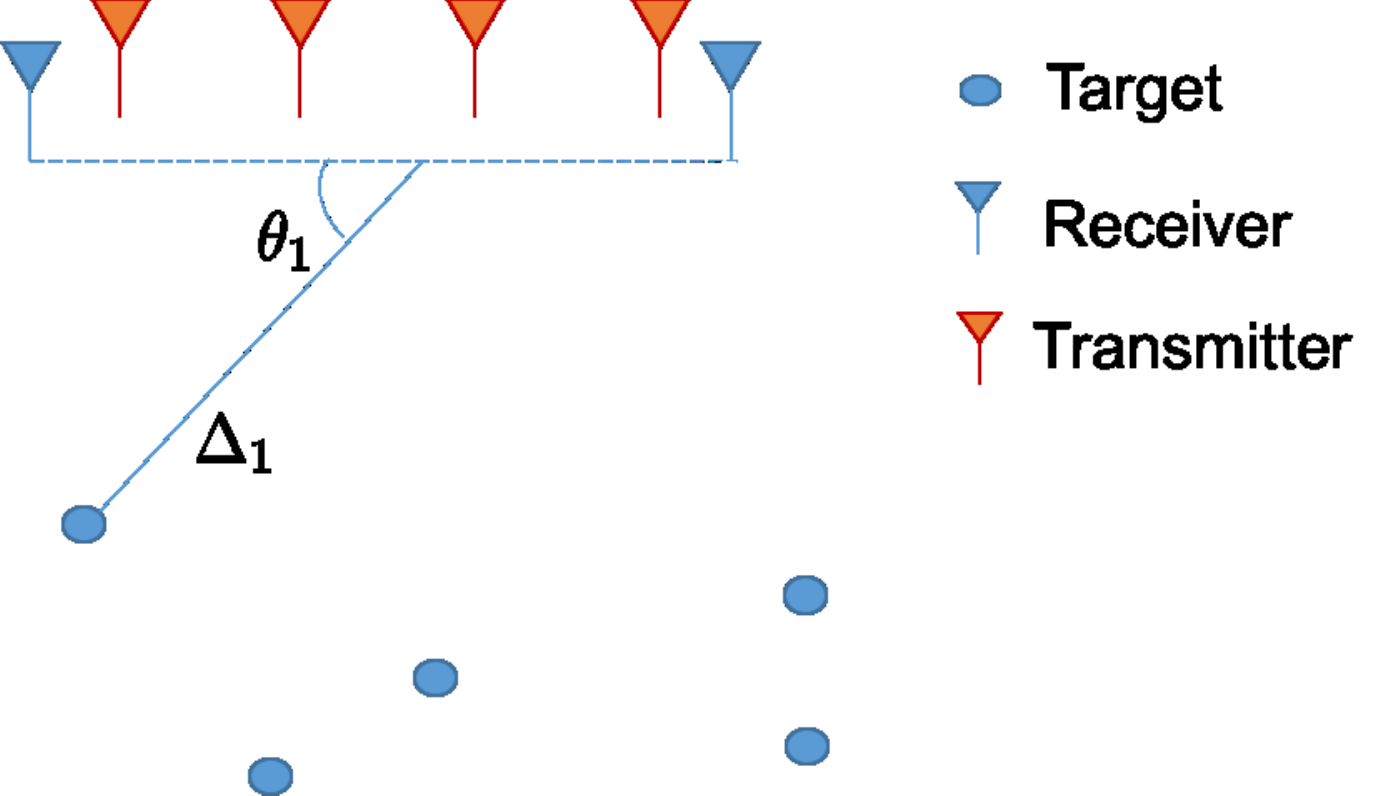}
\caption{\label{fig:ProblemSetup}Illustration of the MIMO system with sparse
targets. } 
\end{figure}
This work estimates the range, angle of arrival,  and amplitude of reflectors in the scene using a MIMO radar system with $N_T$ transmitters and $N_R$ receivers. The $i^{th}$ transmitter utilizes a modulated wideband pulse $\phi_i(t)$ of bandwidth $B$ and pulse duration $\tau$. By assuming that the support of the observed delays are known to
lie on an interval $T_u$ (termed as range swath in radar literature), the
received signal at receiver $l$ can be expressed as 
\begin{align}\label{eq:basicModel} 
	y_l(t) = \sum_{\st{k=1}{i=1}}^{\st{k=K}{ i=N_T}} \alpha_{l,i}(\theta_k) \phi_i(t-\Delta_k) x_k + w_l(t), 
\end{align}
where $w_l(t)$ is the additive receiver noise, $\Delta_k$ is the round-trip delay time, $x_k$ is the complex scattering coefficient of $k^{th}$ target, and $\alpha_{l,i}(\theta)$ is the array factor for the $l^{th}$ receiver and the $i^{th}$ transmitter, which is a function of the angle of arrival $\theta_k$ as shown in figure~\ref{fig:ProblemSetup}. For the case of a single transmitter and receiver setup, the problem simplifies to delay estimation and the received signal $y(t)$ is as shown below 
\begin{align}
	y(t) = \sum_{k=1}^{k=K} \phi(t-\Delta_k) x_k + w(t), 
\end{align}
where $w(t)$ is the receiver noise. Conventionally, matched filtering is
performed to estimate the unknown parameters associated with the targets in the scene. However, the matched filter's implementation requires Nyquist rate sampling, which is proportional to the bandwidth of the transmitted signal. This sampling rate, in turn, severely limits the resolution and dynamic range of the Analog to Digital Converter (ADC) needed for direct digital implementation of the radar since the resolution of the ADC is inversely proportional to the maximum sampling rate~\cite{walden1999}. An approximation of the matched filter can be implemented in the analog domain where the number of samples is reduced to $N = BT_u$ to span the support of delays in the unambiguous time interval of $T_u$. For example, if a linear frequency modulated waveform (LFM) $\phi(t)=\exp\pb{j{B}/{\tau}t^2}$ is used on transmit, this approximation of matched filter is implemented by mixing the received signal with a reference LFM waveform using an analog mixer, and subsequently, low-pass filtering the mixer output. 
\begin{figure}
	\centering 
	\includegraphics[width=\linewidth]{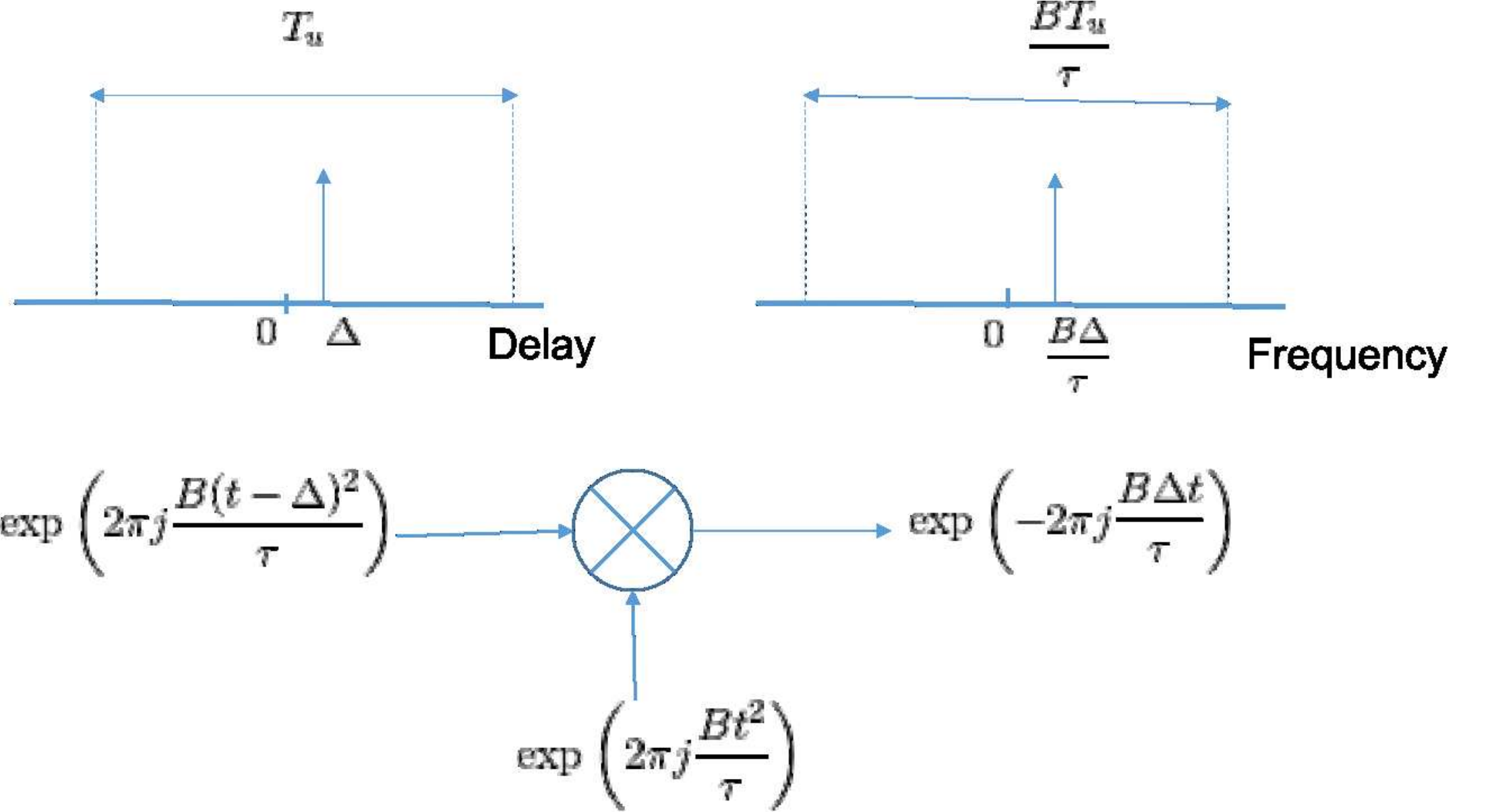}
\caption{\label{fig:StretchProc} Effect of stretch processing, which transforms
the task of delay estimation to a task of spectral estimation. } 
\end{figure}
At the receiver output, the waveform delayed by $\Delta$ 
appears as a sinusoidal tone whose frequency is given by $B\Delta$ as shown in
figure~\ref{fig:StretchProc}. This pre-processing step is termed as stretch
processing \cite{RadarSigProc, Middleton_DechirpMatchedFilter_2012} and can
result in a substantially reduced sampling rate for the ADC used in the receiver if the delay support $T_u$ is smaller than the pulse length $\tau$. Furthermore, the received signal at the stretch processor's output can be written as \[y(t) =
\sum_{n=1}^K x_n \exp\pb{j B \Delta_k t/\tau } +w(t).\] If the scene has $K$ targets much less than the number of delay bins $N$, well-known
results~\cite{DonohoCS_2006,RobustUncertainty_Candes_2006} from compressive
sensing (CS) show that successful estimation of target locations from
sub-Nyquist samples is possible if the number of measurements $M$ scale linearly with the number of targets $K\log N$ up to a logarithmic factor. Similarly, results from \cite{matchedFilter_CS_Romberg_2013} show that
a system that utilizes the LFM waveform for time of arrival estimation,
$\mc{O}(K \log N)$ randomly chosen measurements are sufficient for estimating
the locations and scattering coefficients of $K$ targets. 

Furthermore, there are numerous tractable algorithms,
 with provable performance guarantees, that are either based on convex
relaxation on the discretized space
\cite{Tao_2007_DantzigSel,Tibshirani1996_LASSO}, or the continuous parameter space
in \cite{CS_offGrid_Tang,GridlessMethods_Xie_2015}, or greedy methods
\cite{SignalRecOMP_Tropp_2007,CoSaMP_Tropp_2009} to solve the linear inverse
problem. Motivated by these advances, compressed sensing techniques have been
applied to a variety of problems in Radar~\cite{Potter2010,Eldar_CS_radar:2018,BaranuikRadar_2007}. The problem of range profile
estimation~\cite{TimeDelay_unionSubspaces_Eldar_2010} is solved using filter-banks to acquire low-rate sub-Nyquist samples. In addition to acquiring low-rate samples, the problem of waveform design using
frequency hopping codes for estimation in range, Doppler velocity and angle
domain is solved in~\cite{Chen_MIMO_2008} using mutual coherence as the objective. Similarly, the waveform design problem has been studied in~\cite{matrixDesignMIMO} using a multi-objective
optimization of a combination of mutual coherence and signal to interference
ratio. The conditions for successful recovery of target parameters for single pulse systems utilizing stochastic waveforms are established in~\cite{HighResRadar_Strohmer_2009}. These results are extended to single pulse multiple transmit and receive system for range, Doppler-velocity and azimuth estimation
and target detection in~\cite{SparseMIMORadar,KerdockCodes,Rauhut_SparseMIMO_2015,RemSense_Rauhut_2014}. A similar guarantee for successful estimation of range, angle of arrival, and Doppler-velocity using stepped frequency multi-pulse MIMO radar in each transmitter has been presented in~\cite{Petropulu_CSMIMO_2012,Petropulu_RIP_2014}. A parallel research
thrust~\cite{spatialMIMOCS_Eldar_2012} provided an average case recovery
guarantee for the problem of angle of arrival estimation with randomly located
antenna elements, under the idealized assumption of orthogonality between
received waveform from different range bins. Furthermore, frequency division multiple access based waveforms with sub-Nyquist sampling strategies in fast and slow-time are employed in~\cite{subNyquistFDMA_MIMO_Eldar_2016,subNyquistRadarDoppler_Eldar_2014,eldar_tensor_CS_radar:2019} for estimating the range, angle of arrival and Doppler velocity. The problem of sub-sampling in
array elements is also posed as a matrix completion problem in the grid-less
estimation setting in \cite{chi_MIMOMC_2013,MIMOMC_Bajwa_2015} and a condition
is established on the number of antenna elements that need to be observed in
order to recover the entire low-rank data matrix. 

Radar sensors have also been proposed and implemented 
that acquire compressed samples of the backscattered signal to solve the radar
imaging problem. A common approach based on pure random
waveforms~\cite{NoiseRadar_Shastry_2015} in the time domain and
\cite{RandomConvolution_Romberg2009} in the frequency domain have been
implemented and analyzed. Alternatively, the Xampling
framework~\cite{subNyquistRadar_Eldar_2014,subNyquistRadarDoppler_Eldar_2014}
has also been implemented as a practical system in
\cite{subNyquistFDMA_MIMO_Eldar_2016}. In addition to these schemes, the random
demodulator (RD) \cite{RMPI_Tropp_2010} has also been implemented in practice in
\cite{CSRadarHW_RMPI_2012}. This involves modulation of the received wide-band
signal with pseudo-random sequences followed by a low-pass filter or an
integrator to obtain low-rate sub-Nyquist samples. 
\begin{table*}
	[tbh] 
	\begin{center}
		\begin{tabular}
			{| p{0.324\linewidth} | p{0.22\linewidth} | p{0.21\linewidth} | p{0.13\linewidth} |} \hline Matrix Type of size $M \times N$ & Mutual Coherence & Spectral Norm& Reference \\
			\hline Random matrix $(N M)$ independent random entries& $2\sqrt{\frac{\log N}{M}}$ &$\sqrt{\frac{N}{M}} + 1$ &\cite{cai2011,candesPlanL1,GaussianOpNormConcentration} \\
			\hline Toeplitz block matrix with $(N+M)$ random entries &$\mc{O}\pb{\sqrt{\frac{\log N}{M}}}$ &$\mc{O}\pb{\sqrt{\frac{N}{M}}}$ &\cite{Random_Teoplitz_Bajwa2012} \\
			\hline LFM waveform modulated with $N_c\ll N$ randomly selected tones for single transmitter and receiver& $\mc{O}\pb{\sqrt{\frac{\log N}{M}}}$ &$\mc{O}\pb{ \sqrt{{\frac{N}{M} \log \pb{N}}}}$ & This work\\
			\hline 
		\end{tabular}
	\end{center}
	\caption{Measures that characterize sensing matrices.} 
\label{table:MatrixMeasures} \end{table*}
\begin{table*}
	[tbh] 
	\begin{center}
		\begin{tabular}
			{| p{0.25 
			\textwidth} | p{0.23 
			\textwidth} | p{0.23 
			\textwidth} | p{0.13\linewidth} |} \hline \multicolumn{4}{|p{0.85\linewidth}|}{Recovery Guarantees from noisy measurements with component-wise noise variance $\sigma^2$} \\
			\hline Matrix $\in \mbb{C}^{M \times N}$ & Sparsity condition & Minimum signal & Reference \\
			\hline Random matrix with $(N M)$ independent random entries & $\mc{O}\pb{\frac{M}{\log N}}$ & $\mc{O}\pb{\sigma\sqrt{2\log N }} $ &\cite{candesPlanL1} \\
			\hline Toeplitz block matrix with $(N+M)$ random entries &$\mc{O}\pb{\frac{M}{\log N}}$ & $\mc{O}\pb{\sigma\sqrt{2\log N }} $ &\cite{Random_Teoplitz_Bajwa2012} \\
			\hline LFM waveform modulated with $N_c\ll N$ randomly selected tones for single transmitter and receiver &$\mc{O}\pb{\frac{M}{\log N \log \pb{N + M}} }$ & $\mc{O}\pb{\sigma\sqrt{2\log N }} $ & This work\\
			\hline 
		\end{tabular}
	\end{center}
	\caption{Support recovery guarantees for different sensing matrices.} 
\label{table:SupportRec} \end{table*}
Here we pursue an alternative strategy based on the observation that LFM
waveforms and analog stretch processing convert the range estimation problem into an
equivalent sparse frequency spectrum estimation problem. It is well known that
uniform subsampling in this setting has poor
performance~\cite{SpectralCS_Duarte_2013} and therefore uniform subsampling of a
classical stretch processor cannot be used. While non-uniform random
sub-sampling possess good theoretical guarantees
~\cite{CS_offGrid_Tang,FastFourierSampling_Tropp_2008,SpectralCS_Duarte_2013},
its implementation with commercially available ADCs still require them to be
rated at the Nyquist rate to accommodate close samples. In this paper, we pursue
an alternative strategy and push randomization to the transmit signal structure
to obtain compressive measurements at the stretch processor output. The proposed
scheme can be readily implemented utilizing a small number of random parameters
in waveform generation and uniform sampling ADCs with high analog bandwidth on
the receiver. ADCs whose analog bandwidth exceed their maximum sampling rate by
several factors are readily available commercially and used routinely in
pass-band sampling. This compressive radar structure termed as compressive
illumination was first proposed in \cite{FreqDivWaveform} and utilized a linear
combination of sinusoids to modulate an LFM waveform at the transmitter with
randomly selected center frequencies, while maintaining the simple standard
stretch processing receiver structure. The output of the stretch processor
receiver is given by \[y(t) = \sum_{n=1}^{K} x_n \sum_{k=1}^{N_c}
\exp\pb{j\phi_{n,k}}\exp\pb{ j \pb{B \Delta_n + \omega_k} t },\] where
$\phi_{n,k}$ is a predetermined known complex phase, $N_c$ is the number of
tones modulating the LFM waveform with frequencies $\omega_k$. We observe that under the proposed
compressive sensor design each delayed copy of the transmitted waveform is
mapped to a multi-tone spectra with known structure. We show that this known
multi-tone frequency structure enables recovery from aliased time samples with
provable guarantees complementing previous work with a single transmitter and
receiver~\cite{RecGuarantees_Ertin_2015,sparseTargetMFMChirp} which has shown
good {\em empirical} performance using simulations and practical implementation
in \cite{SISO_cosera_2016}. Table~\ref{table:MatrixMeasures} summarizes the
characteristics of some well-studied random sensing schemes as well as our
proposed scheme. Table~\ref{table:SupportRec} summarizes the support recovery
guarantees for these random sensing schemes as well as our proposed scheme. The
rest of the paper is organized as follows, Section~\ref{sec:Model} states the
signal model, Section~\ref{sec:RecGuarantees} states the main recovery guarantee
and Section~\ref{sec:SimResults} provides simulation
verification of our theoretical results.

\section{System model }\label{sec:Model}
\subsection{System setup} 
\begin{figure}
	\centering 
	\includegraphics[trim=0 0 0 0,clip,width=\linewidth]{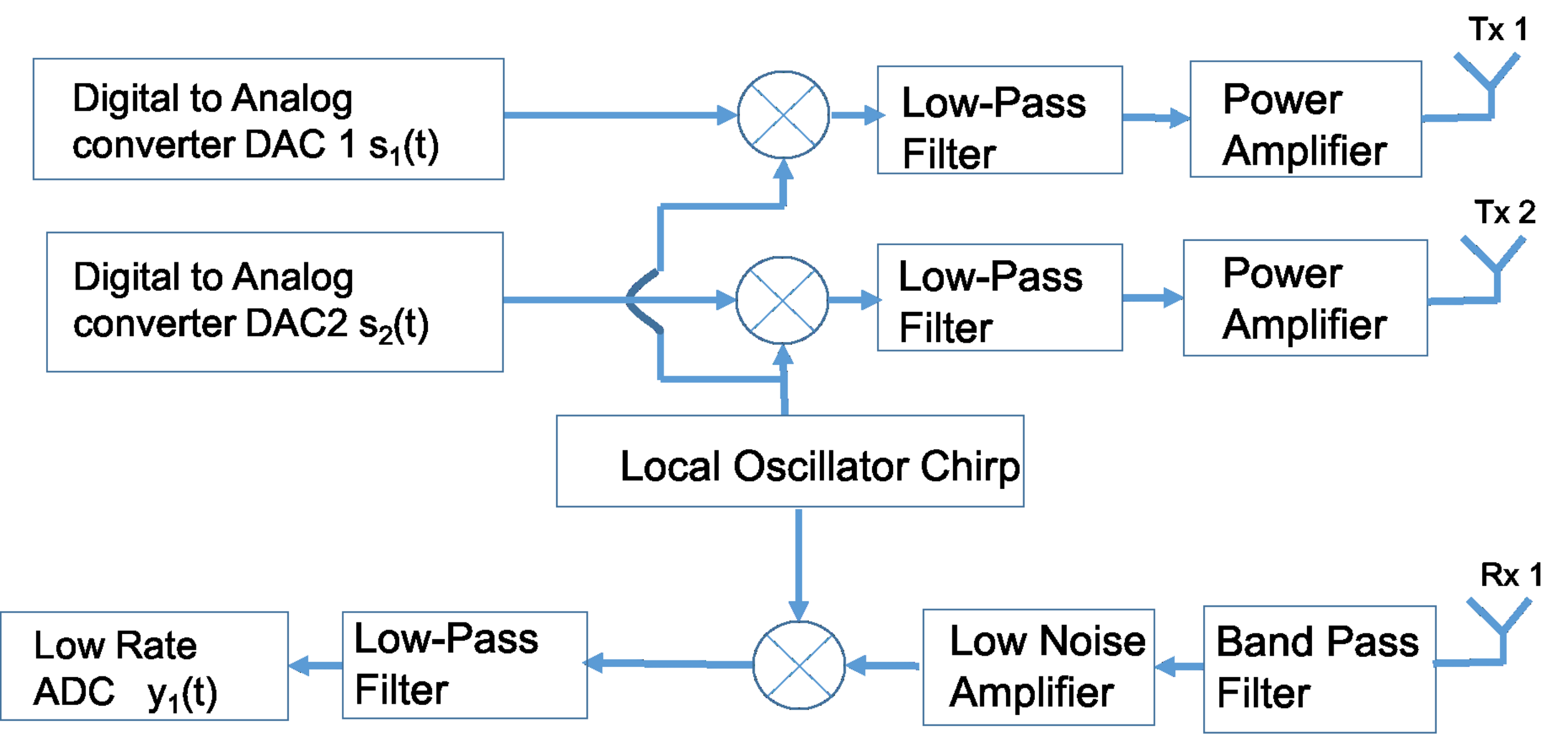} \caption{\label{fig:BlockDiagram} Block diagram of the transmitter and receiver. } 
\end{figure}
\begin{figure}
	\centering 
	\includegraphics[trim=0 0 0 0,clip,width=\linewidth]{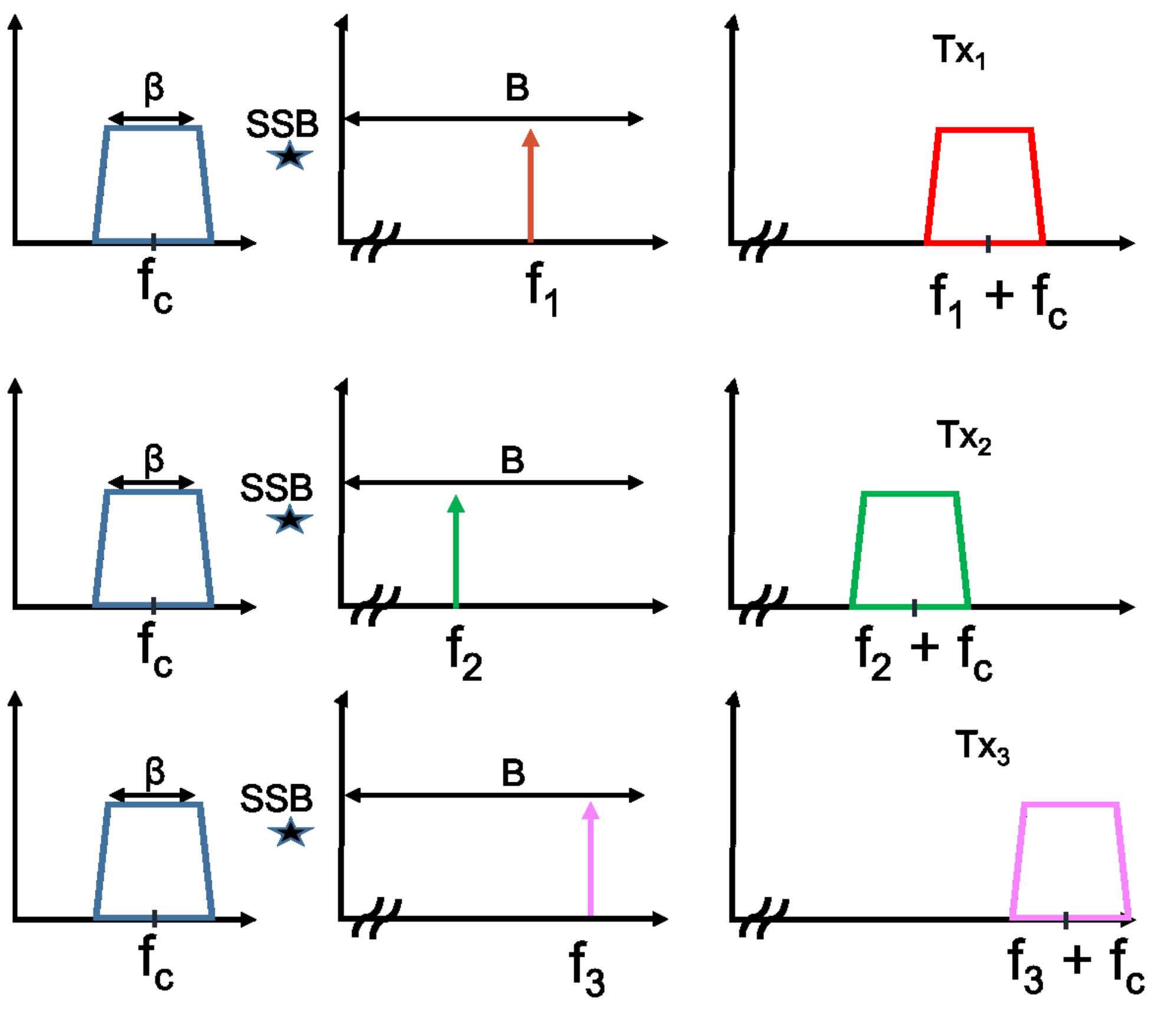} \caption{\label{fig:TxBlockDiagram} Spectra of the transmitted signal obtained by Single side-band (\emph{SSB}) modulation of the chirp waveform with center frequency $f_c$, and bandwidth $\beta$ with sinusoidal signals whose frequencies are chosen at random over a frequency range of $[0,B]$ such that each transmitter utilizes $1$ modulating tone. } 
\end{figure}
\begin{figure}
	[!t] \centering 
	\includegraphics[width=\linewidth]{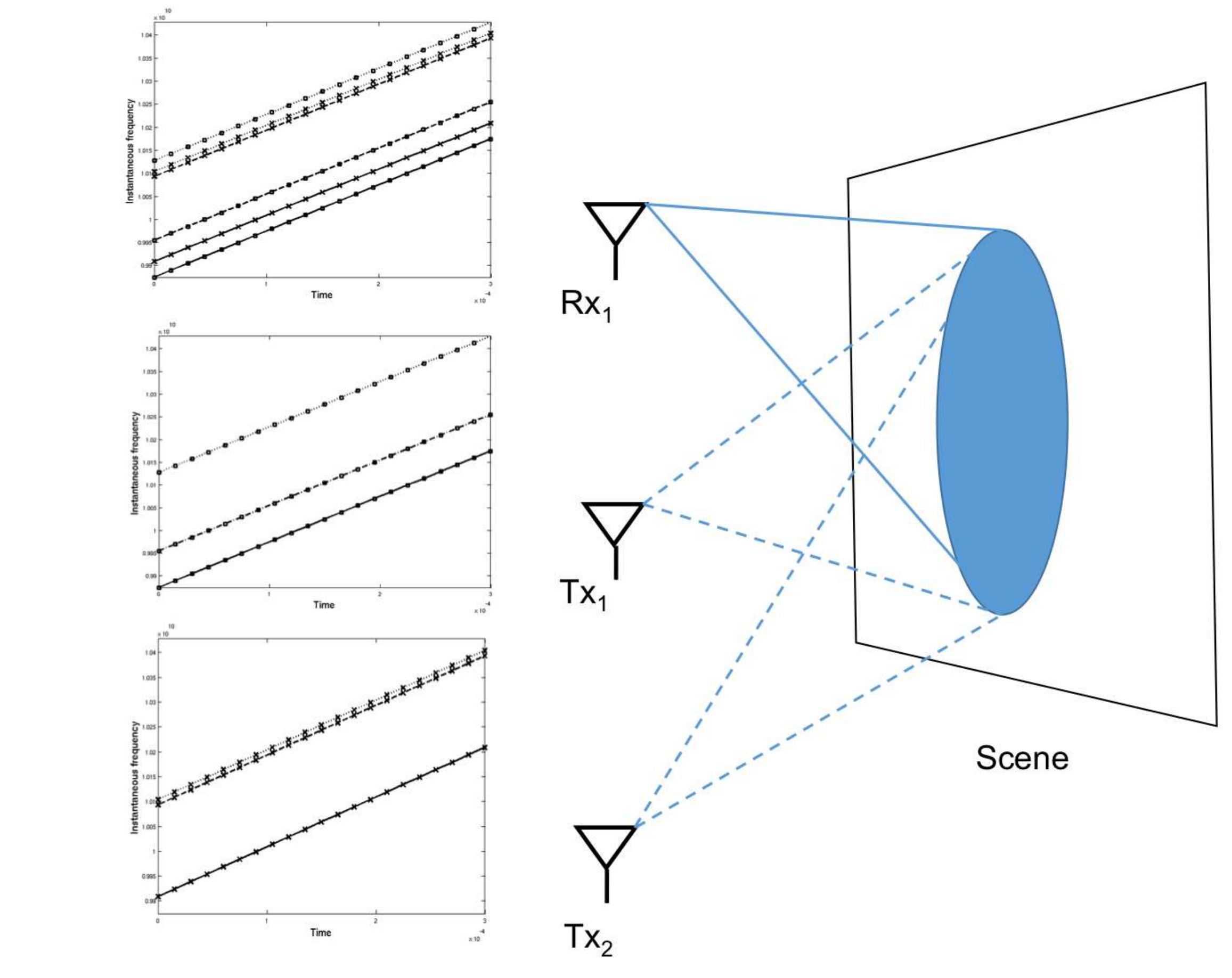} \caption{ The time-frequency representation of the waveform employed by two transmitters $Tx_1,Tx_2$ and received signal at one of the collocated receiver $Rx_1$ are shown. The transmitted waveforms are obtained as a result of modulation of the linear frequency modulated waveform by a set of sinusoids with randomly chosen frequencies. The received signal is a linear weighted combination of delayed versions of the transmitted signals. \label{fig:waveformLFM} } 
\end{figure}
\begin{figure}
	\centering 
	\includegraphics[trim=0 0 0 0,clip,width=\linewidth]{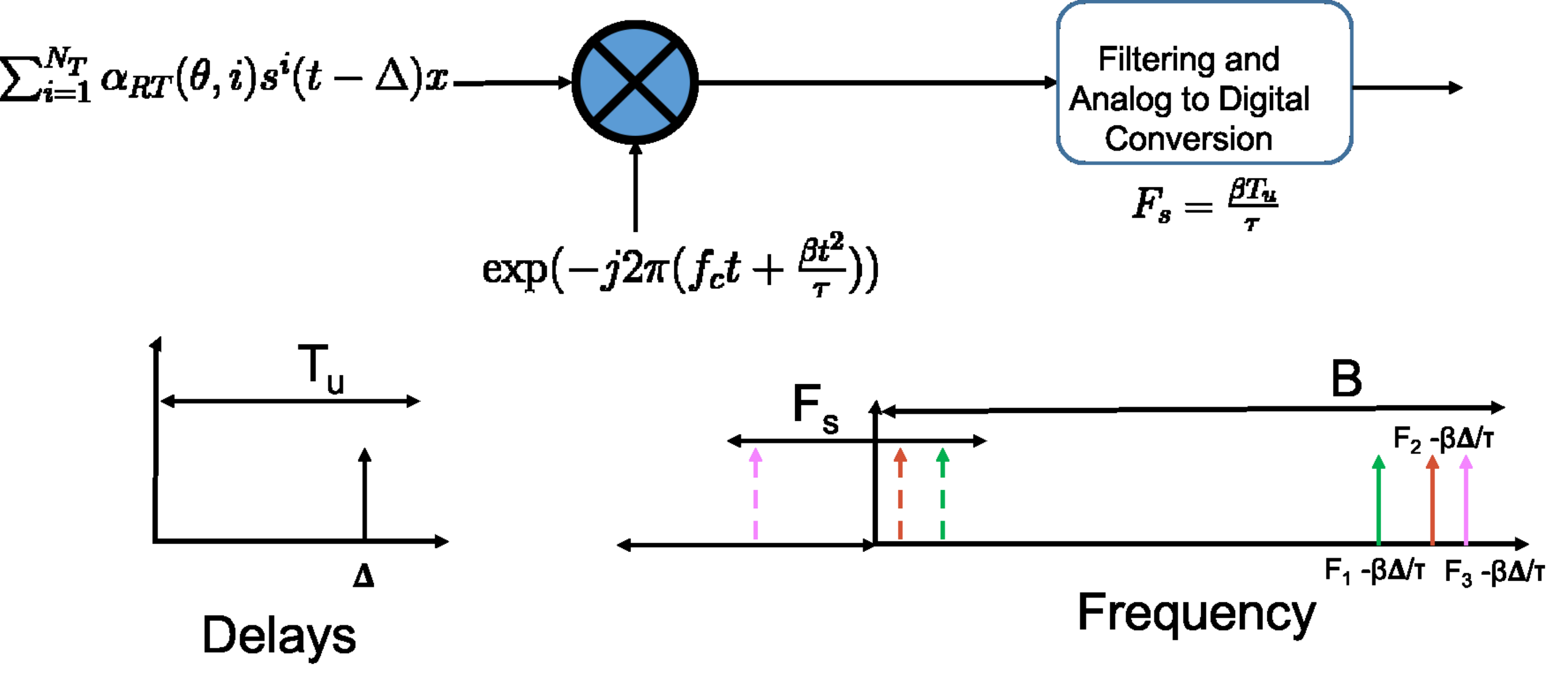} \caption{\label{fig:RxStretch} The figure illustrates the structure of the received signal due to a single scattering center located with range $c \Delta/2$ and angle $\theta$. The stretch processing at the receiver utilizes the transmitted LFM waveform prior to modulation. The effect of this operation recovers the modulating tones, shown in solid lines in the frequency domain, which are further modulated by a complex exponential with a frequency that depends on the range of the scattering center. The sampling rate is set as $F_s = \beta \tau/t_u$, which leads to an aliased spectrum shown in dashed lines. } 
\end{figure}

We consider $N_T$ transmitters and $N_R$ collocated receivers that function as a MIMO radar system. This system employs the compressive illumination framework proposed in \cite{FreqDivWaveform,sparseTargetMFMChirp}, and \cite{ErtinMIMO_2016}, which is extended to the case of multiple transmitters and receivers for estimating the target range and angle of arrival. The transmitter antenna elements are placed with a spacing of $d_T = 0.5$ and the receiver antenna elements are placed with a spacing of $d_R = 0.5N_T$ relative to the wavelength $\lambda_c = c /f_c$ of the carrier signal. We obtain the virtual array with an aperture length $(N_T N_R-1)\lambda_c/2$ meter, where $c$ is the velocity of light in vacuum, and $f_c$ is the carrier frequency. The process used to generate the transmitted signal is shown in Fig.~\ref{fig:TxBlockDiagram}. 
We discretize the frequency range $[0,B]$ into N frequencies $f(1),\cdots, f(N)$, where $N = Bt_u$, $t_u$ is the unambiguous time interval, $B$ is the system bandwidth, and $f(i) = \frac{iB}{N}$. A subset of $N_c N_T$ tones are chosen at random from these $N$ possible frequencies, where $N_c$ is the number of modulating tones used in each transmitter. The chosen tones are used for modulating the LFM waveform with bandwidth $\beta \ll B$, using the Single Side-Band (\emph{SSB}) modulation technique as shown in Fig.~\ref{fig:TxBlockDiagram}. We simplify this selection model for analysis by considering $N$ independent indicator random variables $\hat{\gamma}_i \in\cb{0,1}$ following a Bernoulli distribution with \begin{equation*} 
\hat{\gamma}_i  = \begin{cases} & 1 \text{ with probability }  {N_c N_T}/{N} \\
& 0 \text{ with probability }   1 - {N_c N_T}/{N} \end{cases}
\end{equation*} to select the tones that modulate the LFM waveform such that $N_c N_T$ waveforms are selected on an average. Each chosen LFM waveform is scaled by an independent and identically distributed complex exponential with a uniformly distributed phase such that the probability density function $f_{\Phi}(\phi_i) = {1}/\pb{2\pi} ,\phi_i \in \sqb{0,2\pi}$. We define the sequence of random variables $\cb{\hat{c}_1,\cdots,\hat{c}_N }$ that model this selection process where 
\begin{align}\label{eq:probModel} 
	\hat{c}_i = \hat{\gamma}_i \exp(j\Phi_i). 
\end{align}
Each selected waveform is assigned to one of the $N_T$ transmitters using a deterministic rule.  The transmitted signal from all the $N_T$ transmitters can be written as 
\begin{align}
	s(t) = \sum_{i=1}^{N} & \hat{c}_i\frac{\exp\pb{j2\pi \sqb { \hat{f}(i)t + \frac{\beta}{2\tau} t^2} } }{\sqrt{N_c N_T}} r\pb{\frac{t-\frac{\tau}{2}}{\tau} }, 
\end{align}
where $r\pb{\pb{t-\frac{\tau}{2}}/{\tau} } =1$ if $t\in \pb{0,\tau}$ and $0$ otherwise, and $ \hat{f}(i) = f_c + f(i)$, for $i=1,\cdots,N-1$. The instantaneous frequency of the transmitted and received modulated LFM signal as a function of time is illustrated in Fig.~\ref{fig:waveformLFM} for the case of $N_T=2,N_R =1$. 
Fig.~\ref{fig:RxStretch} shows the stretch processing operation implemented at a particular receiver. 
The sampling rate at the receiver after stretch processing is $F_s = {\beta t_u}/{\tau}$, which leads to $M = \beta t_u$ samples at stretch processor output at each receiver. Since the sampling rate is much lower than the Nyquist rate required for the modulating tones, the multi-tone frequency spectrum corresponding to a target with a delay of $\Delta$ aliases to the range $\sqb{-F_s/2,F_s/2}$. In the following sections, we show that the delay and angle of arrival of a sparse set of targets can be uniquely recovered if a sufficient number of modulating tones are utilized in the transmitter. The $n^{th}$ sample $y_k\pb{n}$ at the stretch processor at receiver $k$ due to a target located at a delay $\Delta \in [0,t_u]$ and an angle of arrival $\theta \in [0,2\pi]$ with amplitude $x \in \mbb{C}$ is given by 
\begin{align}
	&y_k\pb{n} = \sum_{{i=1}}^{{i=N}} \hat{c}_i \exp\pb{-j2\pi f(i)\Delta }\alpha_R(\theta; k) \frac{\alpha_T(\theta;\xi(i)) x}{\sqrt{N_T N_R N_c M}} \nn \\
	&\qquad \exp\pb{j2\pi \pb{f(i) - \frac{\beta \Delta}{\tau} } \frac{n}{F_s}}+ w_{k,n}, \\
    &\alpha_T(\theta;\xi(i)) = \exp\pb{j2\pi{d}_T\xi(i) \theta}, \nn \\
	&\alpha_R(\theta; k) = \exp\pb{j 2\pi{d}_R(k) \theta}, 	\nn
\end{align}
where $w_{k,n}$ is the $n^{th}$ noise sample at receiver $k$, $\alpha_R(\theta; k) $ is the array steering parameter corresponding to receiver $k$, and $\alpha_T(\theta;\xi(i))$ is the steering parameter corresponding to the chosen transmitter specified by the rule $\xi(i)$ for the $i^{th}$ waveform. We present the recovery guarantees for the proposed system by discretizing the range-angle of arrival space.  We also present an algorithm that recovers the range and angle of arrival of a sparse set of targets in the continuum in section~\ref{sec:offGrid}. 

The unambiguous interval from $[0,t_u]$ is discretized at a resolution of ${1}/{B}$ corresponding to the resolution achieved by a system employing a signal of bandwidth $B$ resulting in $N=B t_u$ bins. Each delay bin is denoted as \[\Delta_m = {m}/{B}, m=0,1,\cdots, N-1.\] The angle of arrival characterized by $\cos{\theta} \in \sqb{-1,1}$ is partitioned into $N_{\theta} = N_T N_R$ grids. Each angle bin is denoted as \[\theta_v \in \cb{{2v}/\pb{N_T N_R} \vert v= {-N_T N_R}/{2},\cdots, {N_T N_R}/{2}-1}.\] The receiver and transmitter steering vectors as function of the angle of arrival $\theta_v$ are defined as 
\begin{align}
	\mbf{\alpha_R}(\theta_v) &=
	\begin{bmatrix}
		1 & \cdots & \exp\pb{j \bar{d}_R(N_R - 1) \theta_v} 
	\end{bmatrix}
	^{T}, \text{and} \nn \\
	\mbf{\alpha_T}(\theta_v) &= 
	\begin{bmatrix}
		1 & \cdots & \exp\pb{j\bar{d}_T (N_T - 1) \theta_v} 
	\end{bmatrix}
	^{T}, 
\end{align}
respectively, where $\bar{d}_R = 2\pi d_R$, and $\bar{d}_T = 2\pi d_T$. The normalized sample at the stretch processor output $y_k\pb{n}$ at receiver $k$ due to the targets in the region of interest is given by 
\begin{align}
	&y_k\pb{n} = \sum_{v=0}^{N_T N_R -1} \sum_{\st{m = 0}{i=1}}^{\st{m=N-1}{i=N}} \hat{c}_i \alpha_R(\theta_v; k) \frac{\alpha_T(\theta_v;\xi(i)) x(v,m)}{\sqrt{N_T N_R N_c M}} \nn \\
	& \exp\pb{-j2\pi f(i)\Delta_m }\exp\pb{\frac{j2\pi n}{F_s} \pb{f(i) - \frac{\beta \Delta_m}{\tau} } }+ w_{k,n}, 
\end{align}
where $k = 1,\cdots, N_R$, $n = 0,\cdots M-1$, and $x_{v,m} \in \mbb{C}$ is the scattering coefficient at range bin $m$ and angle of arrival bin $v$. The concatenated output from all the $N_R$ receivers can be compactly written as 
\begin{align}\label{eq:sensingMIMO} 
	\mbf{y} = \mbf{\mc{A}} \mbf{x} + \mbf{w}, 
\end{align}
where the signal is given by 
\begin{align*}
	\mbf{y} &= \sqb{\mbf{y}_1 \cdots \mbf{y}_{N_R}}^T, \mbf{y}_k = \sqb{y_k(0) \cdots y_k(M-1)}^T \in \mbb{C}^{M}. \\
	\mbf{w} &= \sqb{\mbf{w}_1 \cdots \mbf{w}_{N_R}}^T , \mbf{w}_k = \sqb{w_{k,0} \cdots w_{k,M-1}}^T \in \mbb{C}^{M} 
\end{align*}
is the zero mean additive white Complex Gaussian noise with variance $\sigma^2$, and $\mbf{x} \in \mbb{C}^{N N_T N_R}$ contains the complex scattering amplitudes associated with targets at all possible grid locations in the range-angle domain. The sensing matrix $\boldsymbol{\mc{A}} \in \mbb{C}^{N_R M \times N_{\theta}N }$ can be expressed as a series of deterministic matrices with random coefficients as follows 
\begin{align}\label{eq:sensingMatrixMIMO} 
	\boldsymbol{\mc{A}} = \sum_{i=1}^N \hat{c}_i \pb{\mbf{\bar{\alpha}_R \bar{\alpha}_T(\xi(i))} } \otimes \pb{\mbf{H}_i \mbf{\bar{A}} \mbf{D}_i}, 
\end{align}
\begin{align}
	&\text{where, }\mbf{\bar{\alpha}_R} = \sqrt{{1}/\pb{N_R N_T}}
	\begin{bmatrix}
		\mbf{\alpha_R}(\theta_0) &\cdots &\mbf{\alpha_R}(\theta_{N_{\theta}-1}) 
	\end{bmatrix}
	\nn \\
	&\mbf{\bar{\alpha}_T(\xi(i))} = \text{diag}\pb{\exp\pb{j\bar{d}_T\xi(i)\theta_0 \cdots \exp\pb{j\bar{d}_T\xi(i)\theta_{N_{\theta} -1} }}} \nn 
\end{align}
for $i=0,\cdots,N-1$ and $r = 0,\cdots,N-1$. $\mbf{\bar{\alpha}_R \in
\mbb{C}^{N_R \times N_{\theta}}}$ is the matrix consisting of receiver steering
vectors for all the bins of angle of arrival, $\otimes$ represents the Kronecker
product and $\mbf{\bar{\alpha}_T(\xi(i))} \in \mbb{C}^{N_{\theta} \times
{N_{\theta}}}$ is the diagonal matrix with diagonal elements as the $\xi(i)$ transmitter's component of the steering vector for all the angle
bins. The individual components are as follows 
\begin{align}\label{eq:componentsSingleTransmit} 
	&\bar{\mbf{A}} = \frac{1}{\sqrt{M N_c}} 
	\begin{bmatrix}
		\bar{\mbf{A}}(0) &\cdots &\bar{\mbf{A}}(N-1) 
	\end{bmatrix}
	\nn \\
	&\bar{\mbf{A}}(r) = 
	\begin{bmatrix}
		1 &\exp \pb{-2\pi j \frac{r}{N} } &\cdots &\exp \pb{-2\pi j \frac{r(M-1}{N} ) } 
	\end{bmatrix}
	^T , \nn \\
	&\mbf{D}_i = \text{diag}
	\begin{bmatrix}
		1 &\exp\pb{-j 2\pi \frac{i}{N}} &\cdots &\exp\pb{-j 2\pi \frac{i(N-1)}{N}} 
	\end{bmatrix}
	, \nn \\
	&\mbf{H}_i = \text{diag}
	\begin{bmatrix}
		1 &\exp\pb{j 2\pi \frac{i p}{M } } &\cdots &\exp\pb{j 2\pi \frac{i p(M-1)}{M }} 
	\end{bmatrix}
	, 
\end{align}
where $i=0,\cdots,N-1$ and $r = 0,\cdots,N-1$, $\bar{\mbf{A}} \in \mbb{C}^{M \times N}$ are the samples from tones that correspond to each delay bin generated as a result of the de-chirping process in case of a single transmitter and receiver system employing an LFM waveform with bandwidth $\beta$ Hz, $\mbf{H}_i \in \mbb{C}^{M \times M}$ is the shift in frequency due to the $i^{th}$ modulating tone, and $\mbf{D}_i \in \mbb{C}^{N \times N}$ contains the phase term associated with different delay bins due to the $i^{th}$ modulating tone. \\
Each column of the sensing matrix $\boldsymbol{\mc{A}}$ can be written as 
\begin{align}\label{eq:MatrixColumnDef}
	& \boldsymbol{\mc{A}}\pb{m,v} = \pb{\mbf{\alpha_{R}(\theta_v) \otimes \pb{\mbf{E_mFG_m}} }} \mbf{\hat{c}}(v)  \\
	&\label{eq:modRandomVars} \hat{c}_r(v) = \hat{c}_r \alpha_T\pb{\theta_v;\xi(r)} 
\end{align}
where $m,r = 0,\cdots, N-1$, $v=0,\cdots,N_{\theta}-1$. The individual terms are 
\begin{align*}
	&\mbf{E}_m = \text{diag}
	\begin{bmatrix}
		1 &\exp\pb{-j 2\pi \frac{m}{N} } &\cdots &\exp\pb{-j 2\pi \frac{m (M-1)}{N}} 
	\end{bmatrix}
	\\
	&\mbf{F} = \frac{1}{\sqrt{M N_c}} 
	\begin{bmatrix}
		\mbf{F}(0) &\cdots &\mbf{F}(N-1) 
	\end{bmatrix}
	\\
	&\mbf{F}(r) = 
	\begin{bmatrix}
		1 &\exp \pb{2\pi j \frac{r p}{M} } &\cdots &\exp \pb{2\pi j \frac{r p(M-1}{M} ) } 
	\end{bmatrix}
	^T , \\
	&\mbf{G}_m = \text{diag}
	\begin{bmatrix}
		1 &\exp\pb{-j 2\pi \frac{m}{N} } &\cdot\cdot &\exp\pb{-j 2\pi \frac{m (N-1)}{N}} 
	\end{bmatrix}
	, 
\end{align*}
and $\mbf{\hat{c}}(v) = [\hat{c}_0(v) \cdots \hat{c}_{N-1}(v) ]^T \in \mbb{C}^{N}$ is the random vector with independent components that selects the modulating waveform.

\subsection{Target model}\label{sec:tarModel} We consider the statistical model studied in \cite{SparseMIMORadar} for the sparse range profile of targets. We assume that the targets are located at the $N N_{\theta} = N N_R N_T$ discrete locations corresponding to different delay bins and angle bins. The support of the K-sparse range profile is chosen uniformly from all possible subsets of size $K$. The complex amplitude of the non-zero component is assumed to have an arbitrary magnitude and uniformly distributed phase in $\sqb{0,2\pi}$. We also empirically study the performance of the proposed illumination system for targets not located on the grid. For the off-grid problem, we assume a minimum separation between the targets in the delay and angle of arrival domain, which is chosen based on the system resolution in each domain. The minimum separation used in the simulation studies for the delay domain is $\min_{i,j} \abs{\Delta_i - \Delta_j} > {2}/{B}$, and the angle of arrival domain is $\min_{i,j} \abs{\cos \theta_i - \cos \theta_j} \geq {2}/{N_T N_R}$. 

\subsection{Problem statement} Given a sparse scene with targets following the statistical model discussed in previous section, and measurement scheme in \eqref{eq:sensingMIMO} with $M \ll NN_{\theta}$ and sparsity level $K \ll NN_{\theta}$, the goal of compressed sensing \cite{DonohoCS_2006} is to recover the sparse or compressible vector $\mbf{x}$ using minimum number of measurements in $\mbf{y}$ constructed using random linear projections $\boldsymbol{\mc{A}}$. The search for the sparsest solution can be formulated as an optimization problem given below 
\begin{align}\label{eq:l0_minimization} 
	\min_{\mbf{x}} \norm{\mbf{x}}_0, \mbox{ subject to } \norm{\boldsymbol{\mc{A}}\mbf{x} - \mbf{y} }_2 \leq \eta, 
\end{align}
where $\eta^2$ is the noise variance. This problem is NP-hard and hence, intractable as shown in \cite{FoucartMathIntroCS_2013}, and many approximate solutions have been found. One particular solution is to use the convex relaxation technique to modify the objective as an $\ell_1$ norm minimization instead of the non-convex $\ell_0$ norm, which is given by, 
\begin{align}\label{eq:l1Relaxation1} 
	\min_{\mbf{x}} \norm{\mbf{x}}_1 \mbox{ subject to } \norm{\boldsymbol{\mc{A}}\mbf{x} - \mbf{y} }_2 \leq \eta. 
\end{align}
This approach has been shown to recover sparse or compressible vectors successfully~\cite{ChenBP_2001,RIP_compressedSensing_Candes2008} given that the sub-matrices formed by columns of the sensing matrix are well-conditioned. Our analysis is based on LASSO~\cite{Tibshirani1996_LASSO}, which is a related method that solves the optimization problem in \eqref{eq:l1Relaxation1}. It has been shown in~\cite{candesPlanL1} that for an appropriate choice of $\lambda$ and conditions on measurement matrix are satisfied, then the support of the solution of the below-mentioned optimization problem coincides with the support of the solution of the intractable problem in \eqref{eq:l0_minimization}, 
\begin{align}\label{eq:l1Relaxation} 
	\min_{\mbf{x}} \lambda \norm{\mbf{x}}_1 + \frac{1}{2} \norm{\boldsymbol{\mc{A}}\mbf{x} -\mbf{y} }_2^2. 
\end{align}
In this paper, we show that the measurement model formulated in \eqref{eq:sensingMatrixMIMO} satisfies the conditions on mutual coherence given in \cite{candesPlanL1}. Next, we find a bound on the sparsity level of range profile, which guarantees successful support recovery of almost all sparse signals using LASSO with high probability from noisy measurements. Finally, we also provide an estimate of the number of measurements required for the operator representing our scheme to satisfy the restricted isometry property (RIP) of order $K$. We consider the space of K-sparse vector $\mbf{x} \in \mbb{C}^{NN_{\theta}}$ where $\norm{\mbf{x}}_2 \leq 1$ denoted by $\mc{D}_{K,N N_{\theta}}$. The RIP condition of order $K$ is true if the following condition is true for $\mbf{x} \in \mc{D}_{K,N N_{\theta}}$
\begin{align*} 
	&(1 - \delta) \norm{\mbf{x}}_2^2 \leq \norm{\boldsymbol{\mc{A}x}}_2^2 \leq (1 + \delta) \norm{\mbf{x}}_2^2. 
\end{align*}
Equivalently, the condition can be stated as
\begin{align} \label{eq:RIP}
\delta_K = \sup_{\mbf{x} \in \mc{D}_{K,N N_{\theta}}} \abs{\norm{\boldsymbol{\mc{A}x}}_2^2  -\norm{\mbf{x}}^2 }.
\end{align}
In this paper, we bound the random variable $\delta_K$ using the theory for bounding stochastic processes \cite{TalgrandStochasticProc2014} adapted to the CS setting in \cite{Chaos_Rauhut2014,Rauhut_SparseMIMO_2015}. The next section presents the main results of our analysis.

\section{Recovery guarantees}\label{sec:RecGuarantees} 
The following theorems state the recovery guarantee for the proposed MIMO radar system. 
\begin{theorem}\label{thm:MultiChirpsRecGuaranteeMIMO} 
	Consider a compressive MIMO radar system with the measurement model $\mbf{y}= \boldsymbol{\mc{A}}\mbf{x} +\mbf{w}$, where $\boldsymbol{\mc{A}} \in \mbb{C}^{N_RM \times N_R N_T N}$ is defined in \eqref{eq:sensingMatrixMIMO} such that the target scene $\mbf{x}$ is drawn from a K-sparse model with complex unknown amplitudes and observed in i.i.d. noise process $\mbf{w} \sim \mc{CN}(0,\sigma^2I)$. The support of the targets in the scene can be recovered using a LASSO estimator with arbitrarily high probability for a system using $M$ samples at each receiver and $N_c$ tones at each transmitter with $M\sim\mathcal{O}( \log^3( N N_R N_T))$ and $N_c \sim \mathcal{O}(N/N_T)$, if the target scene consists of $K$ targets with $K\sim\mathcal{O}(N_R M /\log^2(2 N N_R N_T) )$ of minimum amplitude 
	\begin{align}\label{eq:minSignalCondition} 
		& \min_{k \in \mbf{S}} \abs{x_k} > \frac{8}{\sqrt{1-\epsilon}}\sigma\sqrt{2 \log\pb{N N_R N_T}}, 
	\end{align}
\end{theorem}
\begin{theorem}\label{thm:RIPConditionMIMO} 
	For the measurement matrix $\boldsymbol{\mc{A}}$ given in \eqref{eq:sensingMatrixMIMO} and any $\delta \in \sqb{0,1}$, the RIP condition in \eqref{eq:RIP} as $\delta_K(\boldsymbol{\mc{A}}) \leq \delta$ is satisfied with high probability if the number of measurements $M$ per receiver satisfies the condition $M \geq  \delta^{-2}K\log (\frac{N_R N_T N}{K}) $ 
\end{theorem}
\subsection{Discussion}
\emph{Theoretical considerations} - There are two types of recovery guarantees in the literature -namely, uniform and non-uniform recovery guarantees. Uniform guarantees imply successful recovery of all $K$-sparse vectors for any realization of the system parameters that are chosen at random. Such guarantees rely on the RIP property. If the sensing operator satisfies the RIP property of order $2K$, given by $\delta_{2K} \leq \delta \approx \sqrt{2} -1$ with high probability then all $K$-sparse vectors are successfully recovered, with a reconstruction error of an oracle estimator that knows the support of the sparse vector or the support of K largest elements~\cite{performanceSparseRec_eldar_2010,RIP_compressedSensing_Candes2008} up to a logarithmic factor of the size of the search space. Non-uniform guarantees imply that almost all realizations of the system successfully recover a fixed $K$-sparse vector. These guarantees impose conditions on the spectral norm and mutual coherence of the measurement operator for successful recovery of a $K$-sparse vector \cite{candesPlanL1}. In the previous section, we established both uniform and non-uniform guarantees for the proposed system. Next, we look at the recovery guarantees for some known linear operators that rely on the estimates of the quantities defined in this section. 

\emph{Baraniuk et. al.} in \cite{RIPRandom_Baraniuk2008} have shown that random matrices with i.i.d entries from either Gaussian or sub-Gaussian probability distribution satisfy the RIP condition, such that for any $\delta \in [0,1]$ $\delta_K \leq \delta$ if number of measurements $M \sim \mc{O}\pb{ K\log \pb{N/K}}$. Although these unstructured random matrices have remarkable recovery guarantees they do not represent any practical measurement scheme, which leads us to consider classical linear time invariant (LTI) systems. 
This leads to a structured measurement matrix that is either a partial or sub-sampled Toeplitz or circulant matrix. The RIP condition of order $K$ for partial Toeplitz matrices in the context of channel estimation was established by \emph{Haupt et. al.} in \cite{Teoplitz_Haupt2010}. They showed that if the number of measurements $M \sim \mc{O}\pb{ K^2{\log N}}$, then $\delta_K \leq \delta $. This quadratic scaling of number of measurements with respect to sparsity was improved in \cite{RandomConvolution_Romberg2009, RIP_toeplitz_Rauhut2012,Chaos_Rauhut2014}. \emph{Romberg} in \cite{RandomConvolution_Romberg2009} considered an active imaging system that used waveform with a random symmetric frequency spectrum and acquired compressed measurements using random sub-sampler or random demodulator at the receiver to estimate the sparse scene. The resultant system is a randomly sub-sampled circulant matrix representing the convolution and compression process. It is shown that for a given sparsity level $K$, the condition that $\delta_{2K} \leq \delta$ is satisfied if the number of measurements $M \geq \alpha_6 \delta^{-2} \min\pb{K (\log N)^6, (K \log N)^2 }$, where $\alpha_6 > 0$ is a universal constant independent of the size of problem and $\delta$. This was extended by \emph{Rauhut et. al.} in \cite{RIP_toeplitz_Rauhut2012}. They consider a deterministically sampled random waveform in time domain with samples following Rademacher distribution, which is modeled as a sub-sampled Toeplitz or Circulant matrix with entries sampled from Rademacher distribution. It was shown that for a given sparsity level K, $\delta_K \leq \delta$ with high probability if the number of measurements $M \geq \alpha_7 \max\pb{\delta^{-1}(K\log N)^{3/2},\delta^{-2} K (\log N \log K)^2}$, where $\alpha_7$ is a universal constant. In the subsequent work by \emph{Krahmer et. al.} in \cite{Chaos_Rauhut2014}, the relation between sparsity level and number of measurements is improved and more general random variables are considered such as vectors following sub-Gaussian distribution to generate the Toeplitz or Circulant matrix. It is shown that, for a given sparsity level K the condition $\delta_K \leq \delta$ is satisfied if the number of measurements $M \geq \alpha_{8} \delta^{-2}K (\log K \log N)^2$, where the constant $\alpha_{8}$ is a function of only the sub-Gaussian norm of the random variables generating the matrix. The system proposed in this work achieves near-optimal scaling in the number of measurements up to an additional logarithmic factor for non-uniform guarantees as shown in table~\ref{table:SupportRec}. We also established that for the sensing scheme to satisfy RIP of order $K$, we need $\mc{O}(K\delta^{-2}\log(N/K))$ measurements. The key advantage of the proposed radar system is that there are only $2 N_c$ parameters comprising of the phase and frequencies of the modulating waveforms. In addition, we require only uniform sampling ADCs operating at low sampling rates.

\emph{Hardware considerations} - \\
\emph{Xampling based acquisition systems}- These systems require
 multiple channels per receiver that perform analog compression and individual
ADCs for each channel to acquire the resultant Fourier coefficients. This approach leads
to a drastic increase in the complexity of receiver design with a system employing multiple
transmitters and receivers. In contrast, the complexity in our implementation
scales gracefully with the increase in the number of receivers and transmitters
and forgoes the orthogonality requirements on the transmitted waveforms.\\
\emph{Random demodulator}- Such systems 
also guarantee successful recovery of multi-tone spectra with high probability.
Generating and mixing with pseudo-random sequences at high rates is a
challenging task and leads to signal dependent uncertainties due to timing
imperfections as studied in \cite{CSsystems_Davies_2012,Xampling_RD_Eldar_2011}.
\\
\emph{Stochastic waveforms}- The measurement operator 
generated from these systems guarantee successful recovery at the expense of
increased design complexities. The memory requirements for generating and
storing these waveform is large due to the high bandwidth requirements. In
addition, the peak to average power ratio (PAPR) of these waveforms are large,
which leads to non-linearity in the operation of the power amplifiers that are
required in practical systems. In contrast, our proposed waveforms derived from
the LFM waveform enjoy bounded PAPR close to $\sqrt{2}$ and at the same time
have low memory requirements for waveform generation.

\section{Simulation Results}\label{sec:SimResults} 
In this section we conduct simulation studies to study the performance of the proposed compressive radar sensor as a function of system parameters. Fixed parameters of the simulations are given in Table~\ref{table:ParamChoice}. 
\begin{table}
	[tp] 
	\begin{center}
		\begin{tabular}
			{| c | c|} \hline Parameter & Value \\
			\hline Bandwidth B & $500 \times 10^6 Hz$\\
			Range Interval & $[0,100]$m \\
			Number of Range Bins $N$ & $334$ \\
			Unambiguous time interval $t_u$ & $6.6\times 10^{-7}$ s\\
			pulse duration $\tau$ & $6.86\times 10^{-5} s$\\
			\hline 
		\end{tabular}
	\end{center}
	\caption{Parameters for simulation results} 
\label{table:ParamChoice} \end{table}

\subsection{Effect of multi-tones on mutual coherence} 
\begin{figure}
	[!t]\centering 
	\includegraphics[width=0.7\linewidth]{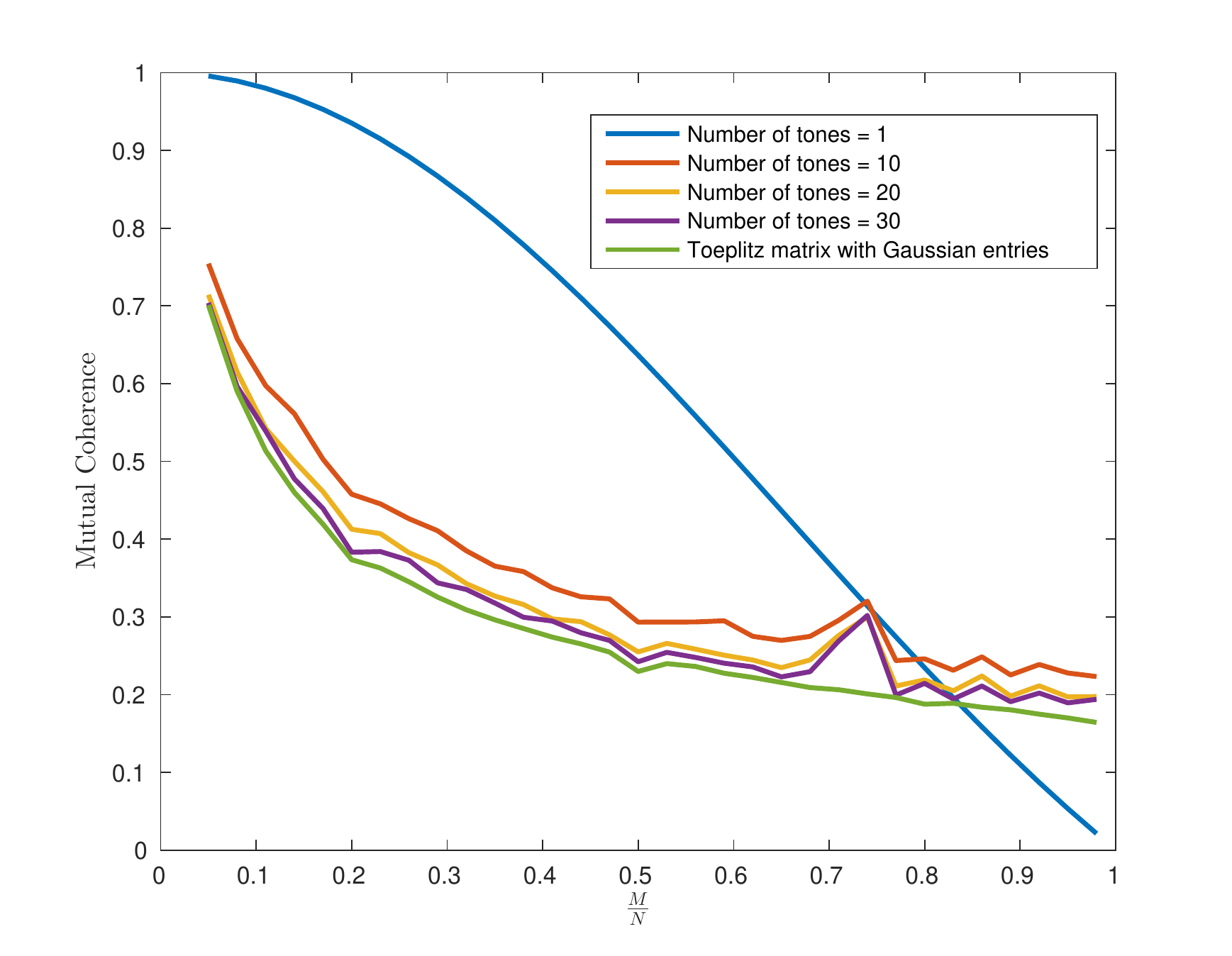} \caption{ Mutual coherence of single transmit system with single receiver as a function of under-sampling ratio ${M}/{N}$ as the number of chirps $N_c$ is increased along with the mutual coherence of the uniformly sub-sampled random Toeplitz matrix. \label{fig:CohGraphSISO} } 
\end{figure}

We first study the effect of increasing the number of tones in a single transmitter and receiver setting. We compare the proposed illumination scheme with a uniformly sub-sampled Toeplitz matrix with independent and identical elements sampled from a complex standard normal distribution. The Toeplitz sensing matrix represents the impulse response of the linear time invariant systen with a randomly distributed waveform with independent entries as input. From figure~\ref{fig:CohGraphSISO}, we observe that the coherence of a system employing a single tone is high for lower sampling rates. Increasing the number of tones improves the mutual coherence and as the number of modulating tones increase, the mutual coherence of the system converges in mean to the mutual coherence of structured random Toeplitz matrix. 
Next, we compare the coherence of the proposed system with a multiple input system employing samples from a Gaussian distribution, which leads to a partial block Toeplitz measurement matrix with random Gaussian entries. 
\begin{figure}
	[!t]\centering 
	\includegraphics[width=0.7\linewidth]{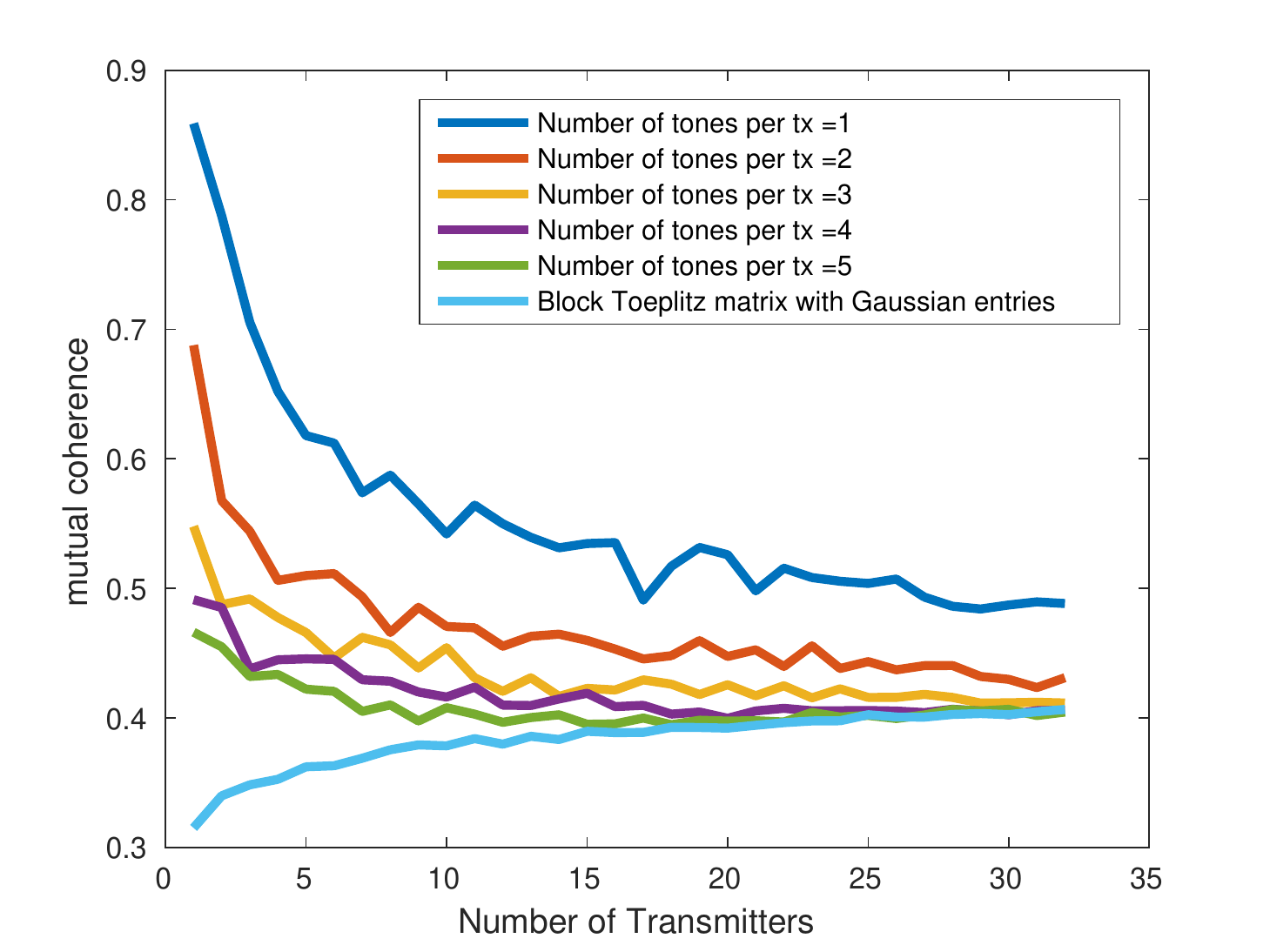} \caption{ Mutual coherence of multiple transmit system with single receiver as a function of number of transmitters $N_T$ as number of chirps $N_c$ is increased along with the mutual coherence of the random block Toeplitz matrix. The under-sampling ratio ${M}/{N}$ is set as $0.3$. \label{fig:CohGraphMISO} } 
\end{figure}
From figure~\ref{fig:CohGraphMISO}, we can see that as the number of transmitters and modulating tones increase, the randomness in the waveform increases and hence the mutual coherence of the system approaches that of a system employing random waveform with independent samples.

\subsection{On-grid recovery} 
\begin{figure*}[!tb]
	\subfigure[Number of tones = $1$]{ 
	\includegraphics[trim=0 0 0 0,clip,width=0.23\linewidth]{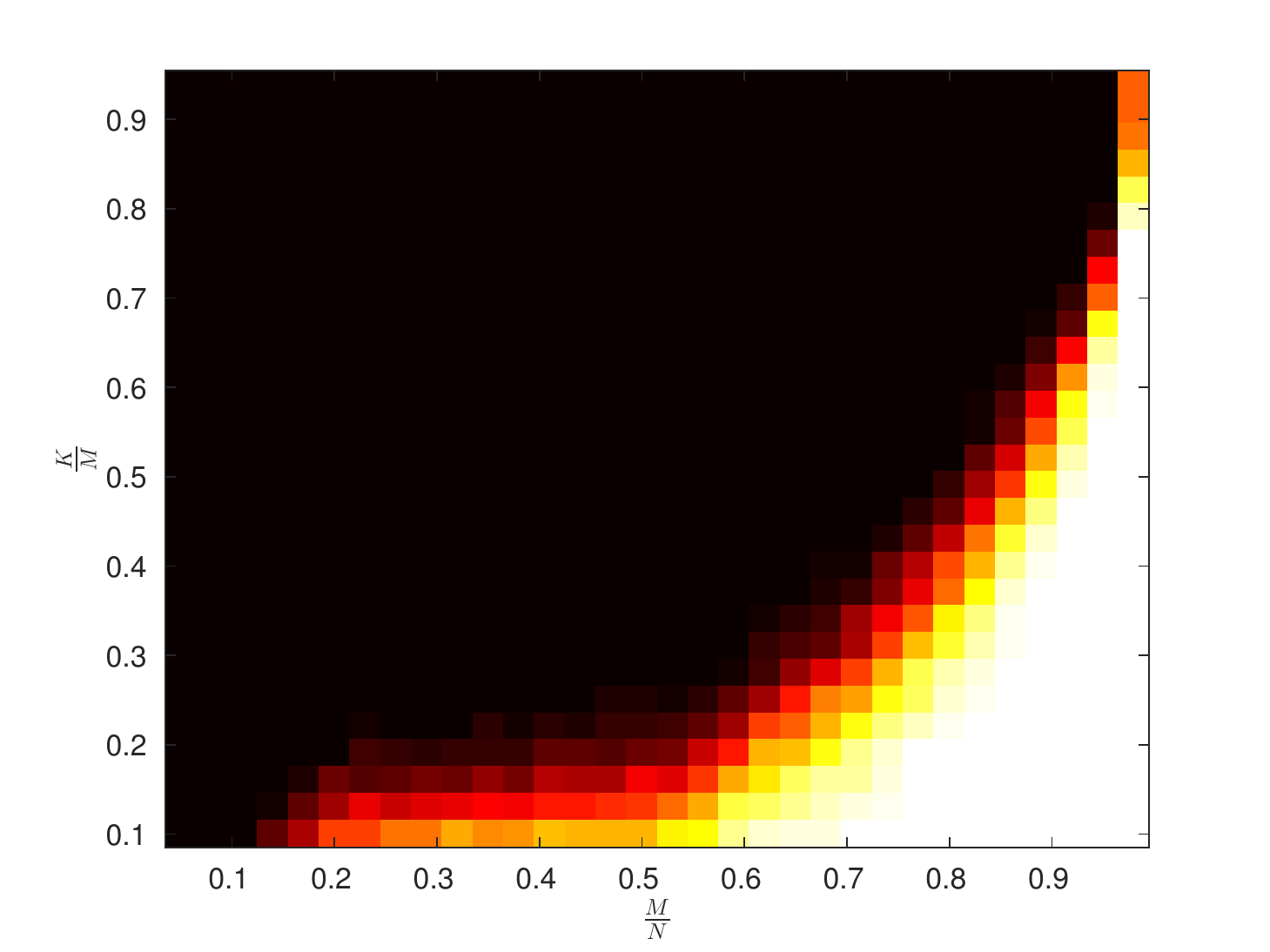} \label{fig:MSE_noiseless_1}} \subfigure[Number of tones = $10$]{ 
	\includegraphics[trim=0 0 0 0,clip,width=0.23\linewidth]{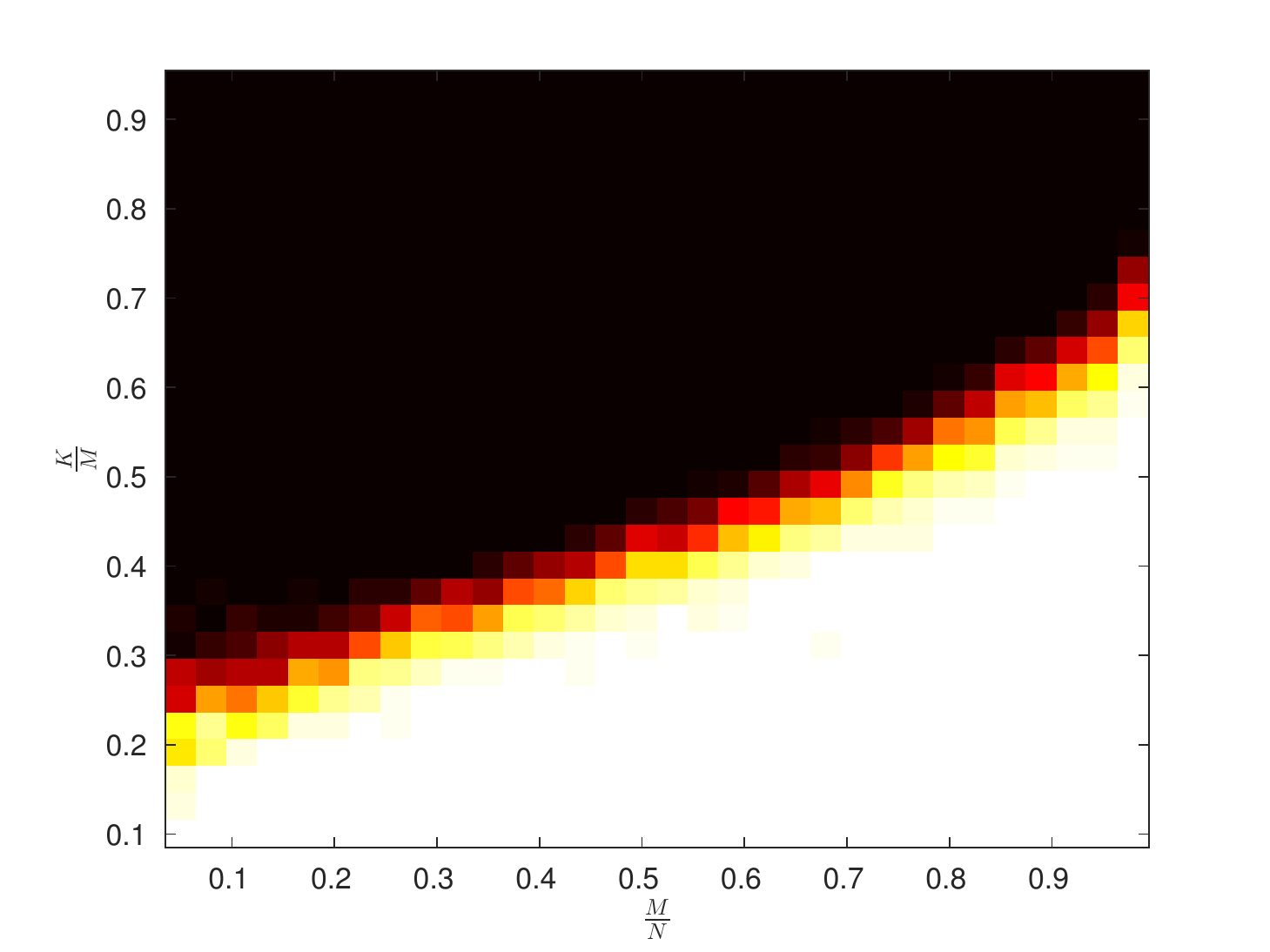} \label{fig:MSE_noiseless_10}} \subfigure[Number of tones = $20$]{ 
	\includegraphics[trim=0 0 0 0,clip,width=0.23\linewidth]{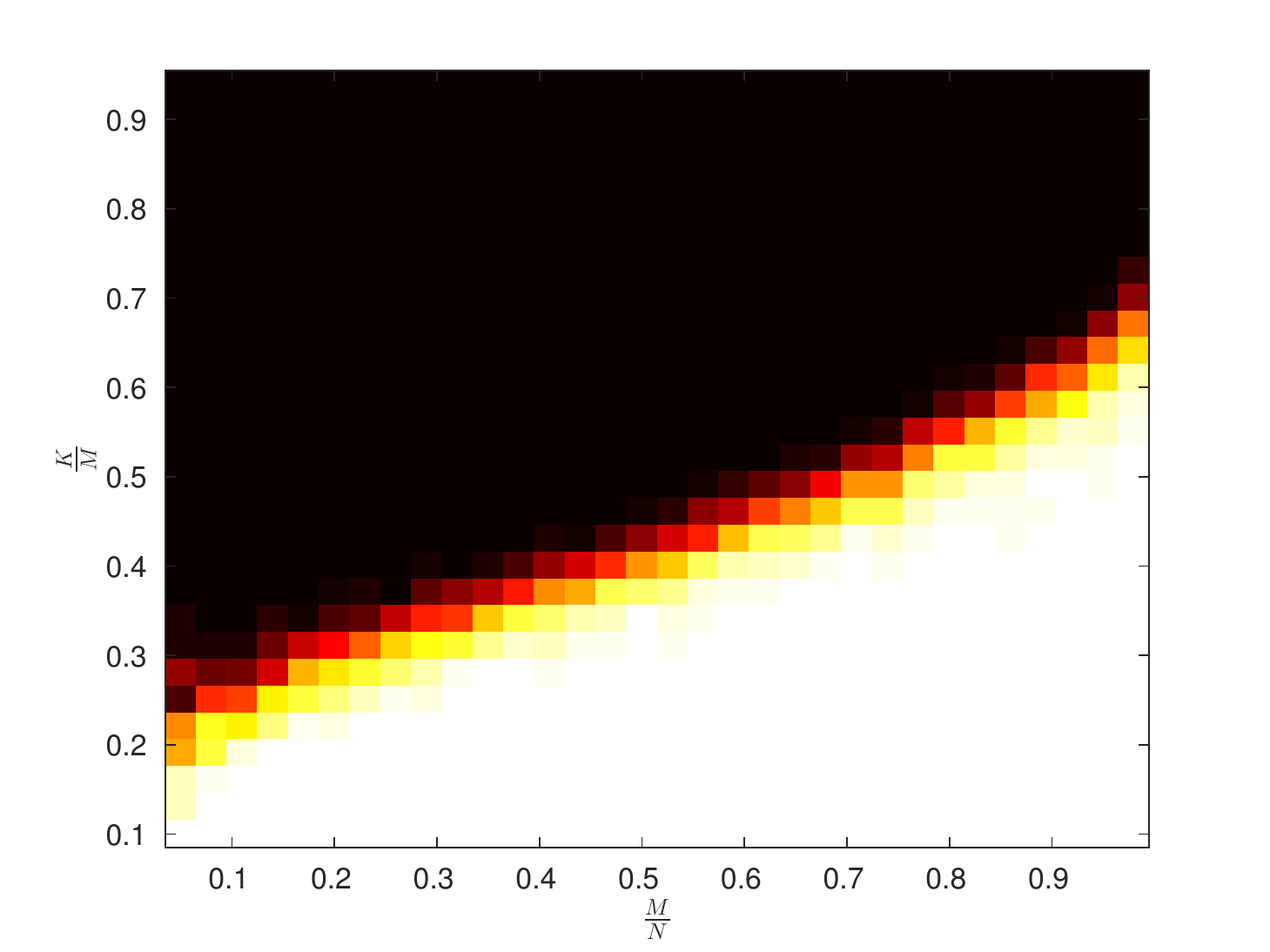} \label{fig:MSE_noiseless_20}} \subfigure[Gaussian Toeplitz system]{ 
	\includegraphics[trim=0 0 0 0,clip,width=0.23\linewidth]{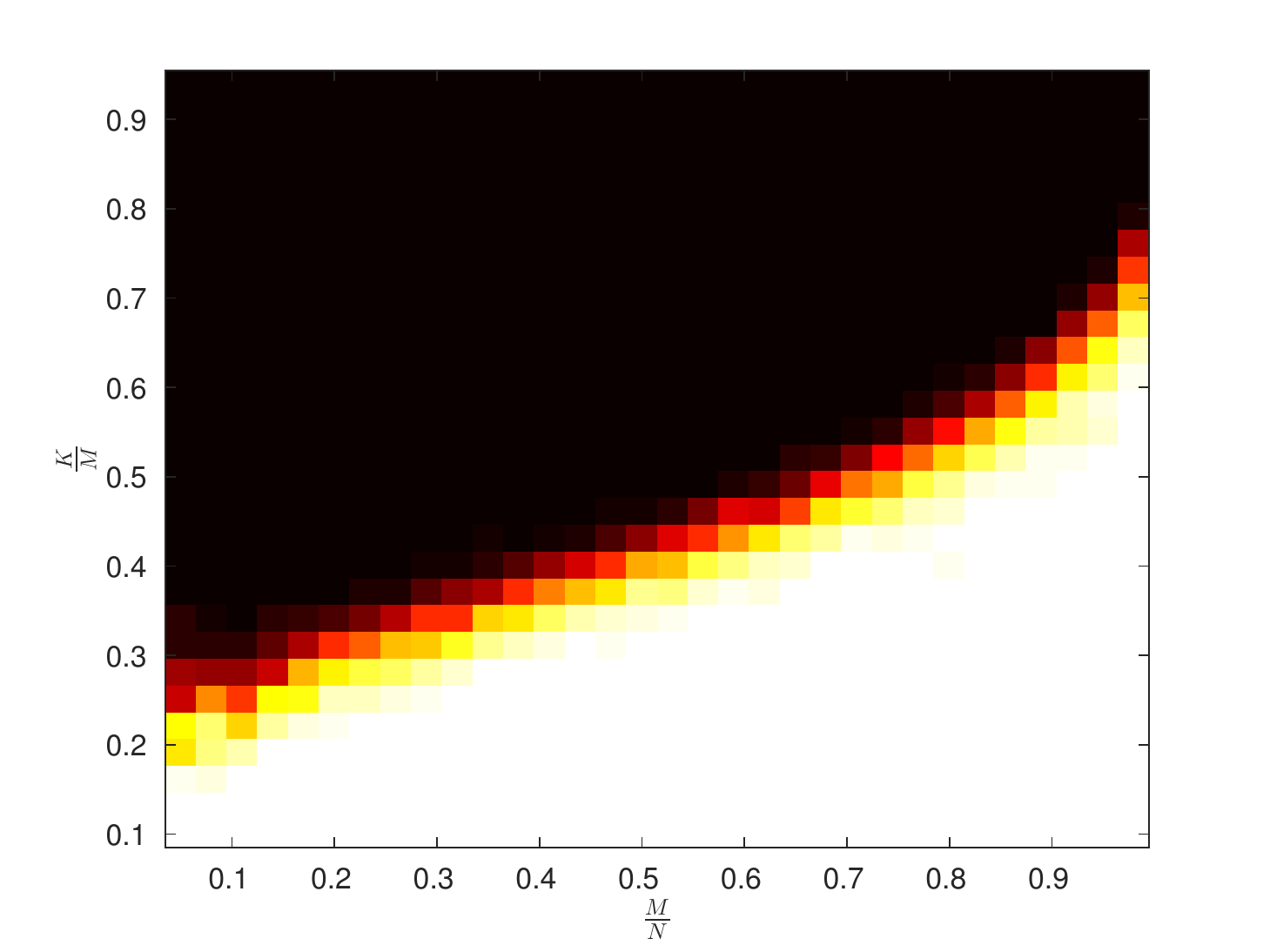} \label{fig:MSE_noiseless_Gaussian}} \subfigure[Number of tones = $1$]{ 
	\includegraphics[trim=0 0 0 0,clip,width=0.23\linewidth]{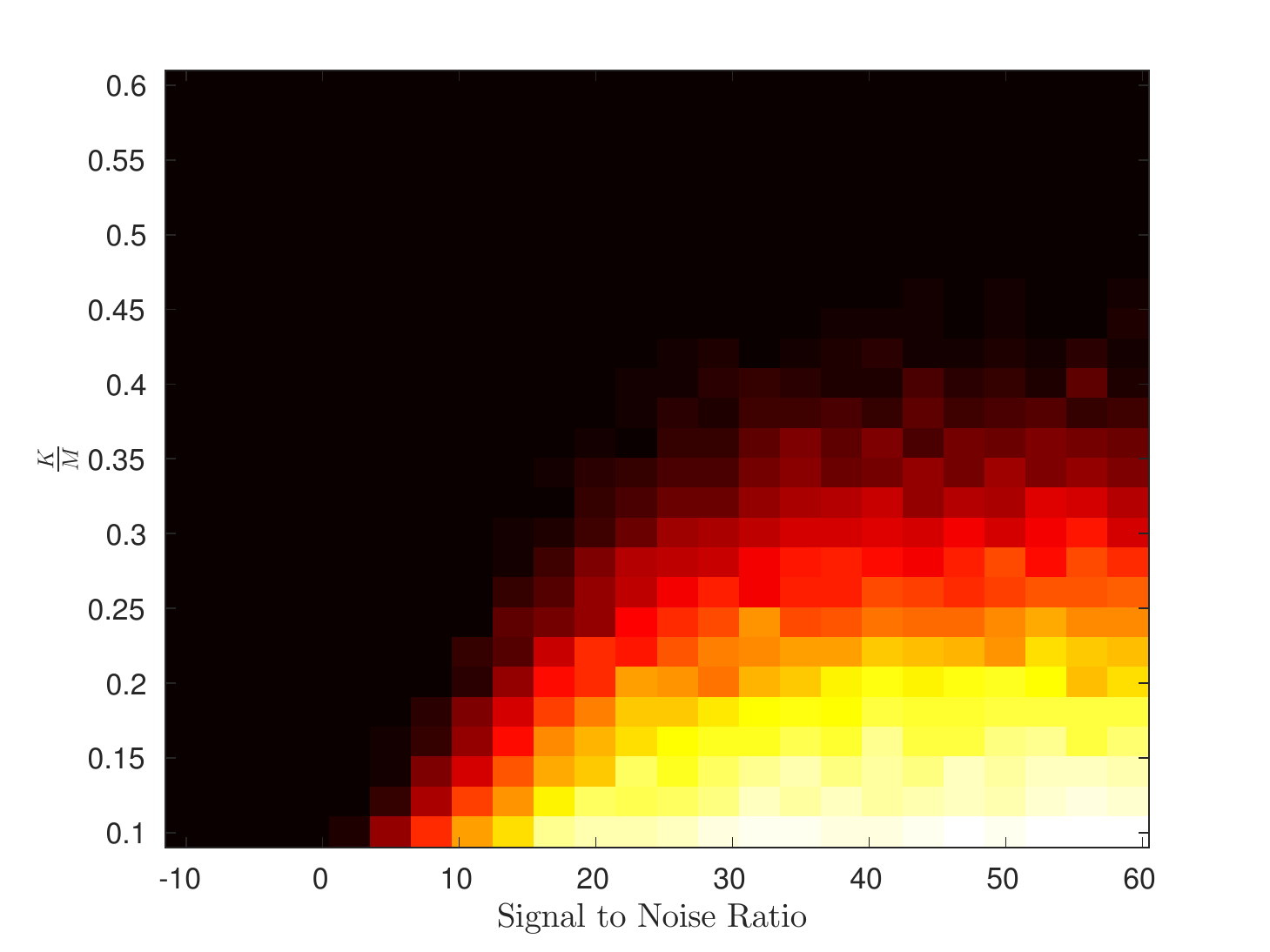} \label{fig:AUC_SNR_1}} \subfigure[Number of tones = $10$]{ 
	\includegraphics[trim=0 0 0 0,clip,width=0.23\linewidth]{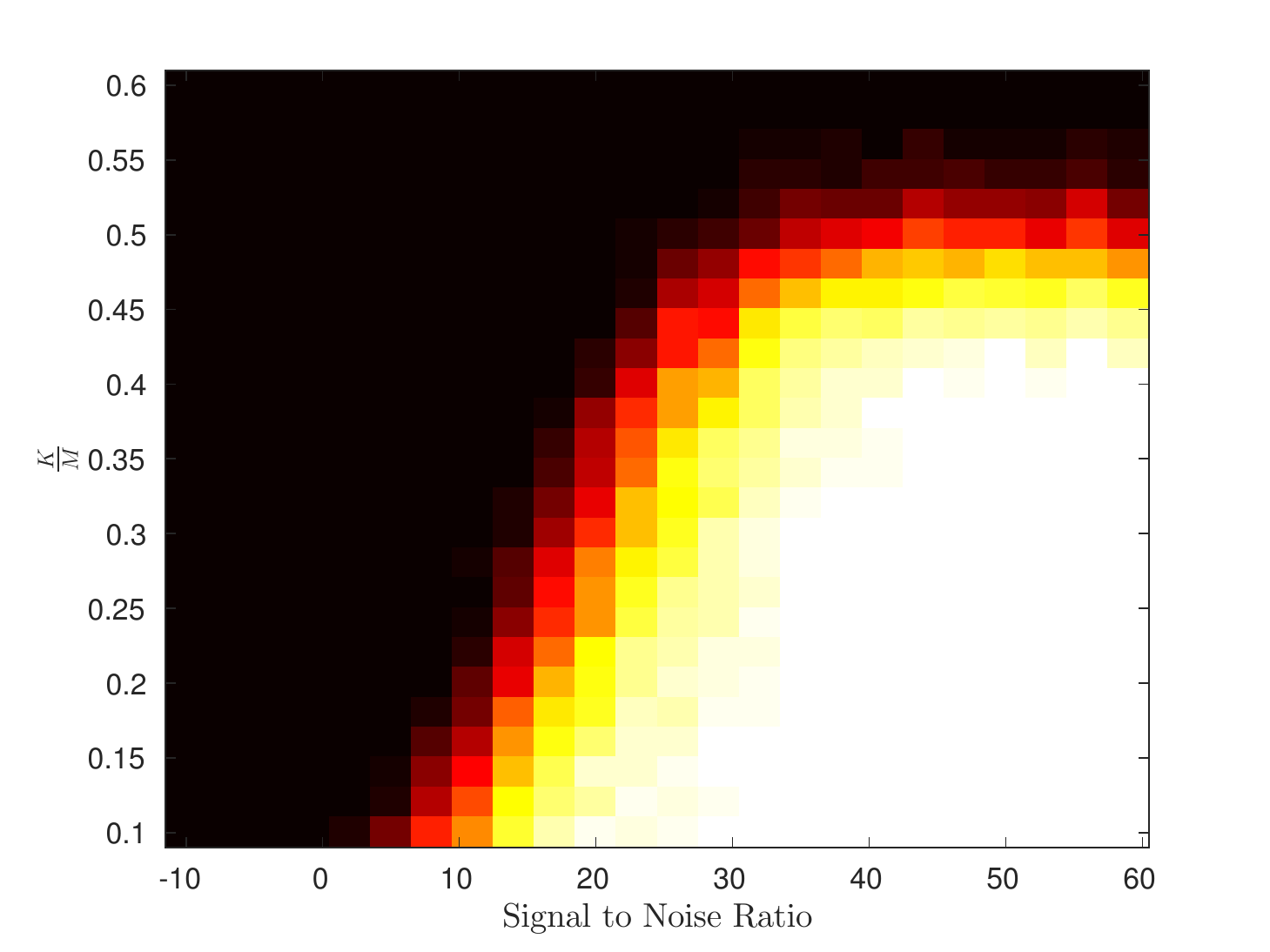} \label{fig:AUC_SNR_10}} \subfigure[Number of tones = $20$]{ 
	\includegraphics[trim=0 0 0 0,clip,width=0.23\linewidth]{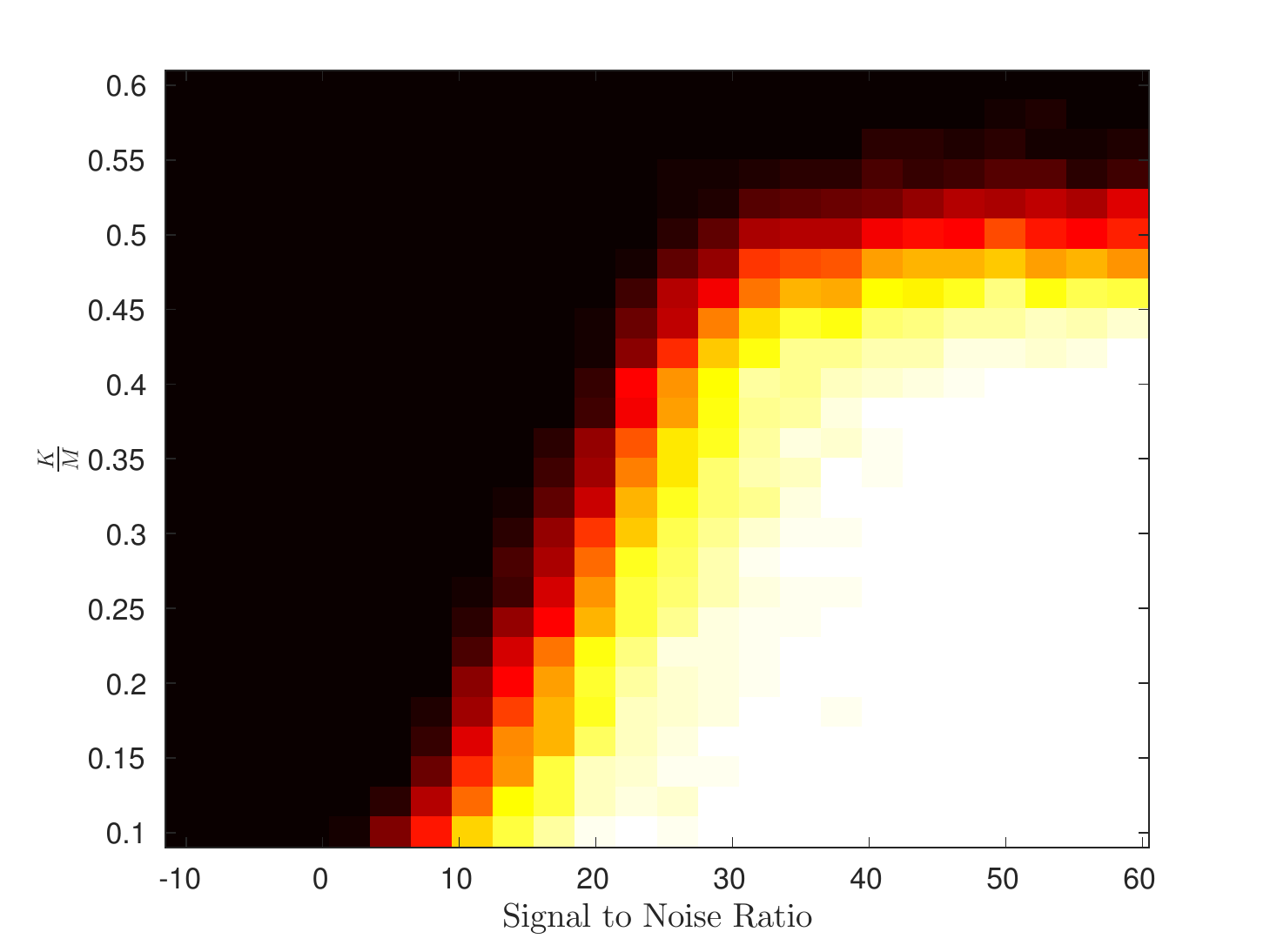} \label{fig:AUC_SNR_20}} \subfigure[Gaussian Toeplitz system]{ 
	\includegraphics[trim=0 0 0 0,clip,width=0.23\linewidth]{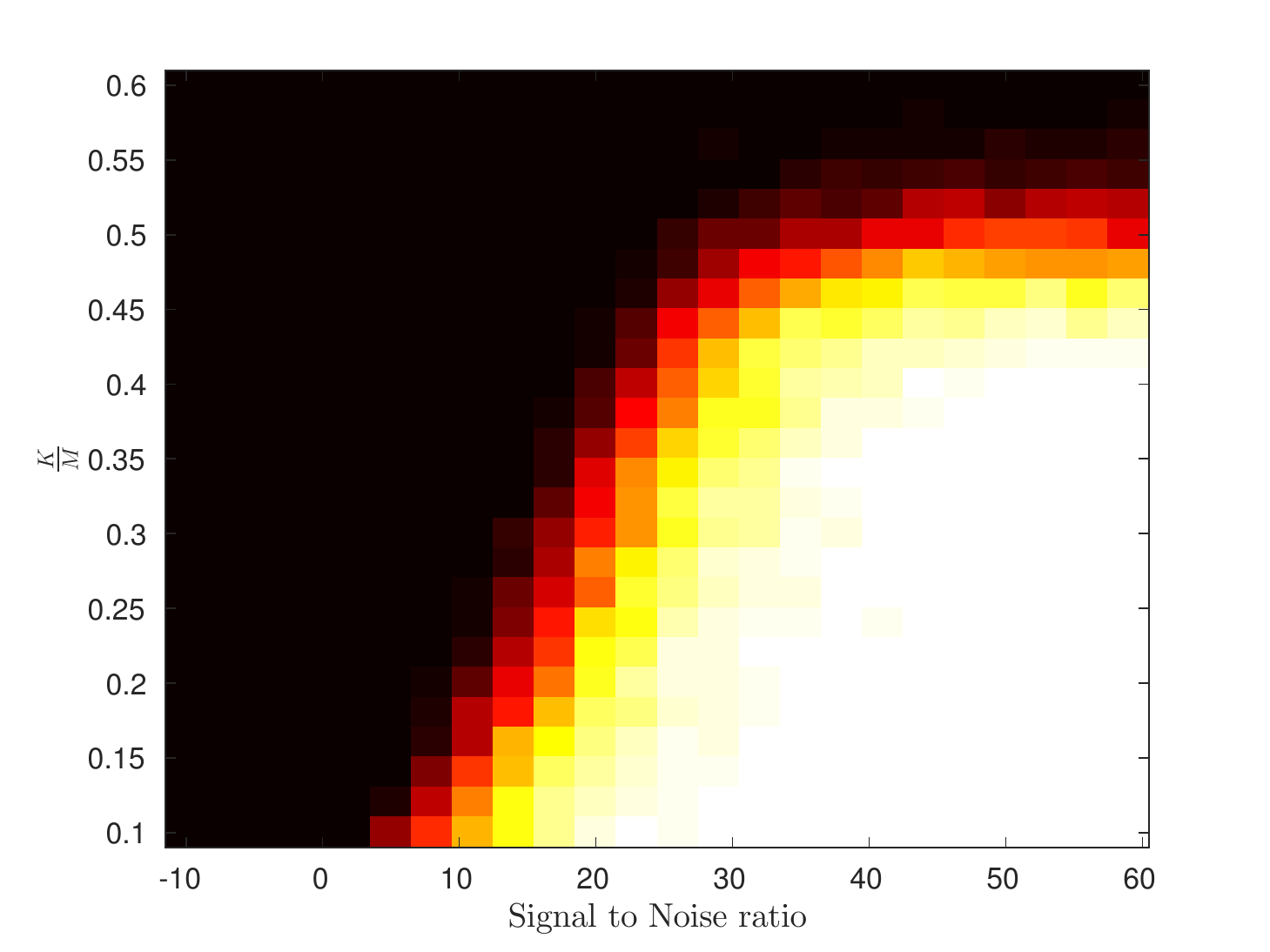} \label{fig:AUC_SNR_G}} \subfigure[Number of tones = $1$]{ 
	\includegraphics[trim=0 0 0 0,clip,width=0.23\linewidth]{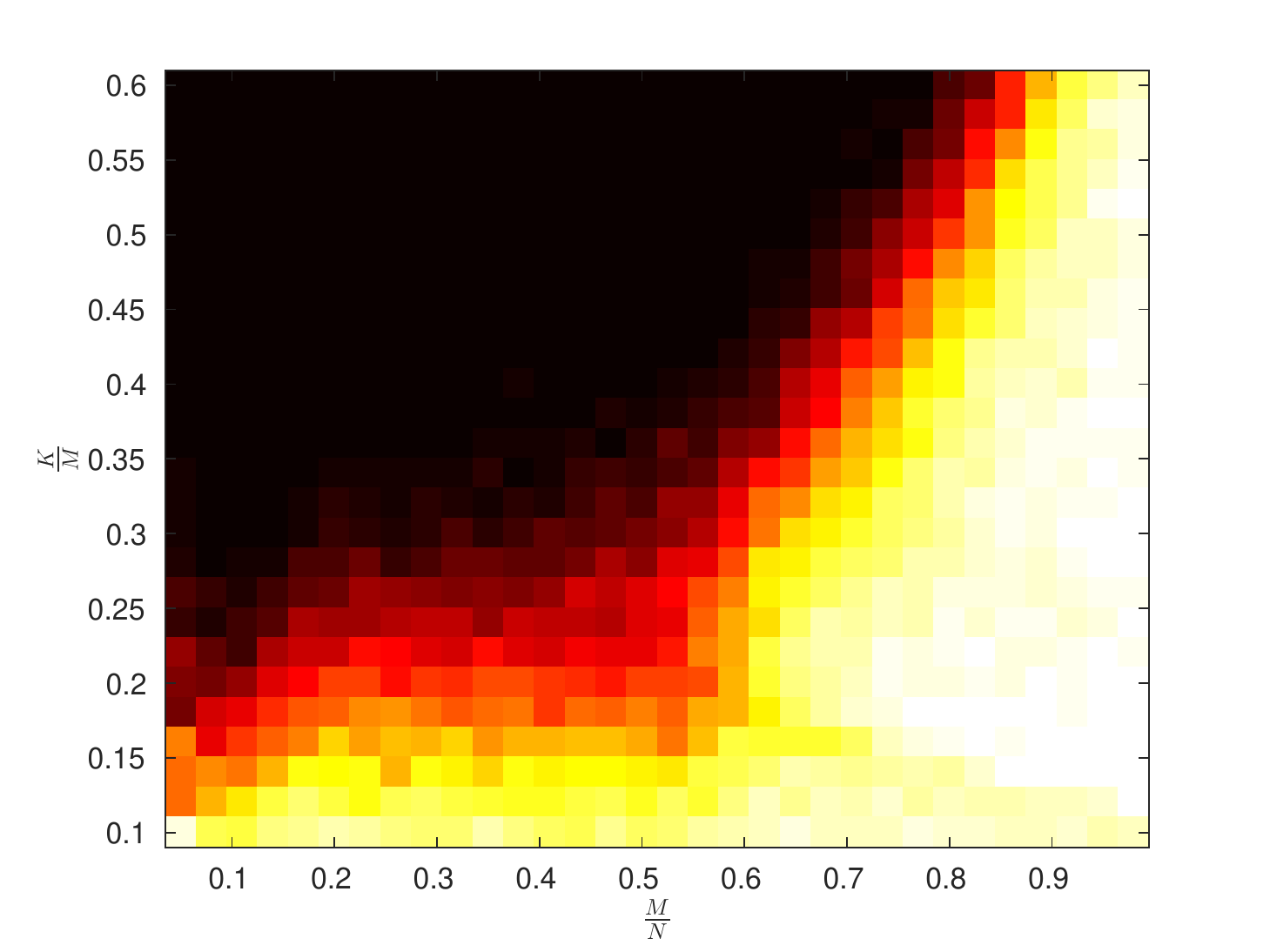} \label{fig:AUC_SNR_1_underSample}} \subfigure[Number of tones = $10$]{ 
	\includegraphics[trim=0 0 0 0,clip,width=0.23\linewidth]{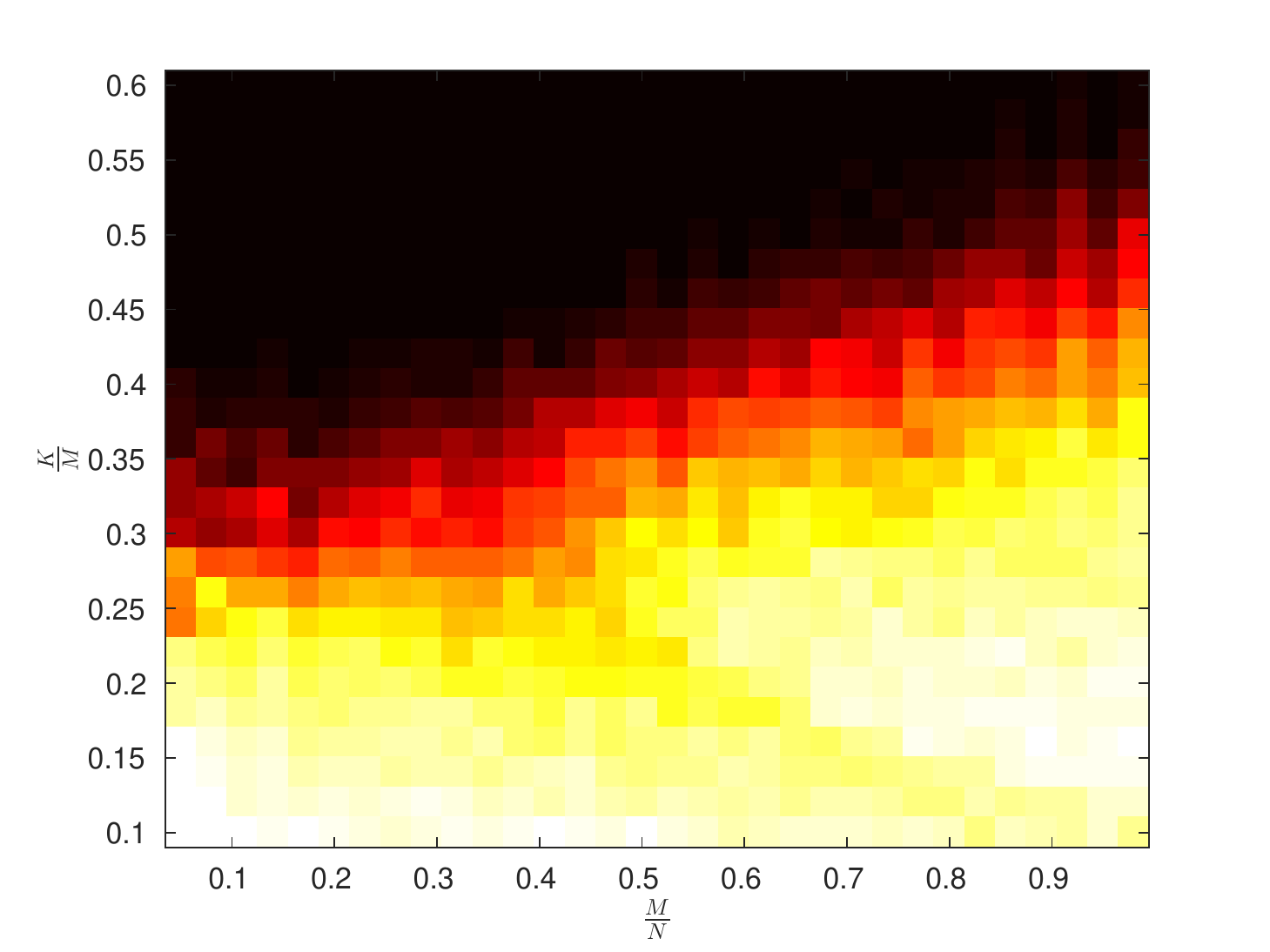} \label{fig:AUC_SNR_10_underSample}} \subfigure[Number of tones = $20$]{ 
	\includegraphics[trim=0 0 0 0,clip,width=0.23\linewidth]{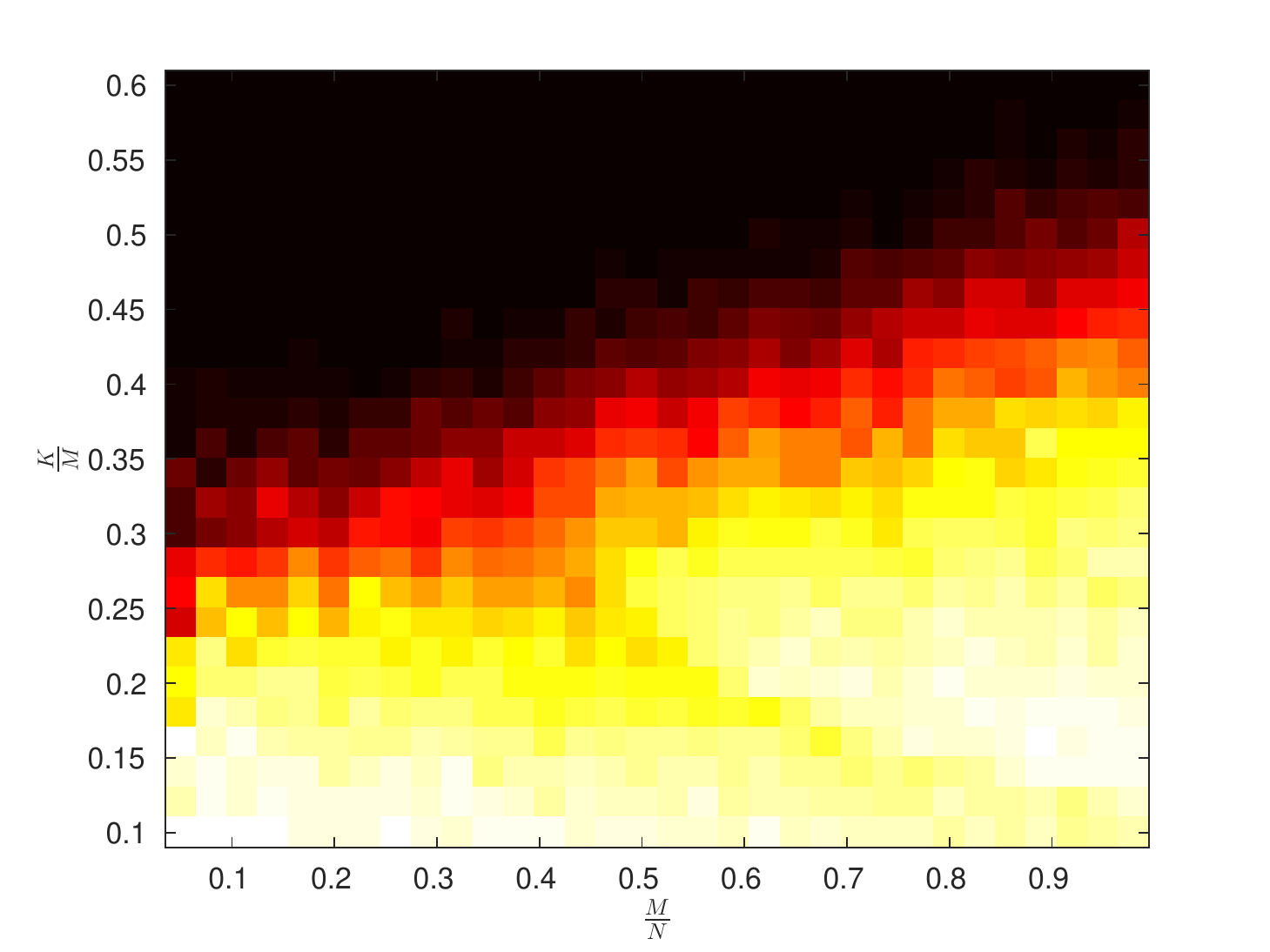} \label{fig:AUC_SNR_20_underSample}} \subfigure[Gaussian Toeplitz system]{ 
	\includegraphics[trim=0 0 0 0,clip,width=0.23\linewidth]{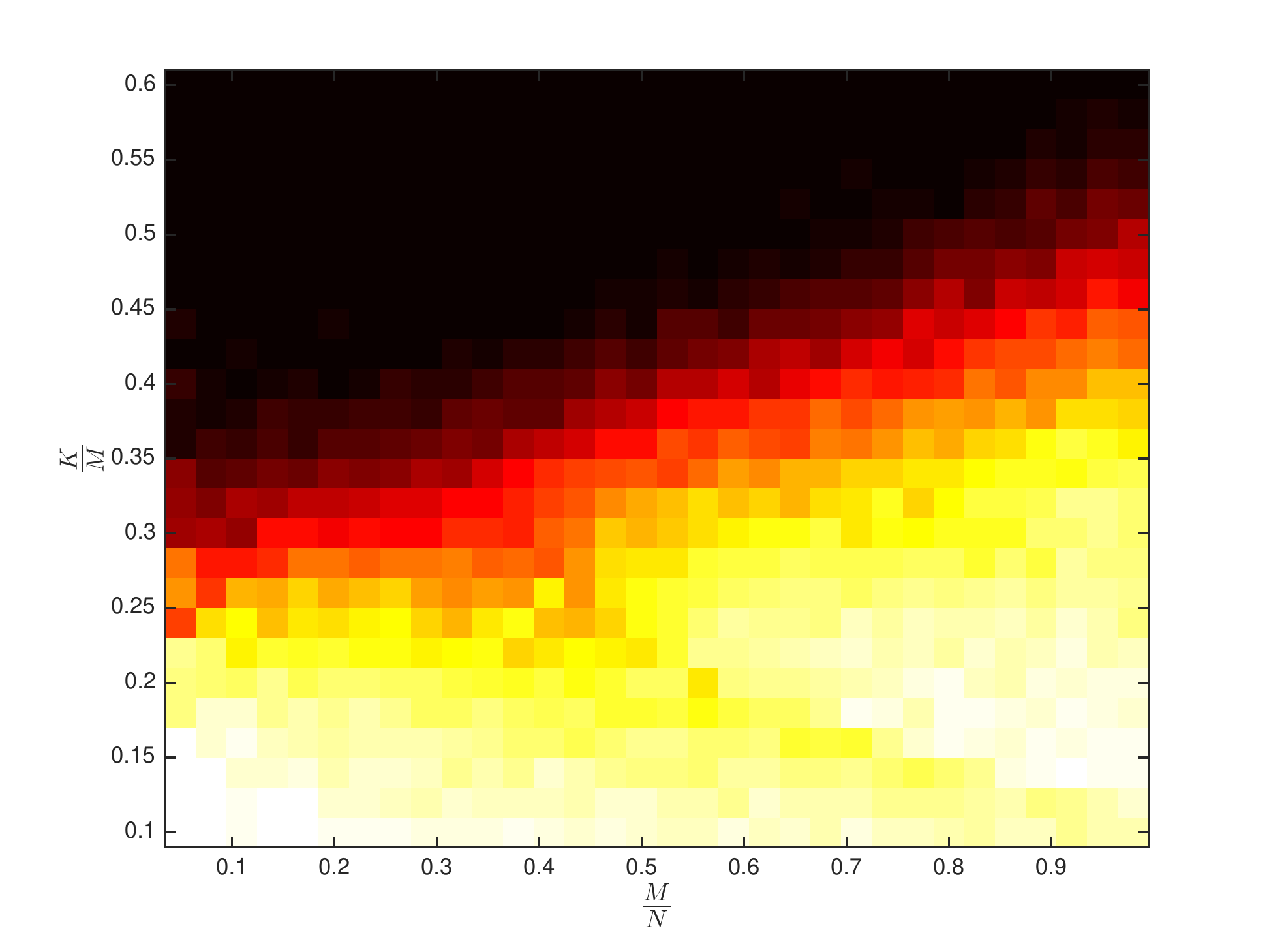} \label{fig:AUC_SNR_G_underSample}} \caption{Figure~\ref{fig:MSE_noiseless_1} to \ref{fig:MSE_noiseless_Gaussian} illustrate the probability of reconstruction error is below $10^{-5} $ in the noiseless setting as a function of under-sampling and sparsity ratio. Figure~\ref{fig:AUC_SNR_1} to \ref{fig:AUC_SNR_G} shows the probability that AUC$> 0.99$ as a function of signal to noise ratio at a fixed under-sampling ratio ${\beta}/{B} = 0.5$. Figure~\ref{fig:AUC_SNR_1_underSample} to \ref{fig:AUC_SNR_G_underSample} shows the probability that AUC$< 0.99$ as a function of under-sampling ratio ${\beta}/{B} ={M}/{N}$ for a fixed SNR of $20dB$. } 
\label{fig:phTransitionOnGrid} \end{figure*}

\begin{figure}
	[!tb]\centering 
	\includegraphics[width=0.8\linewidth]{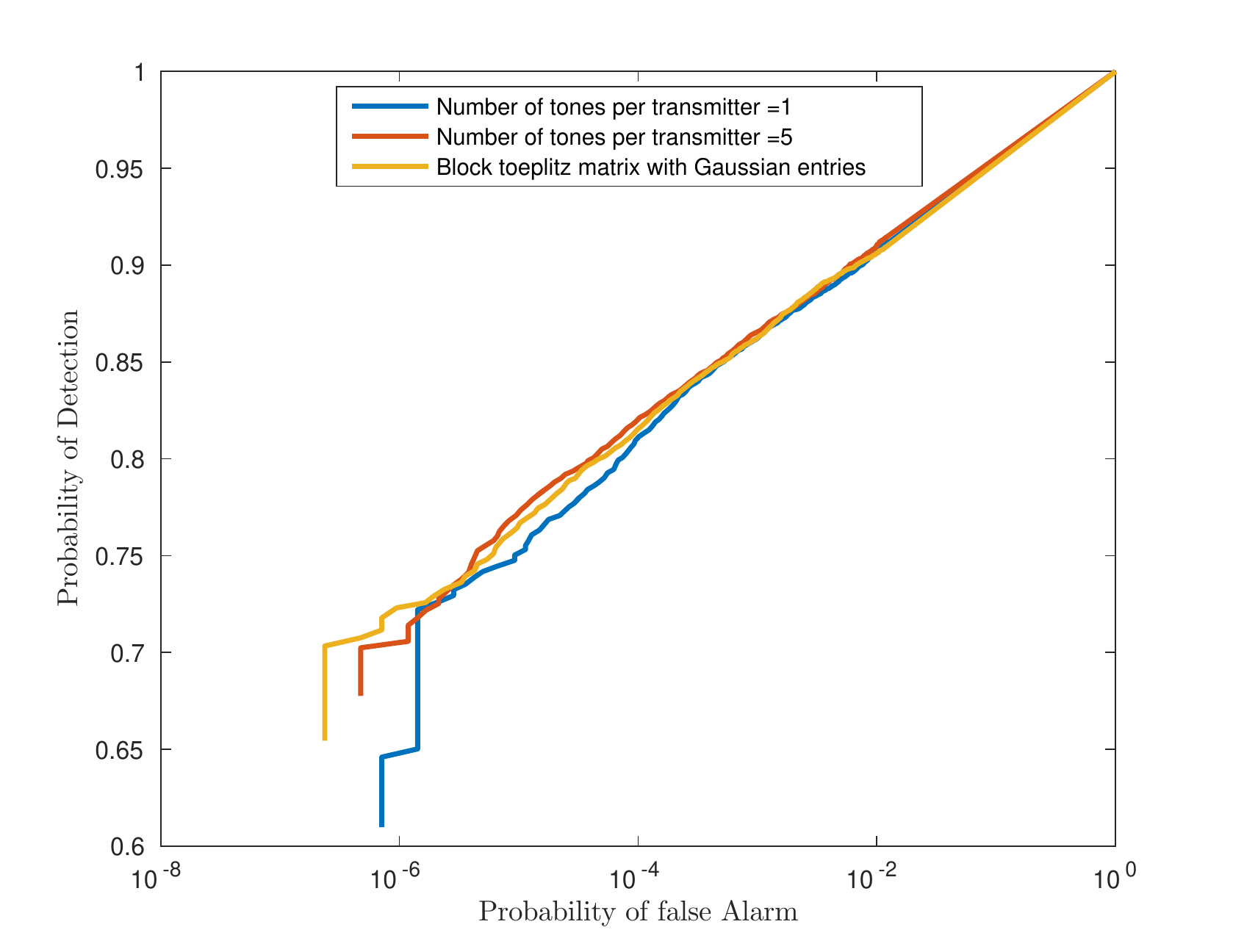} \caption{\label{fig:ROC_120} Receiver operating characteristics with number of targets $K=120$ and SNR = 12dB. The number of transmitters $N_T = 16$, and the number of receivers is $N_R = 8$. The under-sampling ratio is set as $M/N=1/3$. The number of chirps per transmitter is increased and it is compared to the ROC obtained for a system employing random waveform.} 
\end{figure}

First, we consider the noiseless case and consider the reconstruction error as a performance criterion. In figures~\ref{fig:MSE_noiseless_1} to \ref{fig:MSE_noiseless_Gaussian}, the probability of successful recovery (defined as reconstruction error $< 10^{-5}$) is shown as a function of sparsity ratio (the ratio of number of targets in the scene to number of measurements $\frac{K}{M}$) and under-sampling ratio ($\frac{\beta}{B} = \frac{M}{N}$). We observe that for sufficiently high number of modulating tones, the performance characterized by the phase transition plot is similar to that of a system employing stochastic waveforms on transmit. Next, we consider noisy measurements and fix the under-sampling ratio to $0.5$. The perfomance of support recovery is evaluated using probability of detection and false-alarm. The detection is declared true if the recovered signal at a location exceeds the threshold and the target is present at the specified location. All other detections are declared as false positives. The receiver operating characteristics (ROC) curve illustrates the probability of detection and false alarm parametrized by the threshold. We characterize the performance criterion for successful support recovery (defined by the area under the curve (AUC) of ROC exceeding a threshold of 0.99) as a function of the signal to noise ratio (SNR) and sparsity ratio. The results in figures~\ref{fig:AUC_SNR_1} to \ref{fig:AUC_SNR_G} illustrate the probability of successful recovery. Next, we fix the SNR as $20dB$ and study the criterion for support recovery (defined by area under ROC (AUC) exceeding a threshold of 0.99 ) as the under-sampling ratio and the sparsity levels are varied. The probability of successful recovery is shown in figures~\ref{fig:AUC_SNR_1_underSample} to \ref{fig:AUC_SNR_G_underSample}. It can be seen that the performance of the system approaches the performance of the system employing random waveforms as the number of tones is increased . 
Next, we characterize the performance of the MIMO system for support recovery using the Receiver operating characteristics for successful support recovery in figure~\ref{fig:ROC_120}. The noise is generated from a complex Gaussian distribution with the variance that guarantees the SNR of $12dB$. The target support is generated uniformly from the grid and the amplitudes are sampled from a complex Gaussian distribution. We fix the number of targets $K=120$ and consider a MIMO system with $N_T=16$ transmitters and $N_R=8$ receivers and vary the number of modulating tones per transmitter. We set the under-sampling ratio as $M/N = 1/3$. We also compare the ROC curve for a system employing random waveform with samples from a Gaussian distribution. 
It can be seen that the performance of the proposed reduced complexity compressive system approaches the performance of the random code system for small number of modulating tones at each transmitter. 
\begin{figure*}[!tb]
	\subfigure[$\mbf{A}$ with $N_c=1$ Chirps]{ 
	\includegraphics[trim=0 0 0 0,clip,width=0.3\linewidth]{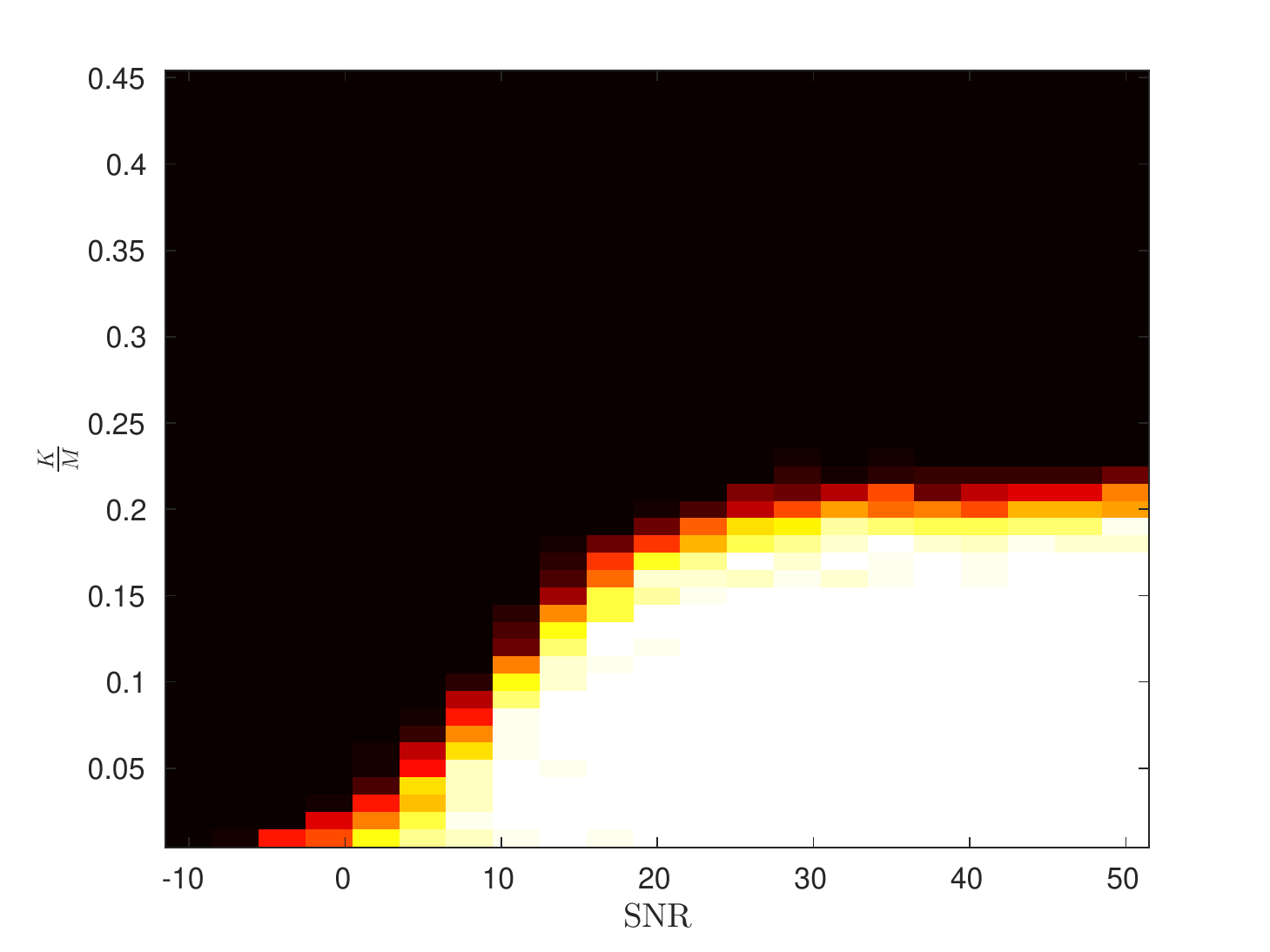} \label{fig:AUC_1Chirps_95}} \subfigure[$\mbf{A}$ with $N_c=5$ Chirps]{ 
	\includegraphics[trim=0 0 0 0,clip,width=0.3\linewidth]{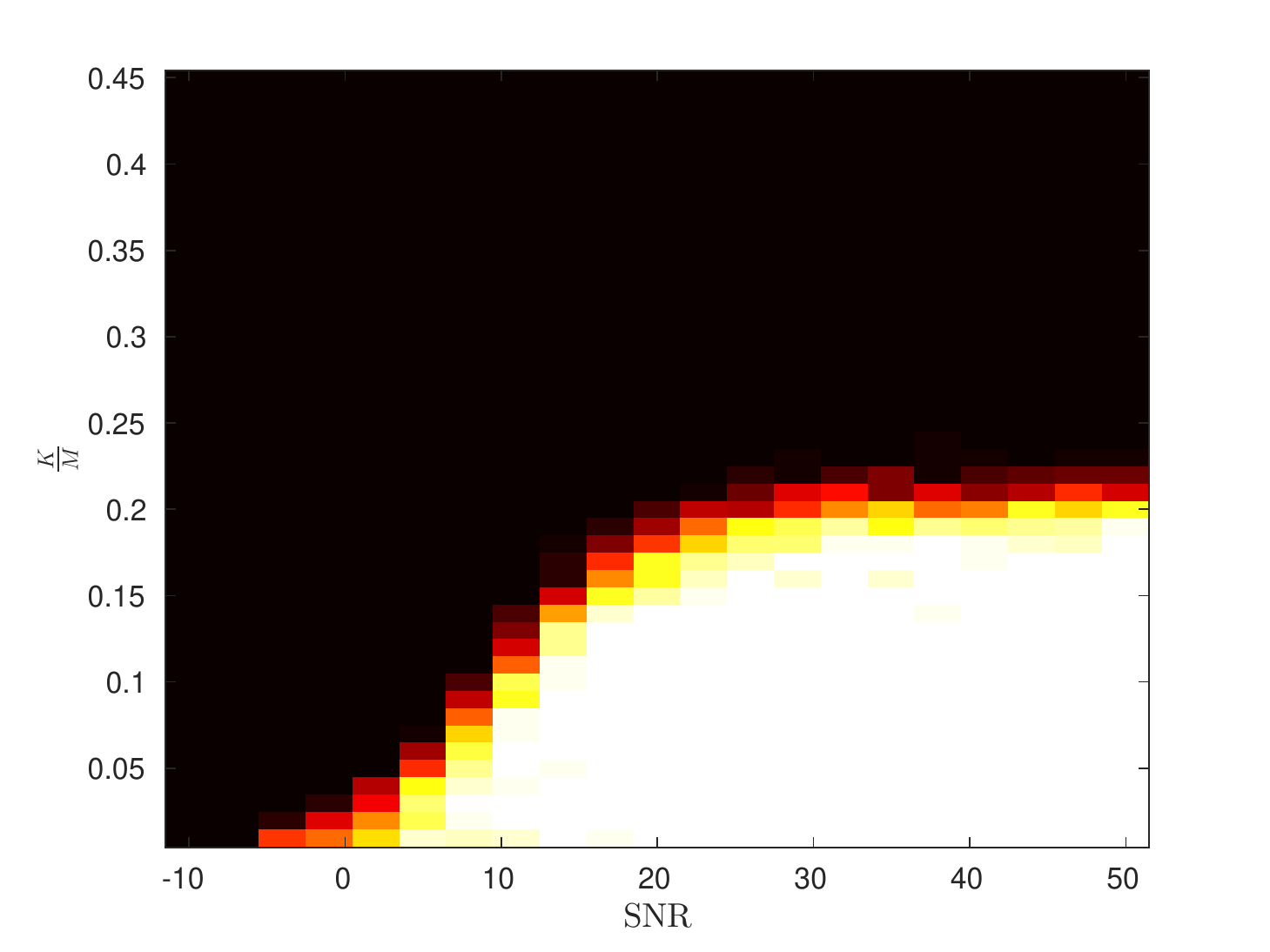} \label{fig:AUC_5Chirps_95}} \subfigure[Teoplitz sensing matrix ]{ 
	\includegraphics[trim=0 0 0 0,clip,width=0.3\linewidth]{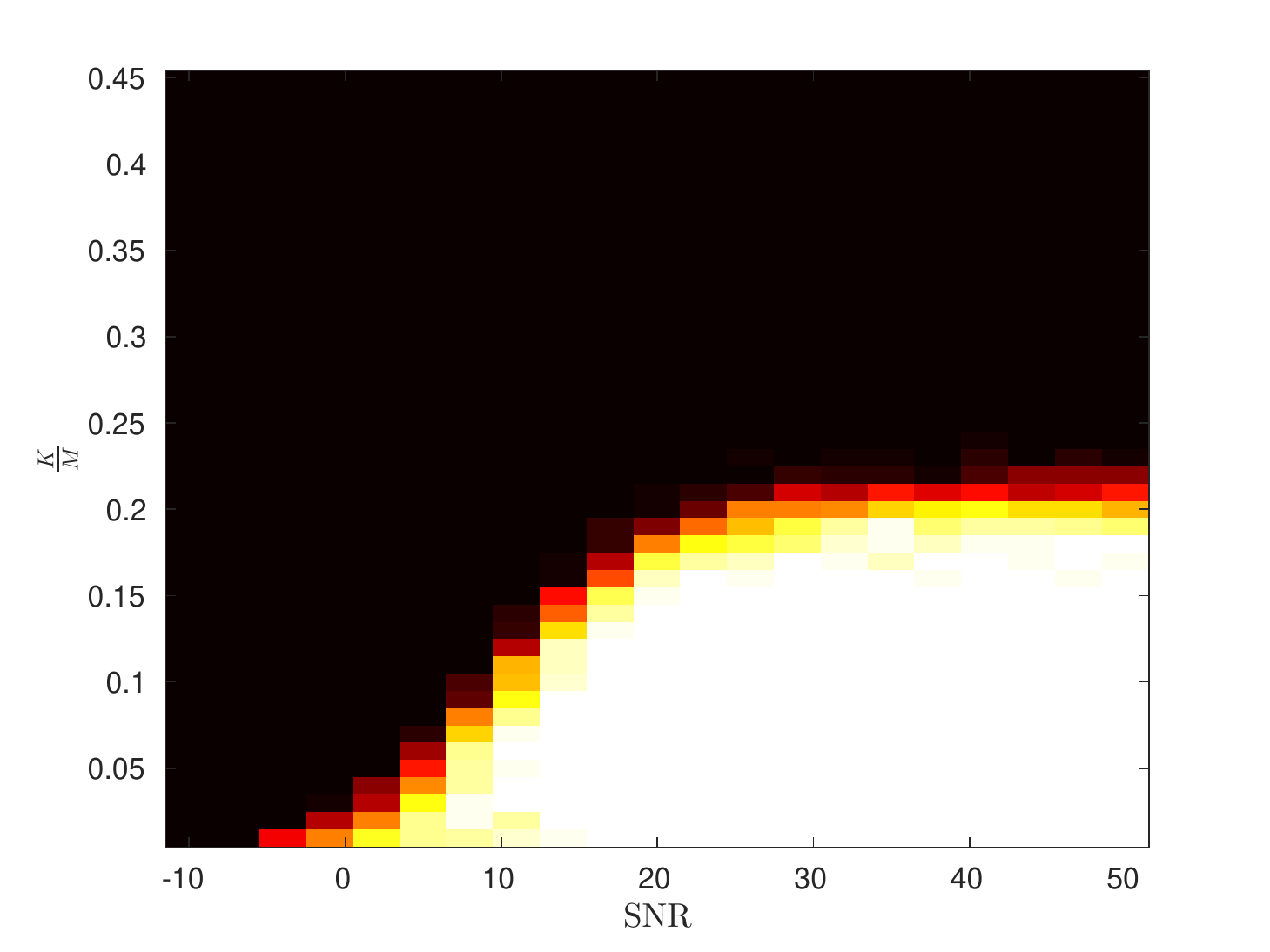} \label{fig:AUC_toeplitz_95}} \caption{The intensity values of the images is 1 if the Area under the curve for receiver operating characteristics curve exceeds 0.95 and 0 otherwise as the SNR and number of targets in the scene is varied. The under-sampling ratio is set as $M/N=1/3$. The number of transmitters and receivers in the system are $N_T= 16$, and $N_R =8$, respectively. } 
\label{fig:RecoveryGuaranteeSNR} \end{figure*}
We further show the effect of noise variance on the support recovery guarantee in the form of a phase transition diagram in figure~\ref{fig:RecoveryGuaranteeSNR}, where the criterion used is even that the area under the ROC curve exceeds $AUC > 0.95$. We observe a region in noise variance and number of targets where the support recovery is guaranteed with high probability.

\subsection{Off-grid recovery} \label{sec:offGrid} Next, we consider targets
that lie in the continuous range and angle-of-arrival domain. We define the parameter space $\Omega = \left\{ \left(\Delta,\theta \right) \vert \Delta \in (0,t_u), \theta \in (-1,1) \right\}$ The samples at the stretch processor's output at receiver $k$ due
to a target with a time of arrival given by $\Delta$ and angle of arrival
$\theta$ as stated in \eqref{eq:atomsMIMO}. For a scene containing $K$
scattering centers, the measurements are given by 
\begin{align}
	&\mbf{y} = \sum_{k=1}^K x_k \boldsymbol{\Psi}\pb{\Delta_k,\theta_k} +
\mbf{w}, \label{eq:atomsMIMO} \\
 &\boldsymbol{\Psi}\pb{\Delta_k,\theta_k} = \sum_{v=1}^{N_T}\sum_{i=1}^{N_c}\exp \left(-j2\left(\pi d_T v \theta_k+ \phi_{v,i}\right)\right) \boldsymbol{\alpha_R}\left(\theta_k\right)\nonumber\\
 & \otimes \exp \left( j2\pi \left[ f_{v,i}\left( \frac{\mbf{m}}{F_s} -\Delta_k \right) + \frac{\beta \Delta_k \mbf{m}}{\tau F_s}\right]\right)
\end{align}
where $f_{v,i}$ is the frequency of the modulating tone utilized in transmitter $v$, $\mbf{w} \sim \mc{CN}\pb{0,\sigma_n^2 \mbf{I}}$ is the receiver noise
following a complex Gaussian 
distribution, $x_k$ are the complex scattering coefficients, $\Delta_k,\theta_k$
are the delay and angle of arrival for each scattering center,$\mbf{m} =[0,1,\cdots,M-1]$ denote the $M$ time samples at each receiver, and
$\boldsymbol{\Psi}$ is the known structured response parametrized by the time
and angle of arrival of the scattering center due to the proposed illumination
scheme. We utilize the differentiability of the measurement model in the unknown
range of the targets in the scene and adopt the method proposed in
\cite{ADCGM_Recht_2017} to solve the sparse estimation problem in the continuum.

Algorithm~\ref{algo:SolutionOffGrid} provides the details
of the method used to solve the estimation problem with sparsity constraints. The method first selects the most explanatory choice of parameters in the parameter space using the residual as shown in \eqref{eq:matchingPursuit}. Next, the weights and the support are refined jointly. This non-convex problem of jointly estimating the weights and the parameters is solved by an alternating minimization approach. The weights are estimated by solving the finite dimensional problem on the detected support set by enforcing the $\ell_1$ constraint on the weights.  The support set is pruned such that only non-zero points in the support set are retained. Next, the support set is refined using the gradient information with the steepest descent method with line search. We consider the convergence condition as a combination of the residual error and the reduction in the loss function. 
\begin{algo}{Alternating descent conditional gradient method\cite{ADCGM_Recht_2017}}
\label{algo:SolutionOffGrid}
\begin{tabular}{|p {0.9\linewidth}|}
\hline\\
    \textbf{Input:} $\mbf{y}$,$\tau$, $\boldsymbol{\Psi}$, $\nabla_{\Theta \in \Omega} \boldsymbol{\Psi}$, $\Omega$, and $K_{max}$.  \\
     \textbf{Return:}  complex weights $\mbf{x}$, delay and angle of arrival of scattering centers $\cb{\boldsymbol{\theta,\Delta}} \in \Omega$.  \\
     \hline
      \textbf{Initialize} $k=0$, support set $S = \cb{\emptyset}$ \\
      \textbf{while} (Convergence condition is not satisfied or $k \leq K_{max}$)     
      { \begin{align}        
        &\text{Residual: } \mbf{r}_k = \mbf{y} - \sum_{i=1}^{k-1} \boldsymbol{\Psi}(\Delta_k,\theta_k) x_k ,\label{eq:matchingPursuit} \\
        &\text{Gradient of loss function: } \mbf{g}_k(\mbf{r}_k) = \nabla_\mbf{r} \pb{0.5 \norm{\mbf{r}_k }_2^2} \\
         &\cb{\Delta_k,\mbf{\theta}_k} = \argmax_{\cb{\Delta,\theta} \in \Omega} \abs{\inProd{\Psi\pb{\Delta,\theta}}{\mbf{g}_k}} ,\\
        &S = S \bigcup \cb{\Delta_k,\mbf{\theta}_k}
        \end{align}}
      \qquad \textbf{while} (Convergence condition) 
      { \begin{align} 
       & \text{ Compute weights: } \argmin_{\st{\mbf{x}}{\norm{\mbf{x}}_1 \leq \tau }} \norm{\boldsymbol{\Psi}_S\mbf{x} - \mbf{y}}^2\\
        &\text{ Prune Support: If } \abs{x_k} =0 \quad S = S \setminus \cb{\Delta_k,\theta_k} \\   
       &\text{ Refine support: } S = S - \nabla_S \norm{\boldsymbol{\Psi}_S\mbf{x} - \mbf{y}}^2 
       \end{align} }
    \qquad \textbf{end} \\
    \qquad $k= \abs{S}$ \\ 
    \textbf{end} \\ \hline
\end{tabular}
\end{algo}

\begin{figure*}
	\centering 
	\subfigure[$m_1$]{ 
	\includegraphics[trim=0 0 0 0,clip,width=0.3\linewidth]{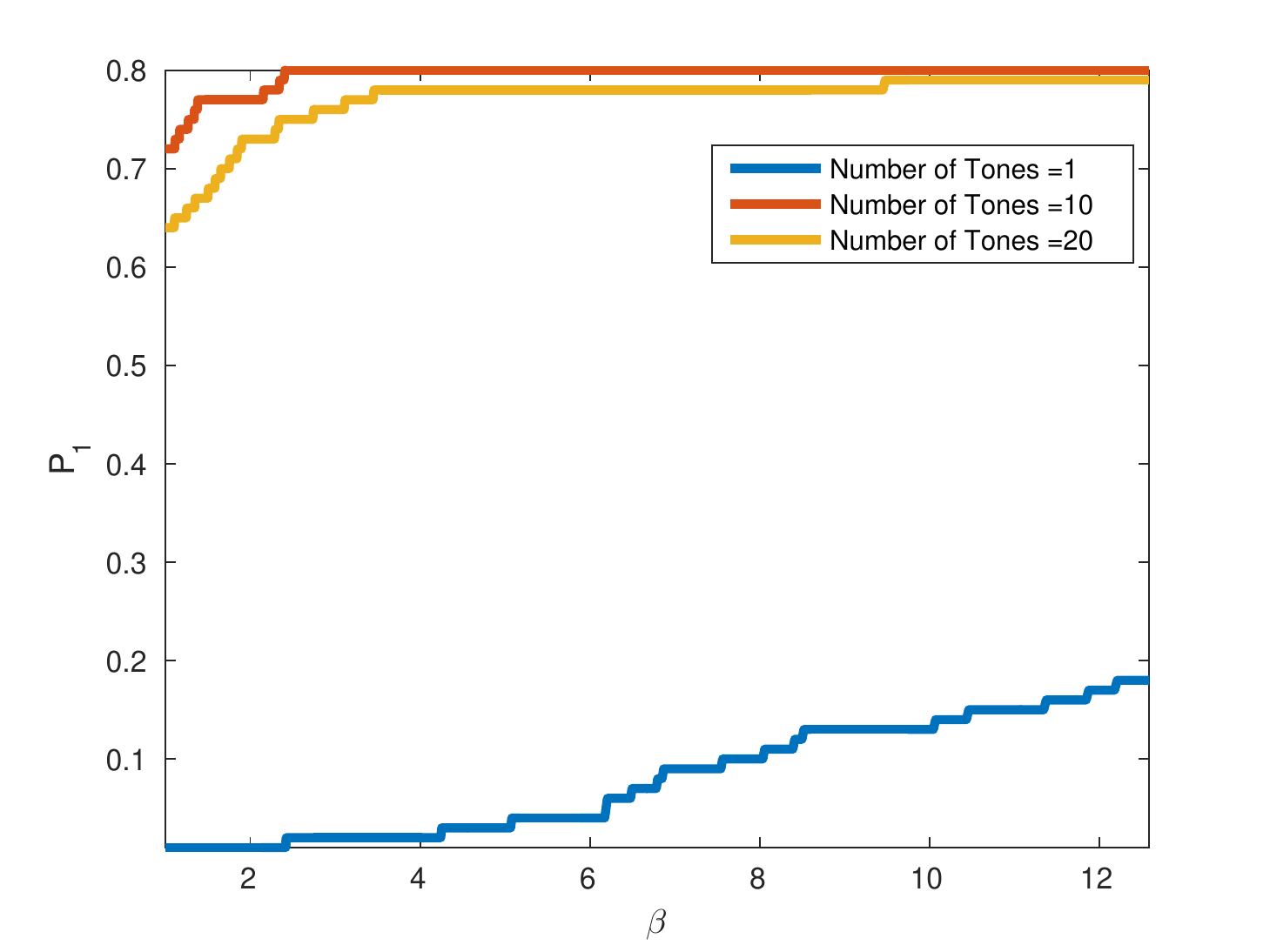} \label{fig:p1_offGrid}} \subfigure[$m_2$]{ 
	\includegraphics[trim=0 0 0 0,clip,width=0.3\linewidth]{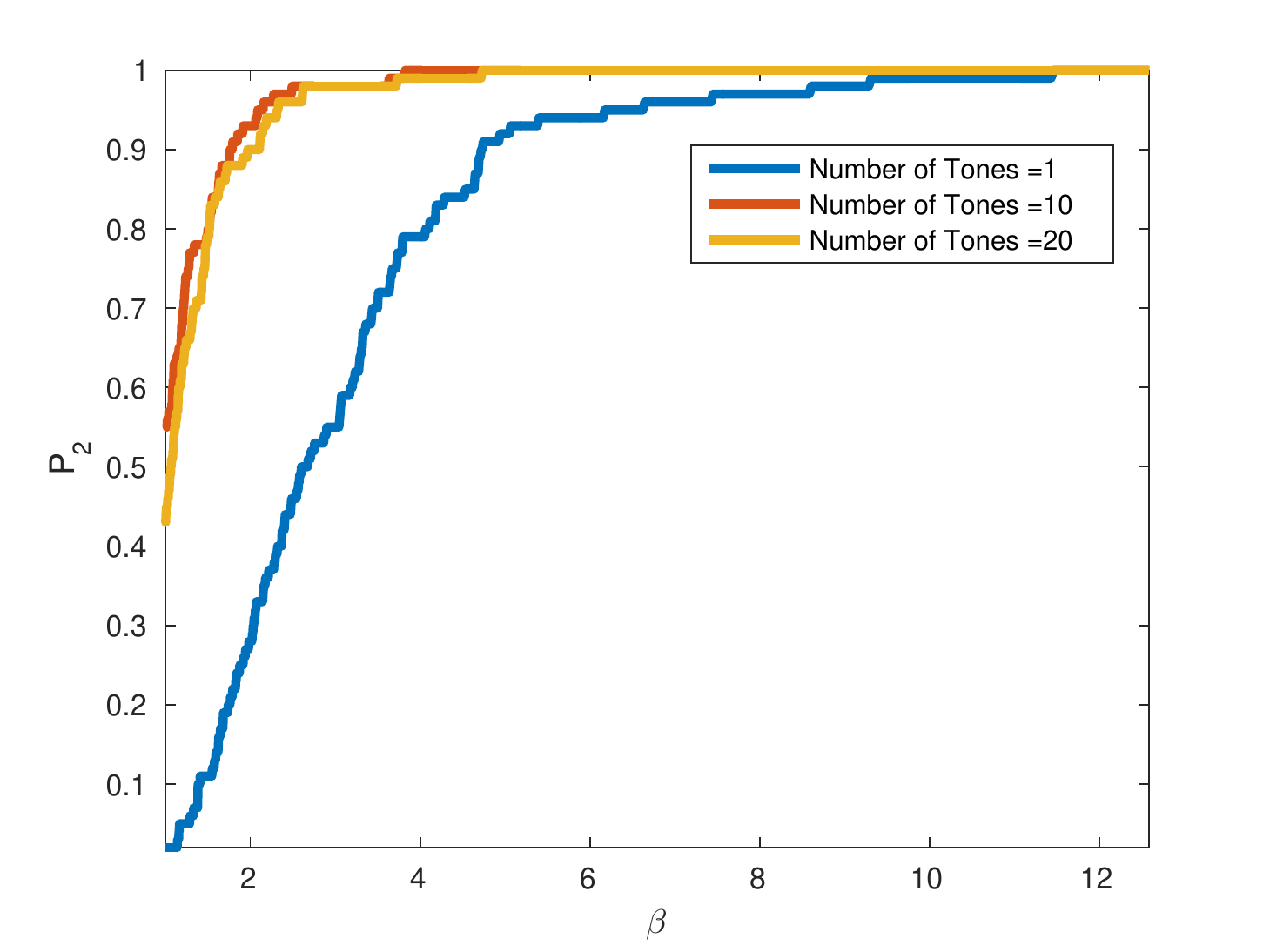} \label{fig:p2_offGrid}} \subfigure[$m_3$]{ 
	\includegraphics[trim=0 0 0 0,clip,width=0.3\linewidth]{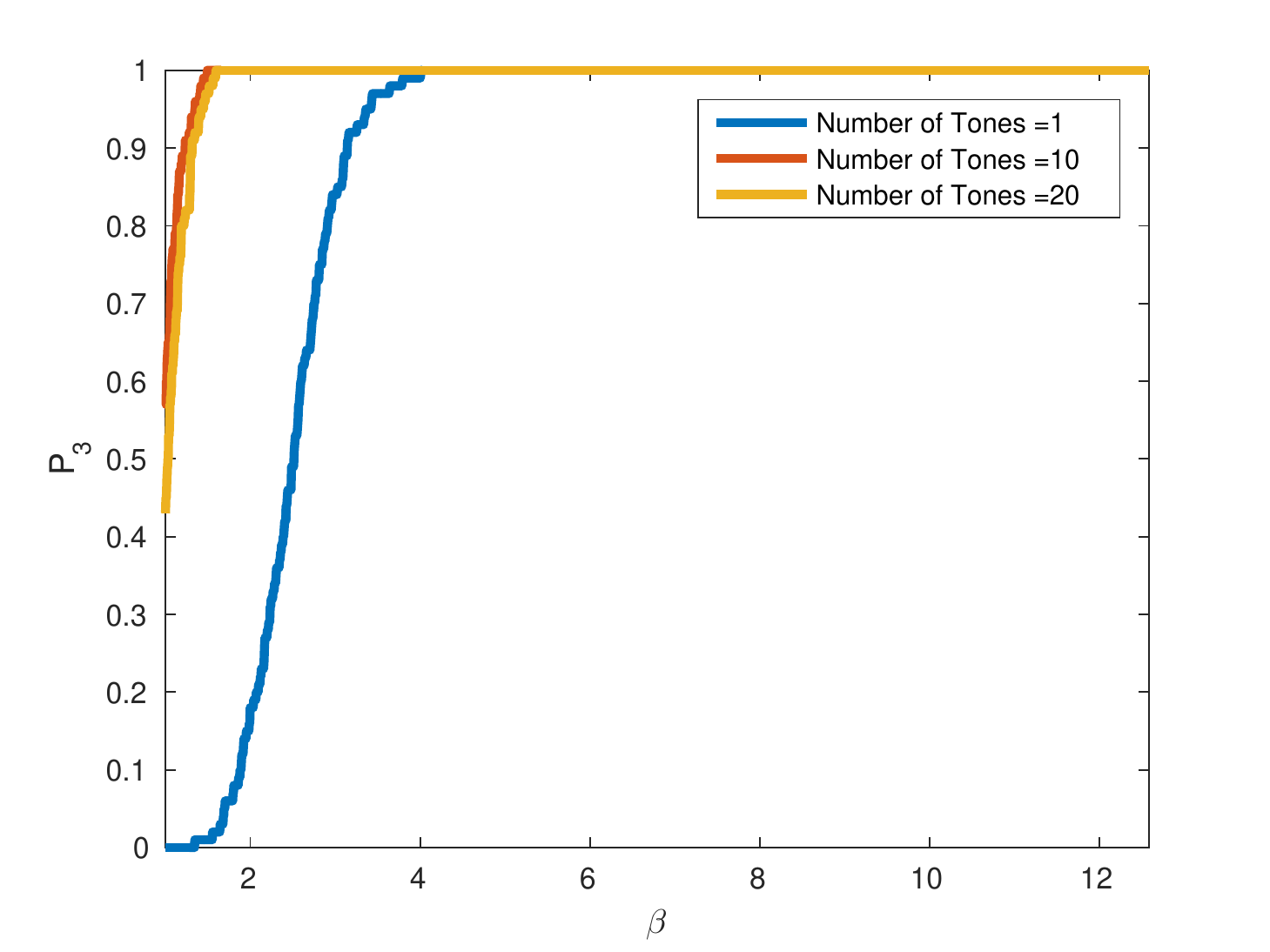} \label{fig:p3_offGrid}} \caption{Performance profile for different number of modulating tones with under-sampling ratio ${\beta}/{B} = {1}/{3}$, $K=20$ and $SNR=12dB$. Figures~\ref{fig:p1_offGrid},~\ref{fig:p2_offGrid}, and ~\ref{fig:p3_offGrid} show the perfomance profile corresponding to metrics indicating the false positives, localization error, and approximation error, respectively. } 
\label{fig:rec_off_grid_range} \end{figure*}

We consider a single input single output system for estimating the range and evaluate the performance of the illumination scheme with off-grid targets. We conduct the simulations with an under-sampling ratio as $\frac{M}{N} = 1/3$ and the SNR of $12dB$.
The number of targets in the scene is $20$ using the model specified in section~\ref{sec:tarModel}. We compare the performance of the system as the number of modulating tones is varied using the metrics defined in \cite{LineSpectralEstimation_tang}. We define the set of true range as $\mc{T}= \cb{r_i} \subset \Omega$ with complex scattering coefficients $\cb{x_i}$ for $i=1,\cdots,K$, where $K$ is the number of targets in the scene. We define $N_{r_i}$ as the set of values of range that are in a neighborhood of the true range $r_i$, such that $N_{r_i} =\cb{r : \abs{r-r_i} \leq 0.2 c/\pb{2B}}.$ We define the region of false detections as $\mc{F} = \Omega \setminus \cb{\cup_i N_{r_i}}$. We consider the following performance measures to evaluate the estimate $\cb{\hat{r}_i}$, and $\cb{\hat{x}_i}$ obtained using the algorithm given by 
\begin{itemize}
	\item error due to false detections given by
	\[m_1 = \sum_{\hat{r}_i \in \mc{F}} \abs{\hat{x}_i},\]
	\item weighted localization error
	\[m_2 = \sum_j \sum_{i : \hat{r}_i \in N_{r_j}} \abs{\hat{x}_i}\min_{r\in \mc{T}} \norm{\hat{r}_i -r }^2, \]
	\item approximation error in the scattering coefficients
	\[m_3 = \max_{j} \abs{x_j - \sum_{l :\hat{r}_l \in N_{r_j}} \hat{x}_l}.\]
\end{itemize}
We evaluate the performance profile, which is studied in \cite{LineSpectralEstimation_tang} to compare the various algorithms for recovery. In our case, we compare the performance of the system by keeping the recovery step fixed and varying the number of modulating tones. The set of tones used is denoted by $\mc{S} = \cb{1,10,20}$. The performance profile is evaluated by repeating the experiment for different realizations of target denoted by the set $\mc{P}$. The performance profile for the system parameter $s \in \mc{S}$, error metric $m_i$, and factor $\eta$, which specifies the ratio $m_i(p,s)/\min_s m_i(p,s)$ is computed as follows 
\begin{align}
	P_s(\eta;i) =\frac{ \mathbf{card}\cb{p\in \mc{P} : m_i(p,s) \leq \eta \min_s m_i(p,s)}}{\mathbf{card}\cb{\mc{P}}}. 
\end{align}
The performance profile of system $s$ indicates the number of realizations such that the error metric $m_i(p,s)$ for the realization $p$ is within a factor of $\beta$ from the error metric corresponding to the best system parameter. Figure~\ref{fig:rec_off_grid_range} shows the profile evaluated for all the error metrics computed using $100$ target realizations. We observe that as the number of modulating tones are increased, the performance improves even if the targets do not lie on the grid. 

\section{Proofs}\label{sec:Proofs} 
In order to obtain the non-asymptotic recovery guarantee for our system, we estimate the tail bounds for mutual coherence and spectral or operator norm of the measurement matrix. We make use of the Matrix Bernstein inequality given in lemma~\ref{lem:MatBernstein} to bound the operator norm of the measurement matrix with high probability given in \eqref{eq:sensingMatrixMIMO}. 
\begin{lemma}\label{lem:opNormTailBound} 
	The operator norm of the sensing matrix in \eqref{eq:sensingMatrixMIMO} is bounded with high probability by 
	\begin{align}\label{eq:MeasMatOpNormMIMO} 
		\norm{\boldsymbol{\mc{A}}}_{op} \leq 2\sqrt{\frac{N_T N}{M} \log\pb{N_RM + N_R N_T N}}. 
	\end{align}
\end{lemma}
\begin{IEEEproof}
	Let $\mbf{P}_i = \hat{c}_i \pb{\mbf{\bar{\alpha}_R \bar{\alpha}_T(\xi(i))}\otimes \pb{\mbf{H}_i \mbf{\bar{A}} \mbf{D}_i}}.$ We obtain the following bounds using the results from lemma~\ref{lem:normBound} and lemma~\ref{lem:varianceBound}, $\norm{\mbf{P}_i} \leq \sqrt{\frac{N}{M N_c}},$ 
	\begin{align*}
		\nu\pb{\boldsymbol{\mc{A}}} &=\max\pb{\norm{\sum_{i=1}^N \Expec{\mbf{P}_i \mbf{P}^{*}_i}}_{op},\norm{\sum_{i=1}^N \Expec{\mbf{P}^{*}_i \mbf{P}_i}}_{op}} \nn \\
		&\leq \frac{N_T N}{M}, 
	\end{align*}
	respectively. The upper bound on the operator norm, which is satisfied with high probability can be obtained by using $t=2\sqrt{N N_T/M \log \pb{NN_TN_R + N_R M }}$ in the bound in lemma~\ref{lem:MatBernstein} as shown below 
	\begin{align*}
		&P\pb{\norm{\boldsymbol{\mc{A}}}_{op} \geq 2\sqrt{\frac{N_T N}{M} \log\pb{N_RM + N_R N_T N}}} \nn \\
		& \leq \pb{\frac{1}{N_R M + N_T N_R N}}^{\alpha_2 -1}, \text{where } \\
		&\alpha_2 = \frac{1}{\frac{1}{3}\sqrt{\frac{1}{N_T N_C} \log\pb{N_R M + N_R N_T N}} + \frac{1}{2}}. 
	\end{align*}
	For the tail probability to decay, we require that $\alpha_2 > 1$, which implies $N_c \geq {4}/\pb{9 N_T} \log\pb{N N_R N_T +MN_R}$. 
\end{IEEEproof}
The following results on the Euclidean norm of columns and the mutual coherence are obtained using concentration inequalities for quadratic forms of random vectors having a sub-Gaussian distribution given in \cite{HansonWrightSubGaussian} and lemma~\ref{lem:HansonWrightCom}.  	
\begin{lemma}\label{lem:minColumnNormTailBound} 
	The minimum of the corollary-Euclidean norm of any column of $\boldsymbol{\mc{A}}$, which is indexed by range bin $m$ angle bin $s$ is bounded by 
	\begin{align}\label{eq:normColumnTailMIMO} 
		\abs{\min_{m,s} \norm{\boldsymbol{\mc{A}}(m,s)}_2^2 -1} \leq \epsilon 
	\end{align}
	with high probability, where $\epsilon \in (0,1)$ is an arbitrary constant. 
\end{lemma}
\begin{IEEEproof}
	The norm of a column indexed by range bin $m$ and angle bin $s$ of the sensing matrix $\boldsymbol{\mc{A}}$ can be written as follows 
	\begin{align*}
		\norm{\boldsymbol{\mc{A}}(m,s)}_2^2 &= \mbf{\hat{c}}^{*} \mbf{B} \mbf{\hat{c}}, 
	\end{align*}
	where $\mbf{B} = \mbf{G}_m^{*} \mbf{F}^{*} \mbf{E}_m^{*} \mbf{E}_m \mbf{F} \mbf{G}_m$, and $\mbf{\hat{c}} \in \mbb{C}^N$ is a sequence of random variables that selects and scales a subset of the modulated waveforms. Since the random variables $\hat{c}_i$ are independent and $\Expec{\hat{c}_i} = 0$, the off-diagonal terms vanish and we get 
	\begin{align*}
		\Expec{\norm{\boldsymbol{\mc{A}}(m,s)}_2^2} = 1. 
	\end{align*}
	Using results from lemma~\ref{lem:FroNormOpNormMatrix} and lemma~\ref{lem:HansonWrightCom} along with the result on sub-Gaussian norm from lemma~\ref{lem:subGaussNormRes}, and using the approximation $\ceil*{\frac{N}{M}} \geq \frac{N}{M}$ we have 
	\begin{align}\label{eq:normTailMIMO} 
		&P\pb{\abs{\norm{\boldsymbol{\mc{A}}(m,s)}_2^2- 1} > \epsilon} \leq \nn \\
		& 8 \exp \pb{-\frac{dM q^{*} \min \pb{\frac{q^{*} \epsilon^2}{\pb{\frac{N_c N_T}{N}}^{\pb{\frac{2}{q^{*}}}-1} } ,\epsilon }}{\pb{\frac{N_c N_T}{N}}^{\pb{\frac{2}{q^{*}}} -1} } }, 
	\end{align}
	equivalently, 
	\begin{align}
		& P\pb{\norm{\boldsymbol{\mc{A}}(m,s)}_2^2 < 1 - \epsilon} \leq \nn \\
		& 8 \exp \pb{-\frac{dM q^{*}\min \pb{\frac{q^{*} \epsilon^2}{\pb{\frac{N_c N_T}{N}}^{\pb{\frac{2}{q^{*}}}-1} } ,\epsilon }}{\pb{\frac{N_c N_T}{N}}^{\pb{\frac{2}{q^{*}}} -1} } }, \nn 
	\end{align}
	where $q^{*} = \max\pb{1,2\log\pb{\frac{N}{N_c N_T}}}$. The concentration inequality for any $\epsilon \in [0,1] $ can be written as 
	\begin{align}
		&P\pb{\min_{m,s} \norm{\boldsymbol{\mc{A}}(m,s)}_2^2 \leq {1 - \epsilon}} \leq \nn \\
		&8 N \exp\pb{-M d \pb{\epsilon \frac{q^{*}}{\pb{\frac{N_c N_T}{N}}^{\frac{2}{q^{*}} -1 }} }^2 }. 
	\end{align}
\end{IEEEproof}
\begin{lemma}\label{lem:mutualCoherenceTailBound} 
	The mutual coherence of the sensing matrix $\boldsymbol{\mc{A}}$ scales as 
	\begin{align}\label{eq:mutualCohTailMIMO} 
		\mu\pb{\boldsymbol{\mc{A}}} \sim \mc{O}\pb{ \sqrt{\frac{\log(\frac{N N_R N_T}{\epsilon})}{M}}}, 
	\end{align}
	with high probability. 
\end{lemma}
\begin{IEEEproof}
	We can express the inner-product between any two columns of sensing matrix indexed by $\cb{m_1,s_1}$ and $\cb{m_2,s_2}$ as 
	\begin{align*}
		\abs{\inProd{\boldsymbol{\mc{A}}(m_1,s_1)}{ \boldsymbol{\mc{A}}(m_2,s_2)}} \leq \abs{\mbf{\hat{c}}^{*} \bar{\mbf{B}}\mbf{\hat{c}}}, 
	\end{align*}
	where $\bar{\mbf{B}}=\mbf{G}_{m_1}^{*} \mbf{F}^{*} \mbf{E}_{m_1}^{*} \mbf{E}_{m_2} \mbf{F} \mbf{G}_{m_2}$ and the inequality is obtained by bounding $\abs{\inProd{\boldsymbol{\alpha_R(\theta_{s_1})}}{\boldsymbol{\alpha_R(\theta_{ s_2})}} } = 1$, if $\theta_{s_1}=\theta_{s_2}$ and $\abs{\inProd{\boldsymbol{\alpha_R(\theta_{s_1})}}{\boldsymbol{\alpha_R(\theta_{ s_2})}} } = 0$, otherwise. By using the fact that $\hat{c}_i$ are zero mean independent random variables, we obtain the following 
	\begin{align*}
		\Expec{\mbf{\hat{c}}^{*} \bar{\mbf{B}} \mbf{\hat{c}}} =0. 
	\end{align*}
	Using the results from lemma~\ref{lem:FroNormOpNormMatrix} and lemma~\ref{lem:HansonWrightCom} along with the sub-Gaussian norm result from lemma~\ref{lem:subGaussNormRes}, and making the approximation $\ceil*{\frac{N}{M}} \geq \frac{N}{M}$ for some universal constant $d>0$, 
	\begin{align*}
		&P\pb{\abs{\inProd{\boldsymbol{\mc{A}}(m_1,s_1)}{\boldsymbol{\mc{A}}(m_2,s_2)}} > t} \leq \nn \\
		& \qquad \qquad 8 \exp \pb{-\frac{dM q^{*} \min \pb{\frac{q^{*} t^2}{\pb{\frac{N_c N_T}{N}}^{\pb{\frac{2}{q^{*}}}-1} } ,t }}{\pb{\frac{N_c N_T}{N}}^{\pb{\frac{2}{q^{*}}} -1} } } , \\
		&P\pb{\max_{\st{m_1,m_2}{s_1,s_2}}\abs{\inProd{\boldsymbol{\mc{A}}(m_1,s_1)}{ \boldsymbol { \mc { A }}(m_2,s_2) } } > t} \leq \nn \\
		& 8 (N_R N_T N)^2 \exp \pb{\frac{-dM q^{*} \min \pb{\frac{q^{*} t^2}{\pb{\frac{N_c N_T}{N}}^{\pb{\frac{2}{q^{*}}}-1} } ,t }}{\pb{\frac{N_c N_T}{N}}^{\pb{\frac{2}{q^{*}}} -1} } } , 
	\end{align*}
	Using $t = \frac{\pb{\frac{N_c N_T}{N}}^{\frac{2}{q^{*}}-1 }}{d q^{*}} \sqrt{\frac{\log\pb{\frac{N_R N_TN}{\epsilon}}}{M}}$, we get 
	\begin{align}\label{eq:inPordTailBoundMIMO} 
		&P\pb{\max_{\st{m_1,m_2}{s_1,s_2}}\abs{\inProd{\boldsymbol{\mc{A}}(m_1,s_1)}{ \boldsymbol { \mc { A}} (m_2,s_2) } } > \kappa_1 \sqrt{\frac{\log\pb{\frac{N_R N_TN}{\epsilon}}}{M}} } \nn \\
		&\qquad \qquad \leq \epsilon, 
	\end{align}
	where $\kappa_1 = \frac{\pb{\frac{N_c N_T}{N}}^{\frac{2}{q^{*}}-1 }}{d q^{*}}$. For a fixed constant $\alpha_3 > 0$, let $\frac{\pb{\frac{N_c N_T}{N}}^{\frac{2}{q^{*}}-1 }}{d q^{*}} \leq \alpha_3$, then we get the relation that, 
	\begin{align}
		\frac{\pb{\frac{N_c N_T}{N}}^{\frac{2}{q^{*}}-1 }}{ q^{*}} \leq d\alpha_3. 
	\end{align}
	For a fixed level $d\alpha_3$, we can choose the ratio $\frac{N_c N_T}{N}$ such that the coherence is small. The concentration inequality for the coherence of matrix $\mbf{\mc{A}}$ is given by 
	\begin{align}\label{eq:mutualCohTailMIMO1} 
		&P\pb{\mu\pb{\boldsymbol{\mc{A}}} \geq \frac{\alpha_3}{1 - \epsilon} \sqrt{\frac{\log\pb{N N_R N_T }}{M}} } \leq \nn \\
		& P\pb{ \max_{ \st{m_1,m_2}{s_1,s_2}}\abs{\inProd{\boldsymbol{\mc{A}}(m_1,s_1)}{\boldsymbol{\mc { A } } (m_2,s_2) } } \geq \alpha_3\sqrt{\frac{\log\pb{\frac{NN_R N_T}{\epsilon} } }{M}} } \nn \\
		&\qquad \qquad \qquad + P\pb{\min_{m,s}\norm{\boldsymbol{\mc{A}}(m,s)}^2_2 \leq 1 - \epsilon } \nn \\
		&\leq \epsilon + 8 N \exp\pb{-M d \pb{\epsilon \frac{q^{*}}{\pb{\frac{N_c N_T}{N}}^{\frac{2}{q^{*}} -1 }} }^2 } . 
	\end{align}
\end{IEEEproof}
\begin{IEEEproof}
	[Proof of Theorem~\ref{thm:MultiChirpsRecGuaranteeMIMO}] Using $M \geq \log(NN_R N_T)^3$ in \eqref{eq:mutualCohTailMIMO1} from lemma~\ref{lem:mutualCoherenceTailBound}, the coherence condition given in \cite{candesPlanL1} is satisfied with high probability as shown below 
	\begin{align}\label{eq:cohBoundFinalMIMO} 
		\mu\pb{\boldsymbol{\mc{A}}} &= \mc{O} \pb{\frac{1}{\log (NN_RN_T)}} \nn\\
		\text{ w.p. } & p_1 \geq 1 -\epsilon - 10N \exp\pb{-d M\bar{\epsilon}^2 }, 
	\end{align}
	where $\bar{\epsilon} = \pb{\epsilon \frac{q^{*}}{\pb{\frac{N_cN_T}{N}}^{\frac{2}{q^{*}} -1 }} }, \epsilon \in (0,1) $. 
	
	The measurement matrix in our analysis is normalized to have unit norm columns to apply results from \cite{candesPlanL1}. Let $\mbf{D} \in \mbb{R}^{N_R N_T N \times N_R N_T N}$ diagonal matrix with diagonal entries corresponding to the norm of the column of $\boldsymbol{\mc{A}}$ given by 
	\begin{equation*}
		D_{i,i} = \norm{\boldsymbol{\mc{A}}(m_i,s_i)}_2. 
	\end{equation*}
	The measurement model can be modified as 
	\begin{align*}
		\mbf{y} = \boldsymbol{\hat{\mc{A}}} \mbf{z} + \mbf{w}, 
	\end{align*}
	where $\boldsymbol{\hat{\mc{A}}} = \boldsymbol{\mc{A}}\mbf{D}^{-1}$ and $\mbf{z} = \mbf{D}\mbf{x}$. Next, we obtain the probability tail bound for the operator norm of the measurement matrix $\boldsymbol{\hat{\mc{A}}}$. Using lemma~\ref{lem:operatorNormBounds} and \ref{lem:opNormTailBound}, we have $\forall \epsilon >0, \epsilon \in (0,1)$, independent of N and M, 
	\begin{align}
		&P\pb{ \norm{\boldsymbol{\hat{\mc{A}}}}_{op} \geq \frac{2}{\sqrt{1 -\epsilon}} \sqrt{\frac{N_T N}{M} \log\pb{N_RM + N_R N_T N}}} \nn \\
		&\leq \pb{\frac{1}{N_R M + N_T N_R N}}^{\alpha_1 -1}+ 8N \exp\pb{-d M\bar{\epsilon}^2 }, \nn 
	\end{align}
	where 
	\begin{align*}
		&\alpha_1 = \frac{1}{\frac{1}{3}\sqrt{\frac{1}{N_T N_C} \log\pb{N_R M + N_R N_T N}} + \frac{1}{2}}, \nn \\
		&N_cN_T \geq \frac{4}{9} \log(NN_R N_T + MN_R), \bar{\epsilon} = \pb{\epsilon \frac{q^{*}}{\pb{\frac{N_cN_T}{N}}^{\frac{2}{q^{*}} -1 }} }. 
	\end{align*}
	Therefore, 
	\begin{align}\label{eq:OpNormBoundFinalMIMO} 
		&\norm{\boldsymbol{\hat{\mc{A}}}}_{op} \leq \frac{2}{\sqrt{1 -\epsilon}} \sqrt{\frac{N_T N}{M} \log\pb{N_RM + N_R N_T N}} \\
		&\text{ w.p. } p_2 \geq 1 - \pb{\frac{1}{N_R M + N_T N_R N}}^{\alpha_1 -1}\nn \\
		&\qquad \qquad \qquad + 8N \exp\pb{-d M\bar{\epsilon}^2 }. \nn 
	\end{align}
	Using the support recovery result from \cite{candesPlanL1}, the maximum number of targets that can be successfully detected is 
	\begin{align}
		K_{max} = \frac{c_0 N_R M}{\log^2(NN_RN_T+MN_R)}. 
	\end{align}
	
	Next, we establish that the measurement matrix does not reduce the absolute value of non-zero entries of the sparse vector $\mbf{x}$ below the noise level. 
	\begin{align}
		&P\pb{\min_i D_{i,i}\abs{x_i} \leq 8\sigma\sqrt{2\log N} } \leq N P\pb{D_{i,i} \leq \sqrt{1 - \epsilon} } \nn \\
		& \leq 8N \exp\pb{-d M\bar{\epsilon}^2 }. \nn 
	\end{align}
	Therefore, we have 
	\begin{align}\label{eq:VectorNormBoundMIMO} 
		&\min_i \abs{z_i} \geq 8\sigma \sqrt{2 \log N} \\
		\text{w.p. } &p_3 \geq 1- 8N \exp\pb{-d M\bar{\epsilon}^2 }. \nn 
	\end{align}
	We define the following events associated with a realization of measurement matrix $\boldsymbol{\mc{A}}$ 
	\begin{align*}
		&\Xi_1 : \mu\pb{\boldsymbol{\mc{A}}} = \mc{O} \pb{\frac{1}{\log N}} \\
		&\Xi_2 : \norm{\boldsymbol{\hat{\mc{A}}}}^2_{op} \leq \frac{c_0 N}{K_{\max}\log N } \\
		&\Xi_3 : \min_i \abs{z_i} \geq 8\sigma \sqrt{2 \log N}. \\
		&\Xi_4 : \mbox{successful support recovery for a fixed sensing matrix}. 
	\end{align*}
	Let $\Xi$ be the event that the sampled measurement matrix satisfies the conditions required for successful recovery and recovers a $K$-sparse vector $\mbf{x}$ selected from the target model. This implies 
	\begin{align}\label{eq:finalUBoundMIMO}
		P\pb{\Xi} &\geq P\pb{\Xi_4 \g \Xi_1 \cap \Xi_2 \cap \Xi_3}\nn \\
		&\qquad \qquad(1-P\pb{\Xi^c_1} - P\pb{\Xi^c_2} - P\pb{\Xi^c_3}). 
	\end{align}
	Using result from \cite{candesPlanL1} for $P\pb{\Xi_4 \g \Xi_1 \cap \Xi_2 \cap \Xi_3}$,\eqref{eq:OpNormBoundFinalMIMO}, \eqref{eq:VectorNormBoundMIMO} and \eqref{eq:cohBoundFinalMIMO} in \eqref{eq:finalUBoundMIMO}, we deduce that successful support recovery is guaranteed with high probability. 
\end{IEEEproof}
The conditions required for RIP of order $K$ to hold are obtained next. We reformulate the system model presented in \eqref{eq:sensingMIMO} and \eqref{eq:sensingMatrixMIMO} by re-scaling the random variables to normalize the variance as follows
\begin{align}
\boldsymbol{\mc{A}} &= \sum_{i=1}^N {c}_i \sqrt{\frac{N_c N_T}{N} }\pb{\mbf{\bar{\alpha}_R \bar{\alpha}_T(\xi(i))} } \otimes \pb{\mbf{H}_i \mbf{\bar{A}} \mbf{D}_i}, \nn \\
\boldsymbol{\mc{A}} &= \sum_{i=1}^N {c}_i \boldsymbol{\mc{A}_i}, \nn \\
 \boldsymbol{\mbf{\mc{A}_i}} &=\sqrt{\frac{N_c N_T}{N} }\pb{\mbf{\bar{\alpha}_R \bar{\alpha}_T(\xi(i))} } \otimes \pb{\mbf{H}_i \mbf{\bar{A}} \mbf{D}_i}
\end{align} 
 where ${c}_i = \sqrt{\frac{N}{N_c N_T}} \hat{c}_i$ such that $E\pb{ \abs{ c_i}^2 } =1$. We define the set \[\mbf{\mc{D}_{K,NN_{\theta}}} = \cb{ \mbf{x} \in \mbb{C}^{NN_{\theta}}: \norm{\mbf{x}}_0 = K, \norm{\mbf{x}}^2_2 \leq 1  }.\] For a $K$-sparse vector $\mbf{x} \in \mbf{\mc{D}_{K,NN_{\theta}}}$, we have
 \begin{align}
 {\mbf{\mc{A}} \mbf{x}} = \mbf{V}_x \mbf{c},   
 \end{align}
  where $ \mbf{V}_x  = \sqb{\mbf{\mc{A}_1 x} \ \ \mbf{\mc{A}_2 x} \ \cdots \ \ \mbf{\mc{A}_N x} } \in \mbb{C}^{MN_R \times N}$ and $\mbf{c}\in \mbb{C}^N$ is the vector comprised of the normalized random variables that select the waveforms. 
\begin{lemma}\label{lem:ExpectNormSensing}
Given the measurement operator $\mbf{\mc{A}}$ and any $\mbf{x} \in \mbf{\mc{D}}_{K,NN_{\theta}}$ we have
\begin{align}
\Expec{\norm{\mbf{\mc{A}x}}^2} = \norm{\mbf{x}}^2.
\end{align}
\end{lemma}
\begin{IEEEproof}
The expectation can be simplified using the independence of the zero-mean random variables $c_i,c_j \forall i,j$ as follows
\begin{align}
&\Expec{\norm{\mbf{\mc{A}x}}^2} = \sum_{i=1}^N \Expec{\abs{c_i}^2} \mbf{x^{*} \mc{A}_i^{*} \mc{A}_i x} + \nn \\
& \sum_{i=1}^N \sum_{\st{j=1}{j \neq i}}^N \Expec{c_i^{*}} \Expec{c_j} \mbf{x^{*} \mc{A}_i^{*} \mc{A}_j x}, \\
&= \frac{N_c N_T}{N} \sum_{i=1}^N \pb{\mbf{\bar{\alpha}_T(\xi(i))^{*} \bar{\alpha}_R^{*}  \bar{\alpha}_R \bar{\alpha}_T(\xi(i))} } \otimes \pb{ \mbf{\mbf{D}_i^{*} \bar{A}^{*}} \mbf{\bar{A}} \mbf{D}_i}, \\
&= \mbf{I}_{N_{\theta} \times N_{\theta}} \otimes \frac{1}{N} \sum_{i=1}^N \mbf{V}_i,  
 \end{align}
It can be seen that 
\begin{align}
V_i(l,m) =\begin{cases}
& 1 , \textbf{ if } l = m\\
&\exp\pb{-j2\pi\frac{i\pb{m-l}}{N}} \inProd{\mbf{\bar{A}}_l }{\mbf{\bar{A}}_m}, \textbf{ otherwise. }
\end{cases}
\end{align}
The term $\frac{1}{N}\sum_{i=1}^N \exp\pb{-j2\pi\frac{i\pb{m-l}}{N}} =0$ if $l \neq m$. This implies that 
\begin{align}
	\Expec{\norm{\mbf{\mc{A}x}}^2} = \mbf{x}^{*} \mbf{I}_{N_{\theta}N \times N_{\theta}N} \mbf{x} = \norm{\mbf{x}}_2^2.
	\end{align}
\end{IEEEproof}
Lemma~\ref{lem:ExpectNormSensing} implies that the RIP constant of order $K$ in \eqref{eq:RIP} can be expressed as a second order chaos process in the random vector $\mbf{c}$ as follows
\begin{align}
	\delta_K &= \sup_{\mbf{x} \in \mc{D}_{K,NN_{\theta}}} \abs{\norm{\mbf{\mc{A}x}}_2^2 - \norm{\mbf{x}}_2^2} \nn \\
	&= \sup_{\mbf{x} \in \mc{D}_{K,NN_{\theta}}} \abs{\norm{ \mbf{V}_x \mbf{c}}_2^2 - \Expec{\norm{\mbf{ \mbf{V}_x \mbf{c}}}_2^2}}. \label{eq:RIP_eq}
	\end{align}
	Therefore, we derive the concentration inequality for the RIP constant using the result in Theorem~\ref{thm:supremaChaoseProc}, which was first established in \cite{Chaos_Rauhut2014}. We define the following terms that are essential components in the result 
\begin{align}
	&\mbb{T} = \cb{\mbf{V_x} = [\mbf{\mc{A}_1 x} \cdots \mbf{\mc{A}_N x}] : \mbf{x} \in \mc{D}_{K,NN_{\theta}}}, \\
	&d_F(\mbb{T}) = \sup_{\mbf{V_x} \in \mbb{T}}\norm{\mbf{V_x}}_F, \qquad 	d_{op}(\mbb{T})= \sup_{\mbf{V_x} \in \mbb{T}} \norm{\mbf{V_x}}_{op}, \\
	&\gamma_2\pb{\mbb{T},\norm{.}_{op}} \leq C\int_{0}^{d_{op}(\mbb{T})} \sqrt{\log\pb{\mbf{N}\pb{\mbb{T},\norm{.}_{op},u}}} du,
	\end{align}
	where $\gamma_2\pb{\mbb{T},\norm{.}_{op}}$ is the Talgrand's chaining functional, which is upper bounded by Dudley's entropy integral \cite{Chaos_Rauhut2014}, $\mbf{N}\pb{\mbb{T},\norm{.}_{op},u}$ is the covering number, which is defined by the number of balls with distance metric $\norm{.}_{op}$ and radius $u$ required to cover the set of matrices $\mbb{T}$ induced by the vector $\mbf{x} \in \mc{D}_{K,NN_{\theta}}$, $C >0$ is a universal constant, and $\log\pb{\mbf{N}\pb{\mbb{T},\norm{.}_{op},u}}$ is defined as the metric entropy. The following lemma derives the estimates for the above defined quantities.
	\begin{lemma}\label{lem:chaosProcessTerms}
		For the set of matrices $\mbb{T}$, we have
		\begin{align}
			d_F(\mbb{T}) &= 1, \\
			d_{op}(\mbb{T}) & \leq \sqrt{\frac{K}{M}}, \\
			\gamma_2\pb{\mbb{T},\norm{.}_{op}} &\leq 	C_1 \sqrt{\frac{K}{M}} \pb{ \sqrt{ \log\pb{\frac{e NN_{\theta}}{K}}} },
			\end{align}
			for some universal constant $C_1 > 0$.
	\end{lemma} 
	\begin{IEEEproof}
	For a particular matrix $\mbf{V_x} \in \mbb{T}$, 
	\begin{align}
		\norm{\mbf{V_x}}_F^2 &= \sum_{i=1}^N\norm{\mbf{\mc{A}_i x } }_2^2 \nn \\
		&= \mbf{x}^{*} \sum_{i=1}^N \mbf{\mc{A}_i^{*} \mc{A}_i} \mbf{x}  = \norm{\mbf{x}}_2^2.
	\end{align}
	The last equality is due to the result in Lemma~\ref{lem:ExpectNormSensing} where it was shown that the $\sum_{i=1}^N \mbf{\mc{A}_i^{*} \mc{A}_i} = \mbf{I}_{NN_{\theta} \times NN_{\theta}}$. Therefore,
	\begin{align}
		d_F\pb{\mbb{T}} = \sum_{\mbf{V_x} \in \mbb{T}} \norm{\mbf{V_x}}_F^2 = \sup_{\mbf{x} \in \mc{D}_{K,NN_{\theta}}} \norm{\mbf{x}}_2^2 = 1.
	\end{align}
	Next, we consider the spectral or the operator norm of $\mbf{V_x} \in \mbb{T}$. 
	\begin{align}
		&\norm{\mbf{V_x}}_{op}^2 = \norm{\mbf{V_x} \mbf{V_x^{*}}}_{op} = \norm{\sum_{i=1}^N \mbf{\mc{A}_i x x^{*}\mc{A}_i^{*} }}_{op} \nn \\
		&\qquad \qquad=\norm{\sum_{i=1}^N \sum_{\hat{p}=1}^{NN_{\theta}}\sum_{\hat{q}=1}^{NN_{\theta}} x(\hat{p}) x^{*}(\hat{q}) \mbf{\mc{A}_i(\hat{p}) \mc{A}_i^{*}(\hat{q}) }}_{op} \nn \\ 
		&\qquad \qquad\leq \sum_{\hat{p}=1}^{NN_{\theta}} \sum_{\hat{q}=1}^{NN_{\theta}} \abs{x(\hat{p}) x^{*}(\hat{q})}\norm{ \sum_{i=1}\mbf{\mc{A}_i(\hat{p}) \mc{A}_i^{*}(\hat{q})}}_{op}, \nn \\
		&\sum_{i=1}\mbf{\mc{A}_i(\hat{p}) \mc{A}_i^{*}(\hat{q})} = \begin{bmatrix} \mbf{\mc{A}_1(\hat{p})} &\cdots &\mbf{\mc{A}_N(\hat{p})}     \end{bmatrix} \begin{bmatrix} \mbf{\mc{A}_1^{*}(\hat{q})} \\
		 \mbf{\mc{A}_2^{*}(\hat{q})}  \\
		 \vdots \\
		 \mbf{\mc{A}_N^{*}(\hat{q})}    \end{bmatrix}  \nn
	\end{align}
	where $x(\hat{p}),x(\hat{q})$ are the elements of $\mbf{x} \in \mbb{C}^{NN_{\theta}}$ indexed by $\hat{p}$ and $\hat{q}$, and $\mbf{\mc{A}_i(\hat{p})} \in \mbb{C}^{MN_R}$ is the column of the matrix $ \mbf{\mc{A}_i} \in \mbb{C}^{MN_R \times NN_{\theta}}$ indexed by $p$. Using the definition of the sensing matrix, we have 
	\begin{align}
	&\begin{bmatrix} \mbf{\mc{A}_1(\hat{p})} &\cdots &\mbf{\mc{A}_N(\hat{p})}     \end{bmatrix}= \nn \\ & \qquad \qquad \sqrt{\frac{N_c}{NN_R}}\mbf{\alpha_{R}(\theta_{\hat{p}})} \otimes \pb{\mbf{E_{\hat{p}}FG_{\hat{p}}}  \mbf{\alpha_{T}}(\xi,\theta_{\hat{p}})}	, \nn \\
		&\mbf{\alpha_{T}}(\xi,\theta_{\hat{p}})= diag\begin{bmatrix} \exp\pb{j \bar{d}_T \xi(1) \theta_{\hat{p}}  } &\cdots & \exp\pb{j \bar{d}_T \xi(N) \theta_{\hat{p}}  }  \end{bmatrix}
		\end{align}
		where $\mbf{\alpha_{R}(\theta_{\hat{p}})}$ and $\mbf{E_{\hat{p}}FG_{\hat{p}}} $ are defined in \eqref{eq:MatrixColumnDef}.  Therefore, we have
		\begin{align}
			&\sum_{i=1}\mbf{\mc{A}_i(\hat{p}) \mc{A}_i^{*}(\hat{q})} = \frac{N_c }{NN_R} \mbf{\alpha_{R}(\theta_{\hat{p}})} \mbf{\alpha_{R}^{*}(\theta_{\hat{q}})} \otimes \nn \\
			& \qquad \qquad \pb{\mbf{E_{\hat{p}}FG_{\hat{p}}}  \mbf{\alpha_{T}}(\xi,\theta_{\hat{p}})} \pb{\mbf{\alpha^{*}_{T}}(\xi,\theta_{\hat{p}})\mbf{G^{*}_{\hat{q}} F^{*} E^{*}_{\hat{q}}} } \\
			&\norm{\sum_{i=1}\mbf{\mc{A}_i(\hat{p}) \mc{A}_i^{*}(\hat{q})} }_{op} \leq  \frac{N_c }{N N_R} \norm{\mbf{\alpha_{R}(\theta_{\hat{p}})}}_2  \norm{\mbf{\alpha_{R}(\theta_{\hat{q}})}}_2  \nn \\
			& \norm{\mbf{E_{\hat{p}}}}\norm{ \mbf{E_{\hat{q}}}} \norm{\mbf{F}}^2\norm{\mbf{G_{\hat{p}}}} \norm{\mbf{G_{\hat{q}}}} \norm{\mbf{\alpha_{T}}(\xi,\theta_{\hat{p}})}_{op} \norm{\mbf{\alpha_{T}}(\xi,\theta_{\hat{p}})}_{op},
			\end{align}
		where $\mbf{\alpha_{R}(\theta_{\hat{p}})} \in \mbb{C}^{N_R}$ such that $\norm{\mbf{\alpha_{R}(\theta_{\hat{p}})}}_2  =\sqrt{N_R}$ and  $\norm{\mbf{\alpha_{T}}(\xi,\theta_{\hat{p}})}_{op} =1$, for all $p$.
		Using the results in lemma~\ref{lem:FroNormOpNormMatrix}, we have 
		\begin{align}
			&\norm{\mbf{E_{\hat{p}}}}\norm{ \mbf{E_{\hat{q}}}} \norm{\mbf{F}}^2\norm{\mbf{G_{\hat{p}}}} \norm{\mbf{G_{\hat{q}}}} \leq \frac{N}{N_c M}, \nn \\
			\implies &\norm{\sum_{i=1}\mbf{\mc{A}_i(\hat{p}) \mc{A}_i^{*}(\hat{q})} }_{op} \leq \frac{1}{M} . \nn \\
			\therefore & \norm{\mbf{V_x}}_{op}^2 \leq \frac{1}{M} \sum_{\hat{p}=1}^{NN_{\theta}} \sum_{\hat{q}=1}^{NN_{\theta}} \abs{x(\hat{p}) x^{*}(\hat{q})} = \frac{1}{M} \norm{\mbf{x}}_1^2, \\
			& \norm{\mbf{V_x}}_{op}^2 \leq \frac{K}{M}\norm{\mbf{x}}_2^2.
			\end{align} 
			By taking the supremum of $\mbf{x}$ over the set $\mc{D}_{K,NN_{\theta}}$, we obtain the result $d_{op}(\mbb{T})  \leq \sqrt{\frac{K}{M}}$.
			
		We estimate the covering number of the set $\mbb{T}$ denoted by  $\mbf{N}\pb{\mbb{T},\norm{.}_{op},.}$. Let $\mbf{x,y} \in \mc{D}_{K,NN_{\theta}}$. We have 
		\begin{align}
			\norm{\mbf{V}_x - \mbf{V}_y }_{op} \leq \norm{\mbf{x - y}}_1 \sqrt{\norm{\sum_{i=1}\mbf{\mc{A}_i(\hat{p}) \mc{A}_i^{*}(\hat{q})} }_{op} }.
			\end{align}
			Since $\mbf{x,y}$ are $K$ sparse vectors, $\mbf{x-y}$ is $2K$ sparse. Therefore, we observe that $\mbf{V_x}$ is a Lipschitz mapping as shown below 
			\begin{align}
				\norm{\mbf{V}_x - \mbf{V}_y }_{op} \leq \sqrt{\frac{2K}{M}}\norm{\mbf{x - y}}_2.
				\end{align}
				Therefore, $\forall u \in [0,\infty]$ the covering number for the set $\mbb{T}$ with the distance metric $\norm{.}_{op}$ is related to the covering number of the set $\mc{D}_{K,NN_{\theta}}$ as shown below \cite{Chaos_Rauhut2014,FoucartMathIntroCS_2013}	
				\begin{align}
				&\mbf{N}\pb{\mbb{T},\norm{.}_{op},u}  \leq \mbf{N}\pb{\mc{D}_{K,NN_{\theta}},\norm{.}_{2},\frac{u\sqrt{M}}{ \sqrt{2K}}}. \label{eq:CoveringNumberBound}
				\end{align}
				 Using \eqref{eq:CoveringNumberBound}, Dudley's integral entropy can be computed as follows
				 \begin{align}
					 &\gamma_2\pb{\mbb{T},\norm{.}_{op}} \leq C\int_{0}^{d_{op}(\mbb{T})} \sqrt{\log\pb{\mbf{N}\pb{\mbb{T},\norm{.}_{op},u}}} du, \nn \\
	&\leq C\int_{0}^{\sqrt{\frac{K}{M}}} \sqrt{\log \pb{\mbf{N}\pb{\mc{D}_{K,NN_{\theta}},\norm{.}_{2},\frac{u\sqrt{M}}{ \sqrt{2K}}} }} du \nn \\
	&\leq 	C\sqrt{\frac{2K}{M}}\int_{0}^{\sqrt{\frac{1}{2}}} \sqrt{\log\pb{\mbf{N}\pb{\mc{D}_{K,NN_{\theta}},\norm{.}_{2},\bar{u}} }}d\bar{u} \nn \\
	&\leq 	C \sqrt{\frac{K}{M}} \pb{ \sqrt{ \log\pb{\frac{e NN_{\theta}}{K}}} + \int_{0}^{1/\sqrt{2}} \sqrt{\log\pb{1 +\frac{2}{\bar{u}}}} d\bar{u} }, \nn \\
	&\gamma_2\pb{\mbb{T},\norm{.}_{op}} \leq 	C_1 \sqrt{\frac{K}{M}} \pb{ \sqrt{ \log\pb{\frac{e NN_{\theta}}{K}}} }, 
					 \end{align}
					 where $\bar{u} = u\sqrt{\frac{2K}{M}}$ and $C_1 > 0$ is the universal constant. The above relation used the covering number estimate $\mbf{N}\pb{\mc{D}_{K,NN_{\theta}},\norm{.}_{2},\bar{u}} \leq \pb{e \frac{NN_{\theta}}{K}}^K \pb{1+\frac{2}{\bar{u}}}^K$ and $\sqrt{a + b} \leq \sqrt{a} + \sqrt{b}$, for all $a,b \geq 0$.
	\end{IEEEproof}
	
\begin{IEEEproof}[Proof of Theorem~\ref{thm:RIPConditionMIMO}] 
Let the restricted isometry constant of order $K$ be $\delta_K$  for the measurement operator, which is obtained in \eqref{eq:RIP_eq}. We can obtain the tail bounds on $\delta_K$ using the results from Lemma~\ref{lem:chaosProcessTerms} in Theorem~\ref{thm:supremaChaoseProc} as follows
\begin{align}
&P\pb{ \delta_K \geq \epsilon_1 E +t } \nn \\
& = P\pb{  \sup_{\mbf{x} \in \mc{D}_{K,NN_{\theta}}} \abs{\norm{ \mbf{V}_x \mbf{c}}_2^2 - \Expec{\norm{\mbf{ \mbf{V}_x \mbf{c}}}_2^2}} \geq \epsilon_1 E +t } \nn \\
&\leq \exp\pb{-\epsilon_2 \min\pb{\frac{t^2}{V^2},\frac{t}{U} }} \leq \eta_1,
\end{align}
where $\epsilon_1,\epsilon_2,E,U,V$ are defined in Theorem~\ref{thm:supremaChaoseProc}, and $\eta_1 \in [0,1]$ is a bound on the tail probability. The constants $E,U,V$ for the measurement operator is given by
\begin{align}
E \leq & C_1^2 {\frac{K}{M}} \pb{ { \log\pb{\frac{e NN_{\theta}}{K}}} } + C_1 \sqrt{\frac{K}{M}} \pb{ \sqrt{ \log\pb{\frac{e NN_{\theta}}{K}}} } \nn \\
& + \sqrt{\frac{K}{M}} \nn \\
\leq & C_2^2 {\frac{K}{M}} \pb{ { \log\pb{\frac{e NN_{\theta}}{K}}} } + C_2 \sqrt{\frac{K}{M}} \pb{ \sqrt{ \log\pb{\frac{e NN_{\theta}}{K}}} } \nn 
\end{align} 
If the number of measurements per receiver 
					 \begin{align}
						 M > 2\frac{C_2^2 \delta^2}{\epsilon_1^2}K \log \pb{\frac{\epsilon N N_{\theta}}{K}},
					 \end{align}
					 then $E \leq \frac{\delta^2}{4\epsilon_1^2} +\frac{\delta}{2\epsilon_1} \leq \frac{c \delta}{2\epsilon_1}$. Given the condition on the number of measurements, the RIP constant $\delta_K$ is bounded as follows using $t = \frac{\delta}{2}$
\begin{align}
	P\pb{ \delta_K \geq \epsilon_1 E +t } &= P\pb{ \delta_K \geq \delta} \nn \\
	 &\leq \exp \pb{-\frac{\epsilon_2}{4}\pb{\frac{M^2\log\pb{\frac{e NN_{\theta}}{K}}}{K^2}\delta^2}}
	\end{align}
This relation establishes the condition the number of measurements required per receiver $M$ for the constant $\delta_K$ to be bounded with high probability. 
\end{IEEEproof}

\section{Conclusion}\label{sec:Conclusion} 
In this work we have presented a compressive acquisition scheme for high resolution radar sensing. We show that the system comprising multi-tone LFM transmit waveforms and uniformly subsampled stretch processor results in a structured random sensing matrix with provable recovery guarantees for delay and angle of arrival estimation in sparse scenes. The recovery guarantees for the proposed compressive illumination scheme is comparable to that of random Toeplitz matrices with much larger number of random elements. The proposed scheme is well matched to practical implementation utilizing small number of random parameters and uniform sampling ADCs on receive. Our simulation show targets both on and off the grid can be detected using sparsity regularized recovery algorithms. A potential direction for future research is to investigate the effect of basis mismatch \cite{BasisMismatch_Chi_2011} due to targets not located on the grid locations and extend the theoretical guarantees to off the grid compressed sensing framework proposed in \cite{CS_offGrid_Tang,LineSpectralEstimation_tang} based on the generalization of notion of sparsity in an infinite dictionary setting \cite{ConvexGeometryLinInvProb_Recht} as well as investigate the effects of clutter and system imperfections. 
\begin{appendices}

\section{Additional lemma} 
	\begin{lemma}\label{sec:signalModel} 
		[Matrix Bernstein inequality \cite{MatrixConcentration}] \label{lem:MatBernstein} Let $\mbf{A}_i$ be a sequence of i.i.d. random matrices. For a random matrix expressed as $\mbf{A} = \sum_i \mbf{A}_i$ we have 
		\begin{align}
			&P\pb{\norm{\mbf{A}}_{op} \geq t} \leq \pb{d_1 + d_2} \exp \pb{\frac{-t^2/2}{\frac{Lt}{3} + \nu\pb{\mbf{A}} }}, \\
			& \text{where, }\norm{\mbf{A}_i}_{op} \leq L, \forall i=1,\cdots, D \nn \\
			& \text{and }\nu\pb{\mbf{A}} = \max \pb{\norm{\Expec{\mbf{A}\mbf{A}^{*}}},\norm{\Expec{\mbf{A}^{*}\mbf {A}} }}, \nn 
		\end{align}
		where $\mbf{A}_i \in \mbb{C}^{d_1 \times d_2}$. The expected value of the operator norm of $\mbf{A}$ is bounded by 
		\begin{align}
			\Expec{\norm{\mbf{A}}_{op} }\leq \sqrt{2 \nu\pb{\mbf{A}} \log\pb{d_1 +d_2}} + \frac{L \log\pb{d_1 +d_2}}{3}. 
		\end{align}
	\end{lemma}
	\begin{lemma}\label{lem:operatorNormBounds} 
		Given a matrix $\mbf{A}$, the operator norm of the matrix $\hat{\mbf{A}} = \mbf{A}\mbf{D}^{-1}$ is bounded by the following inequality 
		\begin{align}
			\frac{\norm{\mbf{A}}_{op} }{\min_i D_{i,i}} \geq \norm{\hat{\mbf{A}}}_{op} \geq \frac{\norm{\mbf{A}}_{op}}{\max_i D_{i,i}}, 
		\end{align}
		where $\mbf{D}$ is a diagonal matrix with positive diagonal elements. 
	\end{lemma}
	\begin{IEEEproof}
		For any unit-norm vector $\mbf{v}$, we have 
		\begin{align}
			&\bar{\mbf{A}}\mbf{v} = \frac{\mbf{A}\mbf{D}^{-1}\mbf{v}}{\norm{\mbf{D}^{-1}\mbf{ v}}_2} \norm{\mbf{D}^{-1}\mbf{v}}_2 = \mbf{A} \mbf{u(v)} \norm{\mbf{D}^{-1}\mbf{v}}_2 \nn \\
			\implies &\norm{\bar{\mbf{A}}\mbf{v}}_2 = \norm{\mbf{A}\mbf{u(v)}}_2 \norm{\mbf{D}^{-1}\mbf{v}}_2,\nn 
		\end{align}
		where $\mbf{u(v)} = {\mbf{D}^{-1}\mbf{v}}/{\norm{\mbf{D}^{-1}\mbf{v}}_2}$. Since $\norm{\mbf{v}}_2=1$, we bound the Euclidean norm of $\mbf{D}^{-1}v$ as follows 
		\begin{align}
			&\frac{1}{\min_i D_{i,i}} \geq \norm{\mbf{D}^{-1}\mbf{v}}_2 \geq \frac{1}{\max_i D_{i,i}}.\nn \\
			\implies &\frac{\norm{\mbf{A}\mbf{u(v)}}_2}{\min_i D_{i,i}} \geq \norm{\bar{\mbf{A}} \mbf{v}}_2 \geq \frac{\norm{\mbf{A} \mbf{u(v)}}_2}{\max_i D_{i,i}}.\nn 
		\end{align}
		By taking the supremum over $\mbf{v}$ in the space of unit norm vectors we obtain the desired result. 
	\end{IEEEproof}
	\begin{lemma}\label{lem:normBound} 
		For the matrix $\hat{c}_i \pb{\mbf{\bar{\alpha}_R \bar{\alpha}_T(\xi(i))}}\otimes \mbf{H}_i \bar{\mbf{A}} \mbf{D}_i \in \mbb{C}^{N_RM\times NN_{\theta}}$ defined in \eqref{eq:sensingMatrixMIMO}, and \eqref{eq:componentsSingleTransmit}, we have 
		\begin{align}\label{eq:chirpMatOpNorm} 
			\norm{\hat{c}_i \pb{\mbf{\bar{\alpha}_R \bar{\alpha}_T(\xi(i))}}\otimes \mbf{H}_i \bar{\mbf{A}} \mbf{D}_i }_{op} \leq \sqrt{\frac{N}{M N_c} }, 
		\end{align}
		$\forall i=1,\cdots, N$ 
	\end{lemma}
	\begin{IEEEproof}
		Using the property of the operator norm and Kronecker product, and the sub-multiplicativity property of the operator norm we have, 
		\begin{align*}
			&\norm{\hat{c}_i \pb{\mbf{\bar{\alpha}_R \bar{\alpha}_T(\xi(i))}}\otimes \mbf{H}_i \bar{\mbf{A}} \mbf{D}_i }_{op} \\
			&\leq \abs{\hat{c}_i } \norm{\mbf{\bar{\alpha}_R} \norm{\bar{\alpha}_T(\xi(i))}} \norm{ \mbf{H}_i}_{op} \norm{\bar{\mbf{A}}}_{op} \norm{ \mbf{D}_i }_{op}. 
		\end{align*}
		Since $\mbf{H}_i$, $\mbf{D}_i$, and $\bar{\alpha}_T(\xi(i))$ are diagonal matrices with complex exponential entries, it can be shown that $\norm{ \mbf{H}_i}_{op} = \norm{ \mbf{D}_i}_{op} = \norm{\bar{\alpha}_T(\xi(i))}_{op}= 1$. Also, we assume that random variables $c_i$ have a sub-Gaussian distribution with $\abs{c_i} \leq 1$. In order to find the operator norm of $\bar{\mbf{A}}$, we define $\mbf{G}= \bar{\mbf{A}} \bar{\mbf{A}}^{*} \in \mbf{C}^{M \times M }$ since it is full rank and $ \norm{\bar{\mbf{A}}}^2_{op} = \norm{\mbf{G}}_{op}$. The entries of matrix $\mbf{G}$ are as follows 
		\begin{align}
			&G(k,k) = \frac{N}{M N_c} \nn \\
			&G(k,l) = \frac{N}{M N_c} \frac{1}{N} \sum_{m=0}^{N-1} \exp\pb{2\pi j\frac{l-k}{N}m} \nn \\
			&= \frac{N}{M N_c} D_{N}\pb{\frac{l-k}{N}} = 0,\nn 
		\end{align}
		$\forall k,l=0,\cdots, M-1$, such that $k \neq l$, and $D_{N}\pb{\pb{l-k}/{N}} = {1}/{N}\sum_{m=0}^{N-1} \exp\pb{2\pi j\pb{l-k}/{N}m} $ is the discrete Dirichlet Kernel. The second term is zero because the discrete Dirichlet kernel is being evaluated at it's zeros, which are the Fourier frequency bins $(\frac{n}{N}), n \in \mbb{Z}$. Therefore, 
		\begin{equation*}
			\mbf{G} = \frac{N}{M N_c} \mbf{I}. 
		\end{equation*}
		This implies that the $\norm{\mbf{G}}_{op} = \frac{N}{M N_c}$. Let $\mbf{G}_1= \mbf{\bar{\alpha}_R \bar{\alpha}_R^{*}} \in \mbf{C}^{N_R \times N_R }$. The entries of matrix $\mbf{G}_1$ are as follows 
		\begin{align}
			&G_1(k,k) = 1 \nn \\
			&G_1(k,l) = \frac{1}{N_R N_T} \sum_{v=-\frac{N_T N_R}{2}}^{\frac{N_T N_R}{2}-1} \exp\pb{2\pi j\frac{l-k}{N_R}v} \nn \\
			&= \frac{1}{N_T}\sum_{v=1}^{N_T} D_{N_R} \pb{\frac{l-k}{N_R}} = 0,\nn 
		\end{align}
		where the final equation was obtained by re-arranging the terms and using the properties of the Dirichlet kernel. This implies that $\norm{\mbf{G}_1}_{op} =1$ and this leads to the result in \eqref{eq:chirpMatOpNorm}. 
	\end{IEEEproof}
	\begin{lemma}\label{lem:varianceBound} 
		For the matrices $ \mbf{P}_i = \hat{c}_i \pb{\mbf{\bar{\alpha}_R \bar{\alpha}_T(\xi(i))}}\otimes \mbf{H}_i \bar{\mbf{A}} \mbf{D}_i \in \mbb{C}^{N_RM\times NN_{\theta}}$ given by \eqref{eq:componentsSingleTransmit}, we have 
		\begin{align}\label{eq:varBound1}
			\norm{\sum_{i=1}^{N} \Expec{\mbf{P}_i^{*}\mbf{P}_i}} &\leq \frac{NN_T}{M}, \\
			\label{eq:varBound2} \norm{\sum_{i=1}^{N} \Expec{\mbf{P}_i \mbf{P}_i^{*}}} &= \frac{NN_T}{M}. 
		\end{align}
	\end{lemma}
	\begin{IEEEproof}
		First, we compute the norm of $\mbf{P}_i \mbf{P}_i^{*}$ as it is a full-rank matrix using $\mbf{D}_i \mbf{D}_i^{*} = \mbf{I}$, $\mbf{H}_i \mbf{H}_i^{*} = \mbf{I}$, $\pb{\mbf{\bar{\alpha}_R \bar{\alpha}_T(\xi(i))}}\pb{\mbf{\bar{\alpha}_R \bar{\alpha}_T(\xi(i))}}^{*} =\mbf{I}$ and $\bar{\mbf{A}} \bar{\mbf{A}}^{*} = {N}/\pb{M N_c }\mbf{I}$ we have 
		\begin{align}
			\mbf{P}_i\mbf{P}_i^{*} &= \hat{c}_i \hat{c}_i^{*} \frac{N}{M N_c} \mbf{I}. \nn 
		\end{align}
		Using the probabilistic model for $c_i$ given in \eqref{eq:probModel}, we get 
		\begin{align*}
			\Expec{c_i c_i^{*}} = \frac{N_cN_T}{N}. 
		\end{align*}
		We have 
		\begin{align*}
			\sum_{i=1}^{N} \Expec{\mbf{P}_i \mbf{P}_i^{*}} &= \frac{NN_T}{M} \mbf{I} . 
		\end{align*}
		Applying the operator norm yields the result in \eqref{eq:varBound2}. Similarly, using the sub-additivity of the operator norm, and the result \eqref{eq:chirpMatOpNorm}, we get 
		\begin{align}
			&\norm{\sum_{i=1}^{N} \Expec{\mbf{P}_i^{*} \mbf{P}_i} }_{op} \leq \frac{NN_T}{M}. \nn 
		\end{align}
	\end{IEEEproof}
	\begin{lemma}\label{lem:FroNormOpNormMatrix} 
		Let 
		\begin{align*}
			\mbf{B} &= \mbf{G}_m^{*} \mbf{F}^{*} \mbf{E}_m^{*} \mbf{E}_m \mbf{F} \mbf{G}_m,\nn \\
			\bar{\mbf{B}} &=\mbf{G}_{m_1}^{*} \mbf{F}^{*} \mbf{E}_{m_1}^{*} \mbf{E}_{m_2} \mbf{F} \mbf{G}_{m_2}, 
		\end{align*}
		then we have 
		\begin{align}
			& \norm{\mbf{B}}_{op} \leq \ceil*{\frac{N}{M}} \frac{1}{N_c} , & \norm{\mbf{B}}_{F} \leq \ceil*{\frac{N}{M}} \frac{\sqrt{M}}{N_c}, \\
			& \norm{\bar{\mbf{B}}}_{op} \leq \ceil*{\frac{N}{M }} \frac{1}{N_c}, &\norm{\bar{\mbf{B}}}_F \leq \ceil*{\frac{N}{M }} \frac{\sqrt{M}}{N_c}, 
		\end{align}
		where $\ceil{x} =z \in \mbb{Z}$ such that $z \geq x,\ \forall x \in \mbb{R}$. 
	\end{lemma}
	\begin{IEEEproof}
		We note that $Rank(\mbf{B}) = Rank(\bar{\mbf{B}}) = M$, and we can obtain a bound on the Frobenius norm of the matrices as shown below 
		\begin{align*}
			\norm{\mbf{B}}_F &\leq \sqrt{M} \norm{\mbf{B}}_{op}, \\
			\norm{\bar{\mbf{B}}}_F &\leq \sqrt{M} \norm{\bar{\mbf{B}}}_{op}. 
		\end{align*}
		In order to find the bound on the operator norm we see that 
		\begin{align*}
			\norm{\mbf{B}}_{op} &\leq \norm{\mbf{G}_m}_{op}^2 \norm{\mbf{E}_m}_{op}^2 \norm{\mbf{F}}^2, \\
			\norm{\bar{\mbf{B}}}_{op} &\leq \norm{\mbf{G}_{m_1}}_{op}\norm{\mbf{G}_{m_2}}_{op} \norm{\mbf{E}_{m_1}}_{op}\norm{\mbf{E}_{m_2}}_{op} \norm{\mbf{F}}^2. 
		\end{align*}
		Since $\mbf{G}_m$ and $\mbf{E}_m$ are diagonal matrices with complex exponentials along the principal diagonal, it can be shown that $\norm{\mbf{G}_m}_{op}^2 = \norm{\mbf{E}_m}_{op}^2 =1$. In order to estimate the $\norm{\mbf{F}}_{op}$, we see that 
		\begin{align*}
			\mbf{F}^{*} \mbf{F} = \frac{1}{N_c} 
			\begin{bmatrix}
				\mbf{V}_{1,1} &\cdots &\mbf{V}_{1,\ceil*{\frac{N}{M}}} \\
				\mbf{V}_{2,1} &\cdots &\mbf{V}_{2,\ceil*{\frac{N}{M}}} \\
				\vdots &\vdots &\vdots \\
				\mbf{V}_{\ceil*{\frac{N}{M}},1} &\cdots &\mbf{V}_{2,\ceil*{\frac{N}{M}}} 
			\end{bmatrix}
		\end{align*}
		Since $p$ is co-prime with $M$, we observe that the N possible frequency tones circularly get mapped onto M possible aliased sinusoids. Therefore, if neither $i,j \neq \ceil*{\frac{N}{M}}$, 
		\begin{equation*}
			\mbf{V}_{i,j} = \mbf{I}. 
		\end{equation*}
		If either $i = \ceil*{{N}{M}}$ or $j=\ceil*{{N}/{M}}$, then $\mbf{V}_{i,j} \in \mbb{C}^{M \times M}$ is a partial identity matrix. It can easily be verified that $\norm{\mbf{F}}_{op}^2 ={1}/{N_c} \ceil*{{N}/{M }} $. Using this result along-with the estimate on the bound for the Frobenius norm, we get the desired results. 
	\end{IEEEproof}
	\begin{lemma}\label{lem:subGaussNormRes} 
		For the random variables described in \eqref{eq:modRandomVars} with independent phase and magnitude, the Sub-Gaussian norm \cite{RandMatricesNonAsymptotics} of the real and imaginary parts of random variable $\hat{c}_i = \hat{\gamma}_i \exp(j \phi_i) \alpha_T\pb{\xi_i ;\theta } = \hat{\gamma}_i \exp(j \hat{\phi}_i)$ are as follows 
		\begin{align}\label{eq:subGaussNormRes} 
			&\norm{\hat{\gamma}_i \cos(\hat{\phi}_i)}_{\Psi_2}\leq \pb{\frac{N_cN_T}{N}}^{\frac{1}{q^{*}} }\frac{1}{\sqrt{q^{*}}}, \\
			& \norm{\hat{\gamma}_i \sin(\hat{\phi}_i)}_{\Psi_2} \leq \pb{\frac{N_cN_T}{N}}^{\frac{1}{q^{*}} }\frac{1}{\sqrt{q^{*}}}, \\
			&q^{*} = \max\pb{1,2\log\pb{\frac{N}{N_TN_c}}}. \nn 
		\end{align}
	\end{lemma}
	\begin{IEEEproof}
		We consider the sub-gaussian norm of the real part and using the fact that $\hat{\gamma}_i$ and $\hat{\phi}_i$ are independent random variables, we have 
		\begin{align}
			\Expec{\abs{\hat{\gamma}_i \cos(\hat{\phi}_i)}^{q}}^{\frac{1}{q}}\frac{1}{\sqrt{q}} &= \pb{\Expec{\abs{\hat{\gamma}_i}^q} \Expec{\abs{\cos(\hat{\phi}_i)}^{q}}}^{\frac{1}{q}} \frac{1}{\sqrt{q}} \nn \\
			&\leq \pb{\Expec{\abs{\hat{\gamma}_i}^q}}^{\frac{1}{q}} \frac{1}{\sqrt{q}}. \nn 
		\end{align}
		The last inequality is a consequence of $\abs{\cos(\hat{\phi}_i)} \leq 1$ and by taking the supremum on both sides, we have 
		\begin{align}\label{eq:subGaussBound} 
			\norm{\hat{\gamma}_i \cos(\hat{\phi}_i)}_{\Psi_2} &\leq \norm{\hat{\gamma}_i}_{\Psi_2},\nn \\
			\norm{\hat{\gamma}_i \sin(\hat{\phi}_i)}_{\Psi_2} &\leq \norm{\hat{\gamma}_i}_{\Psi_2}. 
		\end{align}
		For the probability model \eqref{eq:probModel}, we have 
		\begin{align*}
			&\hat{\gamma}_i= 
			\begin{cases}
				0 &\mbox{ with probability } 1 -\frac{N_cN_T}{N} \\
				1 & \mbox{w.p. } \frac{N_cN_T}{N}. 
			\end{cases}
			\nn\\
			&\Expec{\abs{\hat{\gamma}_i}^{q}}^{\frac{1}{q}} = \pb{ \frac{N_cN_T}{N}}^{\frac{1}{q}}. \nn \\
			\implies &\norm{\hat{\gamma}_i}_{\Psi_2} = \sup_{q \geq 1}\pb{ \frac{N_cN_T}{N}}^{\frac{1}{q}} \frac{1}{\sqrt{q}}. 
		\end{align*}
		The solution to this optimization problem can be found by taking the logarithm and solving the unconstrained optimization problem, which is given as 
		\begin{align*}
			q^{*} = 2\log\pb{\frac{N}{N_cN_T}}. 
		\end{align*}
		In order to satisfy the constraint, the solution is lower bounded by 1. Using the solution in \eqref{eq:subGaussBound} gives the desired bound in \eqref{eq:subGaussNormRes}. 
	\end{IEEEproof}
	\begin{lemma}\label{lem:HansonWrightCom} 
		Given a zero mean random vector $\mbf{c}$ composed of independent complex random variables $c_i = \gamma_i\exp\pb{j\phi_i}$, such that $\gamma_i, \phi_i$ are independent random variables with sub-Gaussian distributions such that $\norm{\gamma_i}_{\Psi_2} \leq K,$ ,for $i=1,\cdots,N$, we have 
		\begin{align}\label{eq:HansonWrightComplex} 
			&P\pb{\abs{\mbf{c}^{*} \mbf{B} \mbf{c} - \Expec{\mbf{c}^{*} \mbf{B} \mbf{c}}} > t} \leq \nn\\
			& 8 \exp \pb{-d\min \pb{\frac{t^2}{K^4 \norm{\mbf{B}}_F^2} ,\frac{t}{K^2 \norm{\mbf{B}}_{op}} } }, 
		\end{align}
		where $\mbf{B} \in \mbb{C}^{N \times N}$, $\mbf{c} \in \mbb{C}^{N}$, for some absolute constant $d >0$ and $\forall t >0$. 
	\end{lemma}
	\begin{IEEEproof}
		Let $\mbf{B} = \mbf{B}_{R} + j \mbf{B}_{I}$, where $\mbf{B}_R, \mbf{B}_{I} \in \mbb{R}^{N \times N}$. Let $\mbf{c} =\mbf{c_R} +j \mbf{c_I}$, where $\mbf{c}_R, \mbf{c}_{I} \in \mbb{R}^{N}$. 
		\begin{align*}
			P&\pb{\abs{\mbf{c}^{*} \mbf{B} \mbf{c} - \Expec{\mbf{c}^{*} \mbf{B} \mbf{c}}} > t} \leq \nn \\
			&P\pb{\abs{Re\pb{\mbf{c}^{*} \mbf{B} \mbf{c} - \Expec{\mbf{c}^{*} \mbf{B} \mbf{c}}}} > \frac{t}{\sqrt{2}}} +\nn \\
			&P\pb{\abs{Im\pb{\mbf{c}^{*}\mbf{B} \mbf{c} - \Expec{\mbf{c}^{*} \mbf{B} \mbf{c}}}} > \frac{t}{\sqrt{2}}}. 
		\end{align*}
		Using the definition of the random variables $c_i$, we can write 
		\begin{align}\label{eq:coh1} 
			&P\pb{\abs{Re\pb{\mbf{c}^{*} \mbf{B} \mbf{c} - \Expec{\mbf{c}^{*} \mbf{B} \mbf{c}}}} > t/sqrt{2}} \leq \nn \\
			&P\pb{\abs{\mbf{c}_R^T \mbf{B}_R\mbf{c}_R - \Expec{\mbf{c}_R^T \mbf{B}_R \mbf{c}_R^T}} \geq t/\pb{2\sqrt{2}} } + \nn \\
			&P\pb{\abs{\mbf{c}_I^T \mbf{B}_R\mbf{c}_I - \Expec{\mbf{c}_I^T \mbf{B}_R \mbf{c}_I^T}} \geq t/\pb{2\sqrt{2}} }, \\
			\label{eq:coh2} &P\pb{\abs{Im\pb{\mbf{c}^{*} \mbf{B} \mbf{c} - \Expec{\mbf{c}^{*} \mbf{B} \mbf{c}}}} > t/sqrt{2}} \leq \nn \\
			&P\pb{\abs{\mbf{c}_R^T \mbf{B}_I\mbf{c}_R - \Expec{\mbf{c}_R^T \mbf{B}_I \mbf{c}_R^T}} \geq t/\pb{2\sqrt{2}} } + \nn \\
			&P\pb{\abs{\mbf{c}_I^T \mbf{B}_I\mbf{c}_I - \Expec{\mbf{c}_I^T \mbf{B}_I \mbf{c}_I^T}} \geq t/\pb{2\sqrt{2}} }. 
		\end{align}
		Each term in \eqref{eq:coh1}, and \eqref{eq:coh2} can be bounded by using the Hanson-Wright inequality for quadratic forms of real random variables presented in \cite{HansonWrightSubGaussian}. This provides the desired inequality in \eqref{eq:HansonWrightComplex}. 
	\end{IEEEproof}
	\begin{theorem}\label{thm:supremaChaoseProc}
		Let $\mbb{T}  =\cb{\mbf{V_x} : \mbf{x} \in \mc{D}_{K,NN_{\theta}}}$ be a set of matrices, and let $\mbf{c}$ be a random vector whose entries $c_j$ are independent, mean-zero, variance 1, and L-subgaussian random variables. Set
		\begin{align}
			E &= \gamma_2(\mbb{T}, \norm{.}_{op}) \pb{\gamma_2(\mbb{T}, \norm{.}_{op}) + d_F (\mbb{T})} + d_F (\mbb{T})d_{op}(\mbb{T}), \\
			V &= d_{op}(\mbb{T})(\gamma_2(\mbb{T}, \norm{.}_{op}) + d_F (\mbb{T})), \nn \\
			  U &= d^2_{op}(\mbb{T}).
			\end{align}
Then for $ t > 0$,  \begin{align}
		&P\pb{\sup_{\mbf{V_x} \in \mbb{T}} \abs{\norm{ \mbf{V}_x \mbf{c}}_2^2 - \Expec{\norm{\mbf{ \mbf{V}_x \mbf{c}}}_2^2}} > \epsilon_1 E + t}  \nn\\
		\leq &\exp\pb{-\epsilon_2 \min\pb{\frac{t^2}{V^2},\frac{t}{U} }}. 
		\end{align}	
	 The constants $\epsilon_1$, $\epsilon_2$ depend only on $L$.
		\end{theorem}
	Proof is given in Theorem $3.2$ in \cite{Chaos_Rauhut2014}.
\end{appendices}

\section*{Acknowledgment} This research was partially supported by Army Research Office grant W911NF-11-1-0391 and NSF Grant IIS-1231577.


\end{document}